%% file: guajardo-2013-jelechem-accepted_manuscript.tex
\documentclass[a4paper, 10pt, preprint, review, fleqn]{elsarticle} \usepackage[width=167mm]{geometry}

\pdfoutput=1

\usepackage[T1]{fontenc}
\usepackage{amssymb}

\usepackage[english]{babel}

\usepackage{subfigure}
\usepackage{subeqnarray}
\usepackage{color}

\usepackage{doi}  

\usepackage{siunitx}
\usepackage[version=3]{mhchem} 

\usepackage{ccicons}

\makeatletter
\def\ps@pprintTitle{%
\let\@oddhead\@empty
\let\@evenhead\@empty
\def\@oddfoot{\footnotesize\itshape
Accepted Manuscript submitted to 
 \ifx\@journal\@empty Elsevier
\else\@journal\fi \hfill July 11, 2013}%
\let\@evenfoot\@oddfoot}
\makeatother

\begin{document}

  \hypersetup{
    bookmarksnumbered=true,
    linktocpage=true,
    pdfauthor={Cristian Guajardo, Sirimarn Ngamchana, Werasak Surareungchai},
    pdftitle={Mathematical modeling of interdigitated electrode arrays in finite electrochemical cells}
  }

  \newtheorem{observacion}{Remark}[section]
  \newtheorem{teorema}{Theorem}[section]

  \newcommand{\abs}[1]{\left\vert#1\right\vert}
  \newcommand{\set}[1]{\left\{#1\right\}}
  \newcommand{\bigpar}[1]{\left(#1\right)}
  \newcommand{\bigcuad}[1]{\left[#1\right]}

  \newcommand{\parderiv}[2]{\frac{\partial #1}{\partial #2}}
  \newcommand{\ud}[1]{\,\mathrm{d}#1}

  \newcommand{\tramo}[1]{\left\{\begin{array}{cl} #1 \end{array}\right.}

  \newcommand{\marca}[1]{#1}

  \newcommand{\especie}{\sigma}
  \newcommand{\especieDet}{\ell}

  \newcommand{\Mjournal}{Journal of Electroanalytical Chemistry}
  \newcommand{\Mpublication}{Mathematical modeling of interdigitated electrode arrays in finite electrochemical cells}
  \newcommand{\Mvolume}{705}
  \newcommand{\Missue}{-}
  \newcommand{\MdateYYYYMMDD}{2013-09-15}
  \newcommand{\Mdoinumber}{10.1016/j.jelechem.2013.07.014}

  \onecolumn \thispagestyle{empty} \setcounter{page}{0}
  
  \begin{center}
    \begin{tabular}{p{0.16\textwidth}p{0.76\textwidth}}
      {\LARGE \ccbyncnd} & Copyright \textcopyright\ 2013. This manuscript version is made available under the license \url{http://creativecommons.org/licenses/by-nc-nd/4.0}.
    \end{tabular}
  \end{center}
    
  \begin{description}
	\item[NOTICE.] This is the author's version of a work that was accepted for publication in \emph{\Mjournal}. Changes resulting from the publishing process, such as peer review, editing, corrections, structural formatting, and other quality control mechanisms may not be reflected in this document. Changes may have been made to this work since it was submitted for publication. 
	
	A definitive version was subsequently published in \emph{\Mpublication}. \Mjournal, vol. \Mvolume, issue \Missue, \MdateYYYYMMDD. \doi{\Mdoinumber}.
  \end{description}

  \begin{footnotesize}
    See Elsevier's sharing policies at \url{https://www.elsevier.com/about/company-information/policies/sharing}
  \end{footnotesize}
  \clearpage

  \journal{Journal of Electroanalytical Chemistry}
  \include{document}

  
  \include{appendix}

\end{document}

%% file: document.tex

\begin{frontmatter}
  \title{Mathematical Modeling of Interdigitated Electrode Arrays in Finite Electrochemical Cells}

  \author[pdti]{Cristian Guajardo\corref{cor1}}
  \ead{cristian.gua@kmutt.ac.th}

  \author[be]{Sirimarn Ngamchana}

  \author[bep]{Werasak Surareungchai}

  \cortext[cor1]{Corresponding author. Tel: +66~2~4707562; Fax: +66~2~4523455.}

  \address[pdti]{Pilot Plant Development and Training Institute}
  \address[be]{Biochemical Engineering and Pilot Plant Research and Development Unit, National Center for Genetic Engineering and Biotechnology, National Sciences and Technology Development Agency}
  \address[bep]{School of Bioresources and Technology, and Biological Engineering Program}
  \address{King Mongkut's University of Technology Thonburi, 49 Soi Thianthale 25, Thanon Bangkhunthian Chaithale, Bangkok 10150, Thailand}

  \begin{abstract}
    Accurate theoretical results for interdigitated array of electrodes (IDAE) in semi-infinite cells can be found in the literature. However, these results are not always applicable when using finite cells. In this study, theoretical expressions for IDAE in a finite geometry cell are presented. At known current density, transient and steady state concentration profiles were obtained as well as the response time to a current step. Concerning the diffusion limited current, a lower bound was derived from the concentration profile and an upper bound was obtained from the limiting current of the semi-infinite case. The lower bound, which is valid when Kirchhoff's current law applies to the unit cell, can be useful to ensure a minimum current level during the design of the electrochemical cell. Finally, a criterion \marca{was developed defining when the behaviors of finite and semi-infinite cells are comparable}. This allows to obtain higher current levels in finite cells, approaching that of the semi-infinite case. Examples with simulations were performed in order to illustrate and validate the theoretical results.
  \end{abstract}

\end{frontmatter}

\section{Introduction}
Among micro- and nanoelectrodes, the \emph{interdigitated array of electrodes} (IDAE) is one of the most common configurations and has drawn great attention since it can produce high currents from the redox cycling/feedback in between closely arranged generators and collectors \cite{Aoki1993, Cohen2000, Iwasaki1995, Yang2005}.
In order to obtain proper designs of IDAE, fundamental understanding of the transport of electrochemical species in between electrodes is required. Many authors have used numerical simulations to understand this working principle \cite{Yang2005, Aoki1989, Jin1996, Morf2006}.
Also theoretical results are available \cite{Morf2006, Aoki1988, Aoki1990}. The most significant of these results was obtained by Aoki \cite{Aoki1988, Aoki1990}, where exact expressions for the current-potential curves and limiting current in steady state were obtained for reversible and irreversible electrode reactions. Later, Morf and colleagues \cite{Morf2006} did a theoretical revision of Aoki's results for the case of reversible electrode reactions with internal/external counter electrode.

All of the results previously mentioned consider that the IDAE is subject to semi-infinite geometry, which means that the ratio between the `height of the cell' and the center-to-center `separation of the electrodes' is very large. This is not always true, as one can see in the case of some microfluidic devices where `channel height' and `electrodes separation' are of comparable size \cite{Goluch2009a, Lewis2010, Chen2011, Daniel2005}, especially when using low cost fabrication techniques or materials.
Soft lithography and the use of transparency sheet masks are examples of simple and inexpensive techniques commonly used for fabricating microfluidic devices \cite{Duffy1998,Whitesides2001}. When using soft lithography, the channel height of microfluidic devices is determined by the thickness of the photoresist mold, which can vary in between \SI{1}{\micro\metre}--\SI{200}{\micro\metre} \cite{Duffy1998}. When using photolithography and transparency sheet masks, the electrodes are constrained by the resolution of the transparency sheet mask, which can generate features between \SI{20}{\micro\metre}--\SI{50}{\micro\metre} when using a printer operating at \SI{3380}{dpi}--\SI{5080}{dpi} \cite{Duffy1998,Whitesides2001}. Therefore, the ratio between the `height of the cell' and the center-to-center `separation of the electrodes' obtained using these techniques is clearly finite and may vary between $\sim\num{0,01}-\num{10}$.

Electrochemical applications \cite{Chen2011, Daniel2005, Goral2006, Hayashi2003, Kurita2000, Kwakye2006, Amatore2004, Bjorefors2000} and research through simulations \cite{Amatore2010, Anderson1985, Fosdick1986, Ou1988} have been reported for IDAE in {continuous flow} microfluidic devices, which take into account the height of the channel and verify the dependence of the current with respect to the flow rate.
Despite these researches, it is known from  previous reports that signal amplification by redox cycling increases with decreasing flow rate, being most effective with stagnant solutions \cite{Bjorefors2000, Morita1997, Niwa1995}.

Experiments \cite{Goluch2009a, Lewis2010} and simulations \cite{Goluch2009a, Morita1997, Strutwolf2005} have been conducted in microfluidic channels with stagnant solutions, establishing that higher currents are obtained for higher microchannels. The current approaches similar values to the case of semi-infinite cells when the `height of the microchannel' is larger than the `width of the electrodes'. Nevertheless, there is neither mention of analytical equations that can predict the current in small volume cells nor analytical criteria to determine quantitatively when these microfluidic cells can be regarded as semi-infinite.

This report aims to establish a theoretical study of IDAE in a finite geometry cell  with stagnant fluid, which can be useful for static fluid electrochemistry in microfluidic devices. By considering a repeating unit cell with internal counter electrode, transient and steady state Fourier series representations of the concentration profile are obtained as a function of the current density. A criterion to estimate the response time to a current step is also obtained. A simple lower bound expression for the limiting current is calculated, which can help to ensure a minimum current level during the design of the electrochemical cell. Finally, a criterion \marca{is developed establishing the conditions under which finite and semi-infinite cells have comparable behaviors. This} would be useful \marca{in finite cells} to obtain current levels that approach that of the semi-infinite case and also would allow to apply the results in \cite{Morf2006,Aoki1988,Aoki1990}.

\section{Theory}
\subsection{Definition of the problem}
\label{problema}
\begin{figure*}
  \centering
  \subfigure[][]{
    \includegraphics{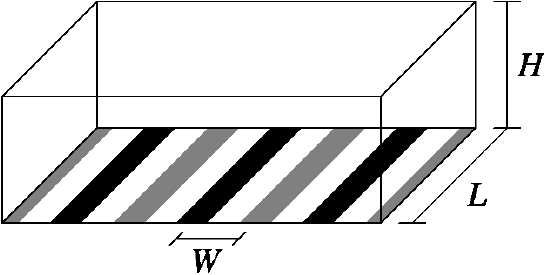}
    \label{problema:fig:celda_ideal}
  }%
  \subfigure[][]{
    \includegraphics{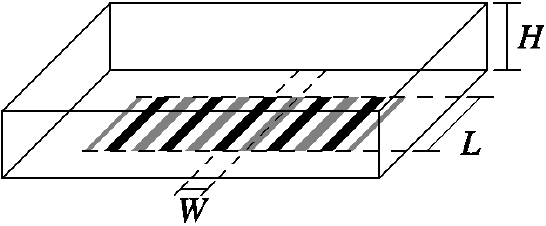}
    \label{problema:fig:celda_practica}
  }%
  \subfigure[][]{
    \includegraphics{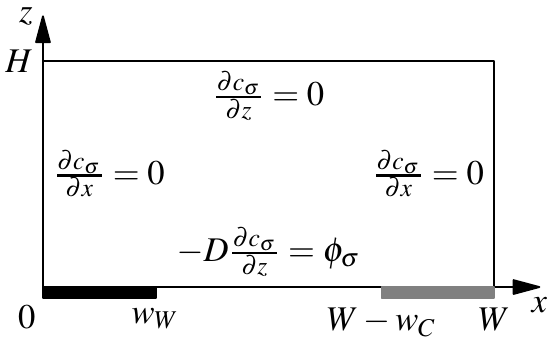}
    \label{problema:fig:celda_unitaria}
  }
  \caption{Conceptual sketch of interdigitated array of electrodes (IDAE) in a finite geometry cell. \subref{problema:fig:celda_ideal} Ideal case where the IDAE fits exactly in the electrochemical cell. \subref{problema:fig:celda_practica} More practical case of an IDAE configuration. \subref{problema:fig:celda_unitaria} Two-dimensional unit cell of finite height $H$, width $W$, and working and counter half electrodes of width $w_{W}$ and $w_{C}$ respectively: Fig \subref{problema:fig:celda_ideal} can be modeled by this 2D unit cell provided that the first and the last microband of the IDAE have half widths. Fig \subref{problema:fig:celda_practica} can be modeled by this 2D unit cell provided that the IDAE consists of a large amount of microbands and the length $L$ of each microband is long enough.}
  \label{problema:fig:celdas}
\end{figure*}

Consider an electrochemical cell with finite height $H$ as illustrated in Fig. \ref{problema:fig:celda_ideal}, where the walls are perfect insulators, and the  working (black) and counter (gray) electrodes are arranged as an \emph{interdigitated array of electrodes} (IDAE). Each microband of the working and counter electrodes has a width of $2w_{W}$ and $2w_{C}$ respectively, the center-to-center separation between consecutive microbands is $W$ and their length is $L$. Inside this cell there is oxidized species $\mathcal{O}$ and reduced species $\mathcal{R}$, which react at the surface of the electrodes according to
\begin{equation}
  \mathcal{O} + n_e\, \ce{e^- <=>} \mathcal{R},\quad \phi_{\mathcal{R}}(x,t) = -\phi_{\mathcal{O}}(x,t)
  \label{problema:ec:reaccion}
\end{equation}
where $\phi_{\especie}(x,t)$ is the generation rate of the species $\especie\in\set{\mathcal{O},\mathcal{R}}$ on the electrodes. Also assume that diffusion is the only available way for transporting the species $\mathcal{O}$ and $\mathcal{R}$, which have the same diffusion coefficient $D$. 

If the first and the last microbands of the IDAE have half width, then the cell in Fig. \ref{problema:fig:celda_ideal} can be regarded as a simple assembly of two-dimensional unit cells, like the one shown in Fig. \ref{problema:fig:celda_unitaria}. This unit cell consists of an upper wall, half microbands of working and counter electrodes at the bottom, and left and right walls representing symmetry boundaries or actual walls. 

The mathematical model for the transport of the species $\especie$ inside the unit cell is given by
\begin{subeqnarray}
  \label{problema:ec:c-PDE}
  \frac{1}{D} \parderiv{c_{\especie}}{t}(x,z,t) &=& \parderiv{^2c_{\especie}}{x^2}(x,z,t) + \parderiv{^2c_{\especie}}{z^2}(x,z,t)
    \slabel{problema:ec:c-PDE-difusion}\\
  c_{\especie}(x,z,0^{-}) &=& c_{\especie,0}(x,z)
    \slabel{problema:ec:c-PDE-ci}\\
  \parderiv{c_{\especie}}{x}(0,z,t) &=& 0, \qquad
  \parderiv{c_{\especie}}{x}(W,z,t) = 0
    \slabel{problema:ec:c-PDE-izda-dcha}\\
  \parderiv{c_{\especie}}{z}(x,H,t) &=& 0
    \slabel{problema:ec:c-PDE-arriba}\\
  f_{\especie}\bigpar{c_{\especie},\parderiv{c_{\especie}}{z},x,t} &=& 0
    \slabel{problema:ec:c-PDE-abajo}
\end{subeqnarray}
where both species must be related by $\phi_{\mathcal{R}}(x,t) = -\phi_{\mathcal{O}}(x,t)$ and each equation represents: transport by diffusion (\ref{problema:ec:c-PDE-difusion}), initial concentration distribution (\ref{problema:ec:c-PDE-ci}), left/right symmetry/insulation boundary (\ref{problema:ec:c-PDE-izda-dcha}), top insulation boundary (\ref{problema:ec:c-PDE-arriba}) and a generic bottom boundary (\ref{problema:ec:c-PDE-abajo}).

For this problem it is also assumed that the initial condition $c_{\especie,0}(x,z)$ comes from a previous steady state, i.e. 
\begin{subeqnarray}
  \label{problema:ec:ci-PDE}
  0 &=& \parderiv{^2c_{\especie,0}}{x^2}(x,z) + \parderiv{^2c_{\especie,0}}{z^2}(x,z)\\
  \parderiv{c_{\especie,0}}{x}(0,z) &=& \parderiv{c_{\especie,0}}{x}(W,z) = 0\\
  \parderiv{c_{\especie,0}}{z}(x,H) = 0, && f_{\especie,0}\bigpar{c_{\especie,0},\parderiv{c_{\especie,0}}{z},x} = 0
    \slabel{problema:ec:ci-PDE-arriba-abajo}
\end{subeqnarray}

In practical cases, the IDAE may not fit exactly in the cell as shown in Fig. \ref{problema:fig:celda_ideal}, but may look like the case in Fig. \ref{problema:fig:celda_practica}. This last case can still be modeled using Eq. (\ref{problema:ec:c-PDE}) provided some conditions \cite{Aoki1988}: (i) The length $L$ of the microbands is long enough so that the problem can be considered in 2D. (ii) The IDAE is composed of a large amount of microband electrodes, so that the edge effects at both ends of the IDAE are negligible and it is still possible to consider a unit cell with symmetry boundary conditions.

\begin{observacion}
\label{problema:obs:ctotal-cte}
  The total concentration at any place in the cell is constant\footnote{here \emph{constant} means that there is no time-dependence and \emph{uniform} means that there is no space-dependence, as it is usual when referring to fields and potentials with these characteristics.} in $t$ and uniform in \marca{$(x,z)$}, i.e.
  $$ c_\mathcal{O}(x,z,t) + c_\mathcal{R}(x,z,t) = c_0,\quad \forall(x,z) \mbox{ and } t\geq 0$$
  where $c_0$ is a real constant. This is due to the fact that both electrochemical species share the same diffusion coefficient and that the sum of the generation rates of both species is zero on the electrodes. Analogous results can be found in \cite{Aoki1988} and \cite[p. 254]{Oldham1994}. See \ref{apendice:c_total} for a general proof.
\end{observacion}

\begin{observacion}
\label{problema:obs:c-promedio}
  In case Kirchhoff's current law is satisfied inside the unit cell $\forall t$ (for example when the unit cell includes a counter electrode), then the `average concentration of the species $\especie$' (along the $x$ axes) is uniform in $z$, constant in $t$ and equal to $\bar{c}_{\especie,0}$
  \begin{displaymath}
    \frac{1}{W} \int_{0}^{W} c_{\especie}(x,z,t) \ud{x} = \bar{c}_{\especie,0},\quad \forall z \mbox{ and } t\geq 0
  \end{displaymath}
  where $\bar{c}_{\especie,0}$ is a real constant and corresponds to the `average of the initial concentration of the species $\especie$' (along the $x$ axes)
  \begin{displaymath}
    \bar{c}_{\especie,0} := \frac{1}{W} \int_0^W c_{\especie,0}(x,z) \ud{x},\quad \forall z
  \end{displaymath}
  and satisfies $\bar{c}_{\mathcal{O},0} + \bar{c}_{\mathcal{R},0} = c_0$. See \ref{apendice:idae-solucion_general}.
\end{observacion}

\subsection{Concentration profile for known current density}
\label{densidad-conocida}
In the problem of Eqs. (\ref{problema:ec:c-PDE}), the bottom boundary condition (\ref{problema:ec:c-PDE-abajo}) contains the equations for the electrodes and insulation that separates such electrodes. Using Nernst or Butler-Volmer equation for the electrodes leads to a problem containing a `mixed bottom boundary', which is more difficult to solve. In order to avoid this `mixture', the current density is assumed to be known, so the complete bottom boundary (electrodes and insulation) can be stated in terms of the concentration gradient. 

When the inward current density $j(x,t)$ is known, the generation rate $\phi_{\especie}(x,t)$ of the species $\especie$ on the surface of the electrodes is also known since $j(x,t) =  F n_{e} \phi_{\mathcal{O}}(x,t) = -F n_{e} \phi_{\mathcal{R}}(x,t)$, thus 
\begin{displaymath}
  \phi_{\mathcal{O}}(x,t) = -\phi_{\mathcal{R}}(x,t) = \tramo{
    \frac{j(x,t)}{F n_{e}} & \mbox{on the electrodes}\\
    0 & \mbox{out of the electrodes}
  }
\end{displaymath}
where $F$ is the Faraday's constant.
Therefore, the bottom boundaries in Eq. (\ref{problema:ec:c-PDE-abajo}) and in Eq. (\ref{problema:ec:ci-PDE-arriba-abajo}) can be written as
\begin{subeqnarray}
  \label{densidad-conocida:ec:cb-abajo}
  f_{\especie}\bigpar{c_{\especie},\parderiv{c_{\especie}}{z},x,t} \hspace{-0.5em}&:=&\hspace{-0.5em}
    D \parderiv{c_{\especie}}{z}(x,0,t) + \phi_{\especie}(x,t) = 0\\
  f_{\especie,0}\bigpar{c_{\especie,0},\parderiv{c_{\especie,0}}{z},x} \hspace{-0.5em}&:=&\hspace{-0.5em}
    D \parderiv{c_{\especie,0}}{z}(x,0) + \phi_{\especie,0}(x) = 0
\end{subeqnarray}

These bottom boundaries define completely the concentration profile in the unit cell. Then the problem in Eqs. (\ref{problema:ec:c-PDE}) and (\ref{problema:ec:ci-PDE}) can be solved using the method of separation of variables, as shown in \ref{apendice:idae-solucion_general} and \ref{apendice:densidad-conocida}. The result for the concentration is stated in the following theorem

\begin{teorema}
  Consider the unit cell defined in \ref{problema}. If Kirchhoff's current law is satisfied in the unit cell $\forall t$, then the concentration $c_{\especie}(x,z,t) = c_{\especie,0}(x,z) + \Delta c_{\especie}(x,z,t)$ is given by the sum of the initial concentration
  \begin{subeqnarray}
    c_{\especie,0}(x,z) &=& 
      \bar{c}_{\especie,0} + \sum_{n=1}^{+\infty} b_{n}^{\especie,0}(z) \cos(n\pi x/W) \\
    b_{n}^{\especie,0}(z) &=& 
      G_{\phi}\bigpar{H-z,n^2\frac{\pi^{2}}{W^{2}}} \cdot \mathcal{I}_n \set{\frac{\phi_{\especie,0}}{D}}\\
    \mathcal{I}_n\set{\cdot} \hspace{-0.5em}&:=&\hspace{-0.5em}
      \frac{2}{W} \int_0^W \set{\cdot} \cos(n\pi x/W) \ud{x} 
      \slabel{densidad-conocida:ec:In}\\
    G_{\phi}(z,s) &=&
      \frac{\cosh(\sqrt{s}\,z)}{\sqrt{s} \sinh(\sqrt{s}\,H)}
      \slabel{densidad-conocida:ec:G_phi}
  \end{subeqnarray}
  and the change in concentration
  \begin{subeqnarray}
    \Delta c_{\especie}(x,z,t) \hspace{-0.5em}&=&\hspace{-0.5em}
      \sum_{n=1}^{+\infty} \Delta b_n^{\especie}(z,t) \cos(n\pi x/W)
      \slabel{densidad-conocida:ec:Dc-xzt}\\
    \Delta b_n^{\especie}(z,t) \hspace{-0.5em}&=&\hspace{-0.5em}
      g_\phi(H-z,Dt)\, \mathrm{e}^{-n^{2}\frac{\pi^{2}}{W^{2}} Dt} D * \mathcal{I}_n\!\set{\!\frac{\Delta\phi_{\especie}}{D}\!}\!(t)\\
    g_\phi(z,t) \hspace{-0.5em}&=&\hspace{-0.5em}
      \frac{1}{H} \bigcuad{1 + 2 \sum_{k=1}^\infty (-1)^k \mathrm{e}^{-k^2\frac{\pi^{2}}{H^{2}} t} \cos\bigpar{k\frac{\pi}{H}z}} 
      \slabel{densidad-conocida:ec:g_phi}
  \end{subeqnarray}
  where $\Delta\phi_{\especie} = \phi_{\especie} - \phi_{\especie,0}$, $*$ represents the time convolution and the Laplace inverse $g_\phi = \mathcal{L}^{-1} \set{G_\phi}$ can be obtained from tables, such as \cite[p.218]{Schiff1999} or \cite[Eq. (20.10.5)]{dlmf2010}, and it is given by the $4^{th}$ \emph{elliptic theta function}.
  
  Here the concentrations of both species have been obtained independently, but they must be related by Remark \ref{problema:obs:ctotal-cte}.
\end{teorema}

In the particular case when the current density is constant in $t$, the generation rate is also constant in $t$ $\phi_{\especie}(x,t) = \phi_{\especie}(x)$ and the coefficient $\Delta b_n^{\especie}(z,t)$ is given by a simpler expression
\begin{equation}
  \Delta b_n^{\especie}(z,t) = 
    \int_0^t g_\phi(H-z,D\tau)\, \mathrm{e}^{-n^{2}\pi^{2} D\tau/W^{2}} D \ud{\tau} \cdot \mathcal{I}_n\set{\frac{\Delta\phi_{\especie}}{D}}
  \label{densidad-conocida:ec:Db_n-z8}
\end{equation}
A `sufficiently long' time after applying this current step \marca{($t \to +\infty$)}, the total concentration stabilizes and reaches the steady state
\begin{equation}
  c_{\especie}(x,z,+\infty) =
    \sum_{n=1}^{+\infty} \mathcal{I}_n\!\set{\frac{\phi_{\especie}}{D}} G_{\phi}\bigpar{H-z,\frac{n^2\pi^{2}}{W^{2}}} \cos\bigpar{\frac{n\pi}{W}x} + \bar{c}_{\especie,0}
  \label{densidad-conocida:ec:c-xz8}
\end{equation}
where $\mathcal{I}_{n}$ and $G_{\phi}$ are defined in Eqs. (\ref{densidad-conocida:ec:In}) and (\ref{densidad-conocida:ec:G_phi}) respectively. 
\marca{This steady state equation applies not only to constant current density, but in general, it relates an steady state value of generation rate (current density) with an steady state value of concentration.} Like before, the validity of this result is subject to the condition that Kirchhoff's current law be satisfied in the unit cell $\forall t$.

The time $T_{ss}^{\phi}$ required to reach the steady state is related to the time constant $\tau_{\phi}$ of the slowest natural mode of $\Delta c_{\especie}(x,z,t)$. The slowest natural mode corresponds to $\exp(\pi^{2} Dt/W^{2})$ as shown in Eq. (\ref{densidad-conocida:ec:Db_n-z8}) when $n=1$ (see \ref{solucion-densidad-constante} for details), therefore
\begin{equation}
  T_{ss}^\phi \propto \tau_\phi = \frac{W^2}{\pi^2 D}
  \label{densidad-conocida:ec:tau_phi}
\end{equation}
This slowest natural mode decays to approximately $\num{1,8}\%$, $\num{0,7}\%$ and $\num{0,2}\%$ for $T_{ss}^\phi$ equal to $4\tau_\phi$, $5\tau_\phi$ and $6\tau_\phi$ respectively.

The error with respect to the steady state can be obtained by using Eqs. (\ref{densidad-conocida:ec:Dc-xzt}) and (\ref{densidad-conocida:ec:Db_n-z8}) and it is summarized below
\begin{teorema}
  Consider the unit cell defined in Section \ref{problema}, where the current density is constant in $t$ and Kirchhoff's current law holds inside the unit cell $\forall t$. If $t>\tau_\phi$ and the aspect ratio satisfies $H/W<1/2$, then the error with respect to the steady state is given by
  \begin{displaymath}
    \Delta c_{\especie}(x,z,t) - \Delta c_{\especie}(x,z,+\infty) \approx -\mathcal{I}_1\set{\frac{\Delta\phi_{\especie}}{D}} \frac{e^{-\pi^{2}Dt/W^{2}}}{H\pi^{2}/W^{2}} \cos\bigpar{\frac{\pi x}{W}}
  \end{displaymath}
  and follows exponential decay given by the time constant $\tau_\phi$, defined in Eq. (\ref{densidad-conocida:ec:tau_phi}). See \ref{solucion-densidad-constante} for details.
\end{teorema}

More precise results can be obtained for $T_{ss}^\phi$ when the unit cell has small aspect ratio and satisfies some symmetry conditions
\begin{teorema}
  Consider the unit cell defined in Section \ref{problema}, where the current density is constant in $t$ and Kirchhoff's current law holds inside the unit cell $\forall t$. If $t>\tau_\phi$, the aspect ratio is small $H/W<1/\pi$, and the microband electrodes have equal width and are located at the ends of the unit cell, then the relative error with respect to the steady state is roughly approximated by
  \begin{equation}
    \frac{\Delta c_{\especie}(x,z,t) - \Delta c_{\especie}(x,z,+\infty)}{\Delta c_{\especie}(x,z,+\infty)} \approx \frac{-e^{-\pi^{2}Dt/W^{2}}}{\cosh(\pi(H-z)/W)}
    \label{densidad-conocida:ec:error_rel-xzt}
  \end{equation}
  See \ref{solucion-densidad-constante} for details.
\end{teorema}

In this case the relative error of the concentration (with respect to the steady state) is maximum at $z=H$ and is approximately $-\num{1,8}\%$, $-\num{0,7}\%$ and $-\num{0,2}\%$ for $T_{ss}^\phi$ equal to $4\tau_\phi$, $5\tau_\phi$ and $6\tau_\phi$ respectively. Depending on the desired precision, $T_{ss}^\phi$ can be chosen as any of the times mentioned previously.

\subsection{Bounds for the limiting \marca{steady state} current}
With the result in Eq. (\ref{densidad-conocida:ec:c-xz8}), it is possible to obtain bounds for predicting the limiting \marca{steady state} current in a finite geometry cell. The limiting \marca{steady state} current is of importance in electrochemistry since it is normally present as plateaus in steady state voltammograms. Thus, these bounds can be useful as criteria for designing electrode configurations and for ensuring a minimum current level in the cell. \marca{The obtention of these bounds is outlined in this section and explained in detail in \ref{apendice:corrente-limite}.}

Consider the unit cell in Section \ref{problema}, where Kirchhoff's current law is satisfied $\forall t$, the electrodes have equal size ($w_{W} = w_{C} = w$) and the species $\especieDet\in\set{\mathcal{O},\mathcal{R}}$ is the species with lowest initial average concentration $\bar{c}_{\especieDet,0} = \min(\bar{c}_{\mathcal{O},0},\bar{c}_{\mathcal{R},0})$. If the unit cell is operating in steady state with the limiting current flowing through it\footnote{Note that the limiting current can be generated by applying extreme potentials at the electrodes}, then the concentration of the species $\especieDet$ is
\begin{displaymath}
  c_{\especieDet}(x,z,+\infty) - \bar{c}_{\especieDet,0} =
    \bar{\phi}_{\especieDet}^{\lim} \sum_{n\:\mathrm{odd}}\! \mathcal{I}_{n}\!\set{\frac{\varphi_{\lim}}{D}} G_{\phi}\!\bigpar{H-z,\frac{n^{2}\pi^{2}}{W^{2}}} \cos\bigpar{\frac{n\pi}{W}x}
\end{displaymath}
where $\mathcal{I}_{n}\!\set{\varphi_{\lim}/D} = 0$ for all even $n$ and
\begin{equation}
  \varphi_{\lim}(x) := \frac{\phi_{\especieDet}^{\lim}(x)}{\bar{\phi}_{\especieDet}^{\lim}},\quad
  \bar{\phi}_{\especieDet}^{\lim} := \frac{1}{w} \int_0^w \phi_{\especieDet}^{\lim}(x) \ud{x}
  \label{corrente-limite:ec:varphi-bar_phi}
\end{equation}
$\varphi_{\lim}(x)$ is the \emph{normalized generation rate} and $\bar{\phi}_{\especieDet}^{\lim}$ is the \emph{average generation rate} on half microband of the working electrode when the limiting current is flowing through the unit cell. This average generation rate is related to the limiting current by
\begin{equation}
  \abs{i_{\lim}} = NLwFn_{e} \abs{\bar{\phi}_{\especieDet}^{\lim}} = N_{W}L\,2wFn_{e} \abs{\bar{\phi}_{\especieDet}^{\lim}}
  \label{corriente-limite:ec:corriente-limite}
\end{equation}
where $N$ is the number of repeating unit cells and $N_{W}$ is the number of microbands of the working electrode.

\begin{figure*}
  \centering
  \includegraphics{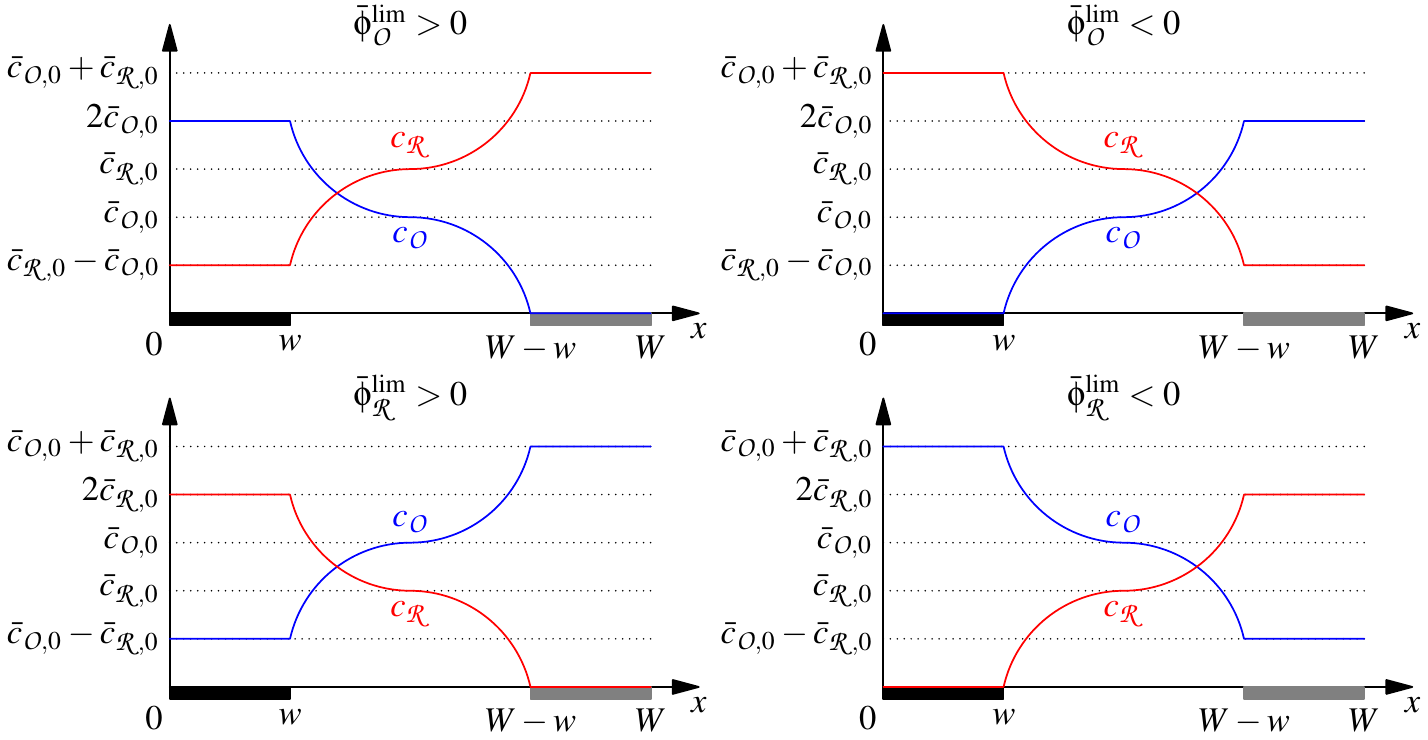}
  \caption{Sketch of the concentrations of oxidized and reduced species in the unit cell at the bottom boundary $z=0$ when the limiting current $\abs{i_{\lim}}$ circulates in the cell. The concentrations must be symmetric with respect to $x=W/2$ due to equal electrode sizes, the horizontal average of the concentration must be $\bar{c}_{\especie,0}$ due to Remark \ref{problema:obs:c-promedio} and the total concentration $c_\mathcal{O}(x,0,t)+c_\mathcal{R}(x,0,t)=c_0=\bar{c}_{\mathcal{O},0}+\bar{c}_{\mathcal{R},0}$ due to Remark \ref{problema:obs:ctotal-cte}. In the left figures $\bar{\phi}_{\especieDet}^{\lim} > 0$. In the right figures $\bar{\phi}_{\especieDet}^{\lim} < 0$. In the top figures, the initial average concentration of the oxidized species is the lowest. In the bottom figures, the initial average concentration of the reduced species is the lowest.}
  \label{corriente-limite:fig:concentracion-limite}
\end{figure*}

Fig. \ref{corriente-limite:fig:concentracion-limite} shows a sketch of the concentrations on the bottom boundary when the limiting current circulates in the cell. The shape of $c_{\especieDet}(x,0,+\infty) - \bar{c}_{\especieDet,0}$ in between the electrodes must be odd symmetric with respect to $W/2$, due to electrodes of equal width and working and counter currents of equal magnitude. The concentrations on the surface of the electrodes are obtained as follows: When $\bar{\phi}_{\especieDet}^{\lim} > 0$ on the working electrode, the concentration of the species $\especieDet$ on the counter electrode reaches the saturation value $0$, whereas the concentration on the working electrode reaches $2\bar{c}_{\especieDet,0}$ due to the average property in Remark \ref{problema:obs:c-promedio} and symmetry of the unit cell with respect to $x=W/2$. Analogously, when $\bar{\phi}_{\especieDet}^{\lim} < 0$ on the working electrode, the concentration of the species $\especieDet$ reaches $0$ on the working electrode.

Once the concentrations on the electrodes are known, one can integrate $c_{\especieDet}(x,0,+\infty) - \bar{c}_{\especieDet,0}$ along the working electrode for the cases where $\bar{\phi}_{\especieDet}^{\lim} > 0$ and $\bar{\phi}_{\especieDet}^{\lim} < 0$. This leads to the following relation
\begin{equation}
  \frac{\bar{c}_{\especieDet,0}w}{\abs{\bar{\phi}_{\especieDet}^{\lim}}} = \sum_{n\:\mathrm{odd}} \mathcal{I}_n \set{\frac{\varphi_{\lim}}{D}} G_{\phi}\bigpar{H,n^2\frac{\pi^{2}}{W^{2}}} \frac{\sin(n\pi w/W)}{(n\pi/W)}
  \label{corriente-limite:ec:bar_c-bar_phi}
\end{equation}
which can be bounded by
\begin{displaymath}
  \frac{\bar{c}_{\especieDet,0}w}{\abs{\bar{\phi}_{\especieDet}^{\lim}}} < \frac{wW}{2D\tanh(\pi H/W)}
\end{displaymath}
Therefore, the following theorem is obtained

\begin{teorema}
  For the unit cell described in Section \ref{problema}, assume that working and counter have microbands of identical width (located at both ends of the unit cell as in Fig. \ref{problema:fig:celdas}) and Kirchhoff's current law is satisfied $\forall t$ inside the unit cell (meaning that there is no external counter electrode). 
  If the unit cell is operating in steady state with limiting current circulating through it, then the limiting average generation rate $\abs{\bar{\phi}_{\especieDet}^{\lim}}$ is bounded from below by
  \begin{equation}
    \label{corriente-limite:ec:phi-cota}
    \abs{\bar{\phi}_{\especieDet}^{\lim}} > \frac{2D}{W}\tanh\bigpar{\pi\frac{H}{W}} \bar{c}_{\especieDet,0}\\
  \end{equation}
  where $\bar{c}_{\especieDet,0} = \min(\bar{c}_{\mathcal{O},0},\bar{c}_{\mathcal{R},0})$ is the average initial concentration of the determinant species $\ell$. Note that this result is independent of whether the bottom boundary is stated in terms of concentration, generation rate or both. For more details see \ref{apendice:corrente-limite}.
\end{teorema}

Due to the assumption that Kirchhoff's current law must be satisfied in the unit cell $\forall t$ and that the microbands of the electrodes have equal width, the limiting generation rate $\abs{\bar{\phi}_{\especieDet}^{\lim}}$ must depend on the initial average concentration $\bar{c}_{\especieDet,0}$ of the species $\especieDet$. The reason is that the current at the working and counter electrodes must be equal in magnitude $\forall t$ but with opposite sign, therefore the deviation of $c_{\especieDet}(x,z,t)$ on the working and counter electrodes with respect to $\bar{c}_{\especieDet,0}$ must be equal but in opposite directions. The higher the current that circulates through the electrodes, the higher the deviation of the concentration with respect to $\bar{c}_{\especieDet,0}$ on the electrodes. For this reason, only the species with lowest initial average concentration ($\especieDet$) must reach zero concentration on one of the electrodes, limiting the current that circulates through the unit cell. Therefore the species $\especieDet$ is the determinant species of the cell, since it is  directly related to the maximum current that the cell can handle. This dependence on the determinant species in the absence of external counter electrodes is also obtained for the case of semi-infinite geometries as shown in \cite[Section 2.3]{Morf2006}.

\begin{observacion}
  \label{corriente-limite:nota:cota-superior}
  Notice that the ratio $\bar{c}_{\especieDet,0}/\abs{\bar{\phi}_{\especieDet}^{\lim}}$ in Eq. (\ref{corriente-limite:ec:bar_c-bar_phi}) depends on the function defined in Eq. (\ref{densidad-conocida:ec:G_phi}) $G_{\phi}(H,n^{2}\pi^{2}/W^{2}) = (n\pi/W)^{-1}\tanh(n\pi H/W)^{-1}$ which decreases as $H/W$ increases. Also the lower bound $\abs{\bar{\phi}_{\especieDet}^{\lim}}$ in Eq. (\ref{corriente-limite:ec:phi-cota}) increases as $H/W$ increases, due to the behavior of the $\tanh(\pi H/W)$ term. These facts support the result obtained through simulations in \cite[Fig. 7]{Strutwolf2005}, which states that the generation rate $\abs{\bar{\phi}_{\especieDet}^{\lim}}$ (limiting current) increases as the unit cell aspect ratio $H/W$ increases. This means that $\lim_{H/W \to +\infty} \abs{\bar{\phi}_{\especieDet}^{\lim}}$ represents an upper bound\footnote{When taking the limit $H/W\to +\infty$, one should fix $W$ to any positive value and let $H\to +\infty$. This is to avoid convergence problems that may be caused by fixing $H$ and letting $W\to 0^{+}$.} for the limiting generation rate $\abs{\bar{\phi}_{\especieDet}^{\lim}}$ of finite aspect ratio cells
  \begin{displaymath}
    \abs{\bar{\phi}_{\especieDet}} \leq \abs{\bar{\phi}_{\especieDet}^{\lim}} \leq \lim_{H/W \to +\infty} \abs{\bar{\phi}_{\especieDet}^{\lim}}
  \end{displaymath}
  
  The value of the limiting generation rate (limiting current) for very high unit cell aspect ratios, was obtained first by Aoki and colleagues \cite{Aoki1988}, which is given approximately by
  \begin{equation}
    \lim_{H/W \to +\infty} \abs{\bar{\phi}_{\especieDet}^{\lim}} \approx 
    \frac{2D}{\pi w} \ln\bigcuad{\frac{8W}{\pi(W-2w)}}\bar{c}_{\especieDet,0}
    \label{corriente-limite:ec:aoki}
  \end{equation}
  and it is accurate within 4\% for $w/W\geq \num{0,4705}$ \cite[Eq. (32)]{Aoki1988}, which correspond to cases of very wide electrodes.
  Later this result was revisited by Morf and colleagues \cite{Morf2006}
  \begin{equation}
    \lim_{H/W \to +\infty} \abs{\bar{\phi}_{\especieDet}^{\lim}} \approx \frac{\pi D\bar{c}_{\especieDet,0}}{2w\ln\bigpar{\frac{4W}{\pi w}}}
    \label{corriente-limite:ec:morf}
  \end{equation}
  and it is accurate within 1\% for $w/W \leq 1/4$ \cite[Section 3.1]{Morf2006}, which correspond to the most relevant cases of electrodes.
\end{observacion}

\subsection{Approximating a semi-infinite geometry cell}
\label{semiinfinita}
The results in Eqs. (\ref{corriente-limite:ec:corriente-limite}), (\ref{corriente-limite:ec:aoki}) and (\ref{corriente-limite:ec:morf}) give a very accurate approximation for the limiting current when the unit cell has 'very high' aspect ratio $H/W$. In other hand, when the cell aspect ratio is not high, the limiting current can be bounded from above using Eq. (\ref{corriente-limite:ec:aoki}) or (\ref{corriente-limite:ec:morf}), and bounded from below using (\ref{corriente-limite:ec:phi-cota}), giving a reasonable estimation of the limiting current.

From the previous facts a key question arises: Which aspect ratio can be considered as `very high' and which not? It is known that semi-infinite cells (very big cells) contain a region of bulk concentration located at the end of the diffusion layer, `very far' from the electrodes. To mimic this in the finite geometry case, the cell should have a region of bulk concentration $\bar{c}_{\especie,0}$ at the furthest location from the electrodes ($z=H$), that means $c_{\especie}(x,H,+\infty)\approx \bar{c}_{\especie,0}$ for all $x$.

An expression for relative error of the steady state concentration with respect to the bulk concentration can be obtained from Eq. (\ref{densidad-conocida:ec:c-xz8}) with $z=H$ and considering equal electrode widths
\begin{displaymath}
  \abs{c_{\especieDet}(x,H,+\infty) - \bar{c}_{\especieDet,0}} =
    \abs{\bar{\phi}_{\especieDet} \sum_{\mathrm{odd}\:n} \mathcal{I}_n\!\set{\frac{\varphi}{D}} G_{\phi}\bigpar{0,\frac{n^2\pi^{2}}{W^{2}}} \cos\bigpar{\frac{n\pi}{W}x}}
\end{displaymath}
where $\especieDet$ is the determinant species, and $\bar{\phi}_{\especieDet}$ and $\varphi$ are defined analogously to Eq. (\ref{corrente-limite:ec:varphi-bar_phi}). The right hand side of this equation can be bounded by using $\abs{\bar{\phi}_{\especieDet}}\leq \lim_{H/W\to +\infty}\abs{\bar{\phi}_{\especieDet}^{\lim}}$ with Eq. (\ref{corriente-limite:ec:morf}), by bounding $\abs{\mathcal{I}_{n}\set{\varphi/D}}<4w/(DW)$ and by approximating $\sum_{\mathrm{odd}\: n} (\pi/W) G_{\phi}(0,n^{2}\pi^{2}/W^{2})$ with the first term of the series. Then an upper bound for the relative error of the concentration with respect to the bulk is obtained in the following theorem

\begin{teorema}
  \label{semiinfinita:teo:error-cota}
  Assume that the unit cell in Section \ref{problema} has working and counter electrodes of equal width (located at both ends of the unit cell) and Kirchhoff's current law is satisfied $\forall t$ (meaning that there is no external counter electrode).
  Then, at $z=H$, the relative error of the steady state concentration of species $\especieDet$ with respect to its bulk value is given by
  \begin{equation}
    \abs{\frac{c_{\especieDet}(x,H,+\infty) - \bar{c}_{\especieDet,0}}{\bar{c}_{\especieDet,0}}}
    \lesssim 2\bigcuad{\ln\bigpar{\frac{4W}{\pi w}} \sinh\bigpar{\pi \frac{H}{W}}}^{-1}
    \label{semiinfinita:ec:error-cota}
  \end{equation}
  when $w/W\leq 1/4$ and $H/W\geq 1/\pi$. More details can be found in \ref{apendice:concentracion-relativa}.
\end{teorema}

From the last theorem, a criterion to determine when a finite aspect ratio cell can be regarded as semi-infinite is obtained and presented below.
\begin{teorema}
  \label{semiinfinita:teo:condicion-semiinfinta}
  Assume that the unit cell in Section \ref{problema} has working and counter electrodes of equal width (located at both ends of the unit cell) and Kirchhoff's current law is satisfied $\forall t$ (meaning that there is no external counter electrode).
  
  If the width of the microband electrodes satisfy $w/W\leq 1/4$, then the finite electrochemical cell can be regarded as semi-infinte when $H/W \geq 3/\pi$, because the value for the concentration at $z=H$ is different in less than $12\%$ compared to the bulk value.
  In case a better approximation is required, less than $\num{4,5}\%$ error with respect to the bulk value is obtained for $H/W \geq 4/\pi$.
\end{teorema}

\section{Results and discussion}
\subsection{Example of a current controlled electrochemical cell}
\label{ejemplo-celda}
The main purpose of this example is to examine whether the transient concentration profile in Eqs. (\ref{densidad-conocida:ec:Dc-xzt}) and (\ref{densidad-conocida:ec:Db_n-z8}), and the steady state concentration profile in Eq. (\ref{densidad-conocida:ec:c-xz8}) are correct. This was achieved by comparing the theoretical results with computer simulations.

Typical dimensions of microfluidic devices were considered for this example: channel height and width of $H=\SI{50}{\micro\metre}$ and $L=\SI{1}{\milli\metre}$ respectively. Also, working and counter electrodes (of moderate size) forming an IDAE pattern with $N=40$ unit cells\footnote{$N=40$ unit cells corresponds to $N_{W}=20$ microbands of working electrode.} were used with electrodes half width of $w=\SI{25}{\micro\metre}$ and center-to-center separation of $W=\SI{100}{\micro\metre}$. The redox couple used in this example is the standard ferri/ferrocyanide
\begin{displaymath}
  \ce{[Fe(CN)6]^{3-} + e^- <=> [Fe(CN)6]^{4-}}
\end{displaymath}
with diffusion constant of $D = \SI{7e-10}{\metre\squared\per\second}$ \cite{Yang2005} and initial concentrations of $c_{\mathcal{O},0}(x,z) = c_{\mathcal{R},0}(x,z) = \SI{0,5}{\mole\per\metre\cubed}$.

This example consisted of applying a constant current of $\abs{i(t)}=\SI{1}{\micro\ampere}$ to the total electrochemical cell. For simplicity in the calculations and simulations, it is assumed that the current density $j(x,t)$ is uniform on the surface of each electrode $\abs{j(x,t)} = {\abs{i(t)}}/({NLw}) = \SI{1}{\ampere\per\metre\squared}$. \marca{This assumption is highly restrictive, since in reality uniform current densities are unlikely to occur except in the limit of very small currents.}

The numerical simulations were carried out by using an exponential mapped mesh, in order to provide higher resolution near the edges of the electrodes. The mesh was incrementaly refined until the first three decimal places of the concentration did not change. See \ref{apendice:s31} for more details on the simulation setup. The concentration profile was obtained for only one of the species $\especie\in\set{\mathcal{O},\mathcal{R}}$, while the concentration profile of the other species can be obtained by using the relation $c_{\mathcal{O}}(x,z,t) + c_{\mathcal{R}}(x,z,t) = \SI{1}{\mole\per\metre\cubed}$ in Remark \ref{problema:obs:ctotal-cte}.

\begin{figure*}
  \centering
  \subfigure[][]{
    \includegraphics{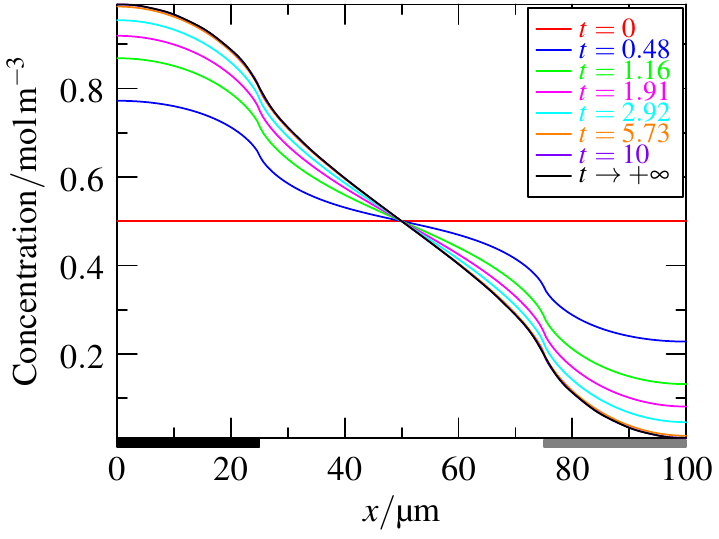}
    \label{ejemplo-celda:fig:conc-electrodos-simu}
  }
  \subfigure[][]{
    \includegraphics{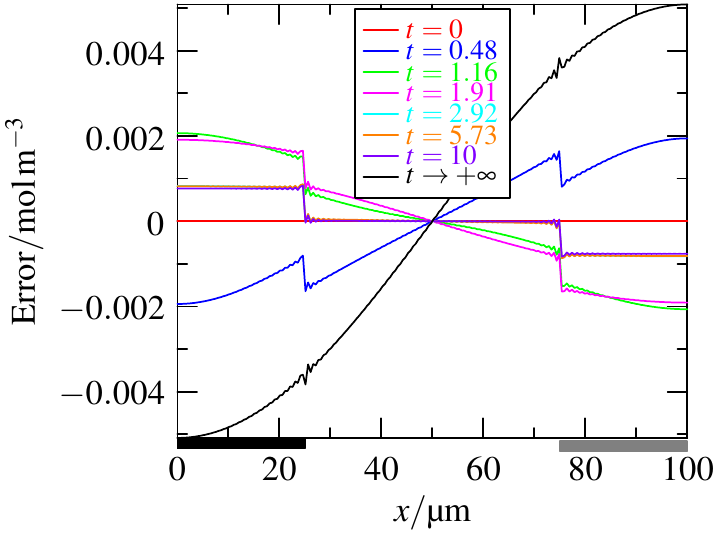}
    \label{ejemplo-celda:fig:conc-electrodos-error}
  }
  \caption{Concentration $c_{\especie}(x,z,t)$ on the electrodes' surface $z=0$ for different values of $t$. \subref{ejemplo-celda:fig:conc-electrodos-simu} Colored lines: Simulations using finite element solver for times between $t=0$ and $t=\SI{10}{\second}$. Black line: Theoretical value for $t\to +\infty$ in Eq. (\ref{densidad-conocida:ec:c-xz8}) using partial sums up to $n=201$. \subref{ejemplo-celda:fig:conc-electrodos-error} Colored lines: Error of the simulations with respect to the theoretical values in Eqs. (\ref{densidad-conocida:ec:Dc-xzt}) and (\ref{densidad-conocida:ec:Db_n-z8}) using partial sums up to $201$ and $200$ for $n$ and $k$ respectively, and times between $t=0$ and $t=\SI{10}{\second}$. Black line: Error of the simulation for $t=\SI{5,73}{\second}$ with respect to the theoretical steady state in Eq. (\ref{densidad-conocida:ec:c-xz8}) using partial sums up to $n=201$.}
  \label{ejemplo-celda:fig:conc-electrodos}
\end{figure*}

Fig. \ref{ejemplo-celda:fig:conc-electrodos} shows the concentration profile on the surface of the electrodes ($z=0$). Fig. \ref{ejemplo-celda:fig:conc-electrodos-simu} was obtained by simulating the time-dependent PDE in Eqs. (\ref{problema:ec:c-PDE}) and shows the evolution of the concentration between $t=0$ and $t=\SI{10}{\second}$ in colored lines, whereas the black line represents the theoretical steady state concentration obtained from Eq. (\ref{densidad-conocida:ec:c-xz8}). Here it is shown that the simulated values reach the theoretical steady state in approximately $\SI{5,73}{\second}$. This time approximately corresponds to $4\tau_\phi$ as it can be checked by Eq. (\ref{densidad-conocida:ec:tau_phi}). 

It is interesting to notice that even though the current density is uniform on the surface of both electrodes, the concentration is not uniform. The reason for this is that the edges of the electrodes are exposed to vertical and horizontal diffusion, in contrast to the centers of the electrodes which present only vertical difussion. This allows the species to escape/reach the edges easier than the center of the electrodes.

Fig. \ref{ejemplo-celda:fig:conc-electrodos-error} shows, in colored lines for $t\in[0,\SI{10}{\second}]$, the differences between the simulated concentrations and their theoretical counterparts obtained from Eqs. (\ref{densidad-conocida:ec:Dc-xzt}) and (\ref{densidad-conocida:ec:Db_n-z8}). These differences decreases as $t$ increases, reaching maximum errors of $\approx\SI{0,002}{\mole\per\metre\cubed}$ and $\approx\SI{0,001}{\mole\per\metre\cubed}$ for $t=\set{\SI{0,48}{\second},\,\SI{1,16}{\second},\,\SI{1,91}{\second}}$ and $t=\set{\SI{2,92}{\second},\,\SI{5,73}{\second},\,\SI{10}{\second}}$ respectively. The black line shows the difference between the simulated concentration for $t=\SI{5,73}{\second}$ and the theoretical steady state in Eq. (\ref{densidad-conocida:ec:c-xz8}) using partial sums up to $n=201$. This difference shows a maximum error of $\SI{0,005}{\mole\per\metre\cubed}$ at $x=0$ and $x=\SI{100}{\micro\metre}$. Also the change in concentration at $x=0$ from the initial value to the steady state corresponds to $\SI{0,990}{\mole\per\metre\cubed}-\SI{0,5}{\mole\per\metre\cubed}$ (see Fig. \ref{ejemplo-celda:fig:conc-electrodos-simu}), therefore
\begin{displaymath}
  \abs{\frac{\Delta c_{\especie}(0,0,\SI{5,73}{\second})-\Delta c_{\especie}(0,0,+\infty)}{\Delta c_{\especie}(0,0,+\infty)}} = \abs{\frac{\num{0,005}}{\num{0,990}-\num{0,5}}} = 1\%
\end{displaymath}
which approximately agrees with the $\num{0,7}\%$ obtained by using the criterion in Eq. (\ref{densidad-conocida:ec:error_rel-xzt}). The difference between the relative errors arises from the fact that the simulated cell has an aspect ratio of $H/W=1/2$ which is higher than the one required in Eq. (\ref{densidad-conocida:ec:error_rel-xzt}). Nevertheless, this $1\%$ relative error indicates that the time $t=\SI{5,73}{\second}$ can be considered as steady state.

From Fig. \ref{ejemplo-celda:fig:conc-electrodos-error} one can notice that the errors present very small oscillations in $x$, this is because the errors are differences of simulated and theoretical concentrations, the later being approximated by truncated Fourier series using partial sums. One can get rid of these oscillations by increasing the upper value of the index $n$ in the partial sums for Eqs. (\ref{densidad-conocida:ec:Dc-xzt}) and (\ref{densidad-conocida:ec:c-xz8}), obtaining more smooth errors.

Colored lines in Fig. \ref{ejemplo-celda:fig:conc-electrodos-error} show that the simulated concentrations are similar to their theoretical counterparts in two decimal places. This error can be reduced when the approximation of the theoretical concentrations is improved, for example by increasing the upper value of the index $k$ in the partial sums for Eqs. (\ref{densidad-conocida:ec:Dc-xzt}), (\ref{densidad-conocida:ec:g_phi}) and (\ref{densidad-conocida:ec:Db_n-z8}), and it can reach three decimal places of accuracy for $t\geq\SI{2,92}{\second}$ when using partial sums up to $k=400$. See \ref{apendice:s31} for aditional figures showing this effect.

Also one can notice from Fig. \ref{ejemplo-celda:fig:conc-electrodos-error} that the errors for $t\in[0,\SI{10}{\second}]$ are discontinuous at the edges of the electrodes, while the error with respect to the steady state (black line) is continuous but has small perturbations at the edges of the electrodes. The reason for this behavior is the use of partial sums for $n$ and $k$ when computing the errors between $t=0$ and $t=\SI{10}{\second}$. However, in the case of the error with respect to the steady state, there are partial sums only in the index $n$. Therefore, by increasing the upper value of the index $k$ in the partial sums, it is possible to decrease the size of the discontinuities, leaving a continuous function in the limit. See \ref{apendice:s31} for aditional figures showing this effect.

\begin{figure*}
  \centering
  \subfigure[][Concentration$/\si{\mole\per\metre\cubed}$ for $t=\SI{10}{\second}$. max: \SI{0,990}{\mole\per\metre\cubed}, min: \SI{0,010}{\mole\per\metre\cubed}.]{
    \includegraphics{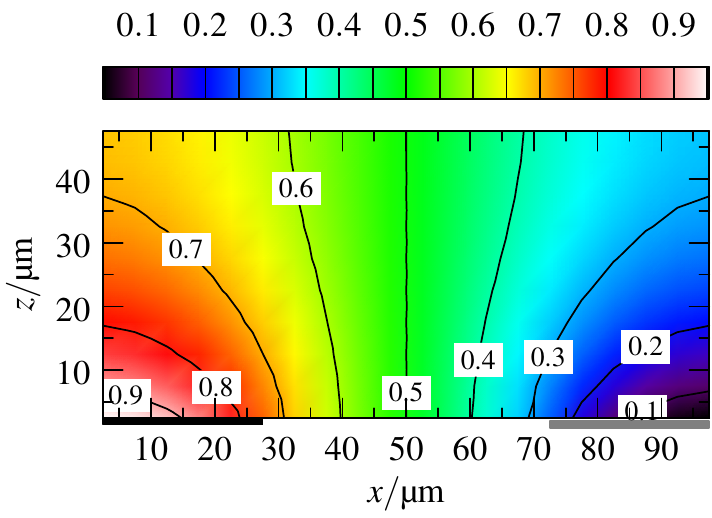}
    \label{ejemplo-celda:fig:conc-simu}
  }\hspace{1em}
  \subfigure[][Error$/\si{\mole\per\metre\cubed}$ between simulation and theoretical concentration. max: \SI{-0,0004}{\mole\per\metre\cubed}, min: \SI{0,0004}{\mole\per\metre\cubed}.]{
    \includegraphics{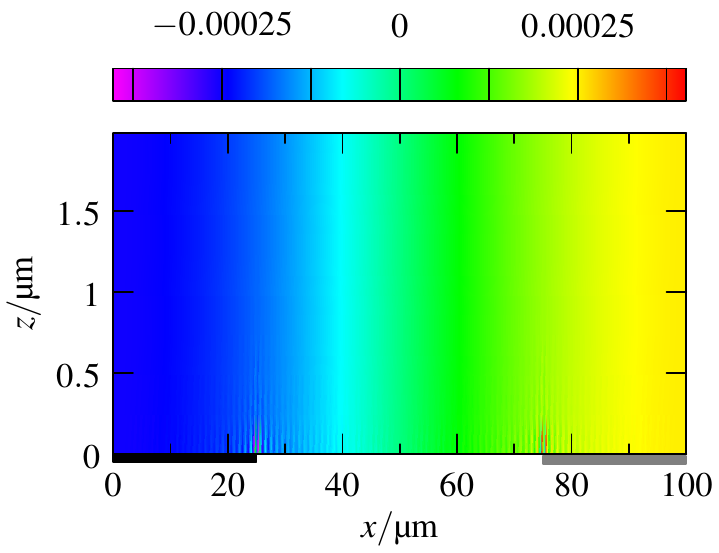}
    \label{ejemplo-celda:fig:conc-error}
  }
  \caption{Contour plot of the concentration profile $c_{\especie}(x,z,t)$ in steady state. \subref{ejemplo-celda:fig:conc-simu} Simulation using finite element solver for $t=\SI{10}{\second}$. \subref{ejemplo-celda:fig:conc-error} Error of the simulation ($t=\SI{10}{\second}$) with respect to the theoretical concentration in steady state Eq. (\ref{densidad-conocida:ec:c-xz8}) using partial sums up to $n=201$.}
  \label{ejemplo-celda:fig:conc}  
\end{figure*}

Fig. \ref{ejemplo-celda:fig:conc-simu} shows the concentration profile of the whole unit cell for $t=\SI{10}{\second}$ obtained by simulation (steady state), which reaches its maximum and minimum on the electrodes' surface. Unlike the cases of semi-infinite geometries, the concentration does not reach the bulk concentration at locations far from the electrodes, due to the low $H/W$ ratio of this electrochemical cell. Fig. \ref{ejemplo-celda:fig:conc-error} shows the difference between the simulation at $t=\SI{10}{\second}$ and the theoretical steady state concentration in Eq. (\ref{densidad-conocida:ec:c-xz8}) for $z\leq\SI{2}{\micro\metre}$ (for $z>\SI{2}{\micro\metre}$ the difference was smaller). Here it is possible to see the presence of small oscillations (as in the case of Fig. \ref{ejemplo-celda:fig:conc-electrodos-error}), which are more evident near the edges of the electrodes. This oscillations arise from the use of partial sums in the index $n$ when computing the steady state concentration, and they can be reduced by increasing the upper value of the index $n$ in the partial sums. See \ref{apendice:s31} for additional figures showing this phenomenon.

\subsection{Effect of the cell geometry in the concentration profile}
\label{ejemplo-concentracion}
The problem in Eq. (\ref{problema:ec:c-PDE}) was normalized to make it parameter independent
\begin{subeqnarray}
  \label{ejemplo-concentracion:ec:escala}
  \xi:=x/W & \tau:=t/\tau_\phi & \gamma_\especieDet:=\frac{c_\especieDet - \bar{c}_{\especieDet,0}}{\bar{c}_{\especieDet,0}} \\
  \zeta:=z/W & \tau_\phi := W^2/(\pi^2 D) & \widehat{\phi}_\especieDet := \frac{W}{\pi^2 D\bar{c}_{\especieDet,0}} \phi_\especieDet
\end{subeqnarray}
where it has been assumed that $\especieDet$ is the determinant electrochemical species in the cell such that $\bar{c}_{\especieDet,0} = \min(\bar{c}_{\mathcal{O},0},\bar{c}_{\mathcal{R},0})$. Therefore, the original problem and the normalized version are equivalent
\begin{eqnarray*}
  \frac{1}{D} \parderiv{c_\especieDet}{t} = \parderiv{^2c_\especieDet}{x^2} + \parderiv{^2c_\especieDet}{z^2} &\Leftrightarrow& \pi^2 \parderiv{\gamma_\especieDet}{\tau} = \parderiv{^2\gamma_\especieDet}{\xi^2} + \parderiv{^2\gamma_\especieDet}{\zeta^2}\\
  D\parderiv{c_\especieDet}{z} = \phi_\especieDet &\Leftrightarrow& \frac{1}{\pi^2} \parderiv{\gamma_\especieDet}{\zeta} = \widehat{\phi}_\especieDet
\end{eqnarray*}

Several simulations were carried out considering that the unit cell consists of only two electrodes, working and counter, both of the same half width and located at both ends of the unit cell. The initial concentration was set to $\gamma_\especieDet(\xi,\zeta,\marca{0^{-}})=0$ and a {constant and uniform generation rate} (current density) $\abs{\widehat{\phi}_\especieDet(\xi,\tau)}=1$ was applied to the electrodes. \marca{As stated previously, the main reasons to choose a uniform generation rate are to facilitate the simulation process and to facilitate the comparison of the simulation results against the theory. However, assuming a uniform generation rate is a severe limitation and practical conclusions cannot be drawn easily.} The rest of the parameters was varied in order to test the unit cell under different geometries and electrode widths. 

\marca{An} exponential mapped mesh was used for the simulations, in order to provide higher resolution near the edges of the electrodes. The mesh was incrementaly refined until the first three decimal places of the relative concentration did not change, see \ref{apendice:s32} for more details on the simulation setup.

The relative concentration $\gamma_\especieDet(\xi,\zeta,\tau)$ was obtained for only one of the species, the determinant species $\especieDet\in\set{\mathcal{O},\mathcal{R}}$, while the relative concentration of the other species can be obtained by $\bar{c}_{\mathcal{O},0}\gamma_{\mathcal{O}}(\xi,\zeta,\tau) = -\bar{c}_{\mathcal{R},0}\gamma_{\mathcal{R}}(\xi,\zeta,\tau)$. See \marca{Remarks} \ref{problema:obs:ctotal-cte} and \marca{\ref{problema:obs:c-promedio}}.

\begin{figure*}
  \centering
  \subfigure[][Relative concentration for $H/W=\num{0,2}/\pi$. $\gamma_{\max}=\num{12,611}$.]{
    \includegraphics{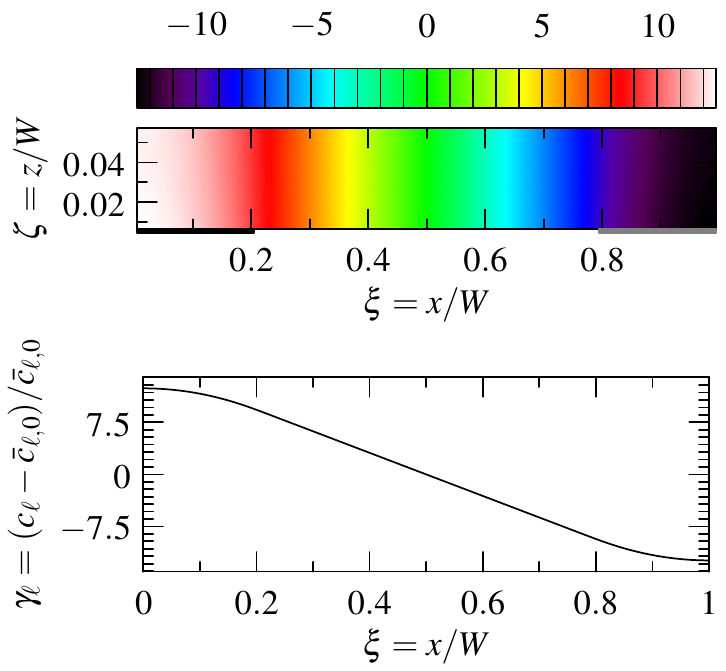}\hspace{1em}
    \label{ejemplo-concentracion:fig:distribucion-estacionaria-chica}
  }
  \subfigure[][Relative concentration for $H/W=\num{5}/\pi$. $\gamma_{\max}=\num{2,698}$.]{
    \includegraphics{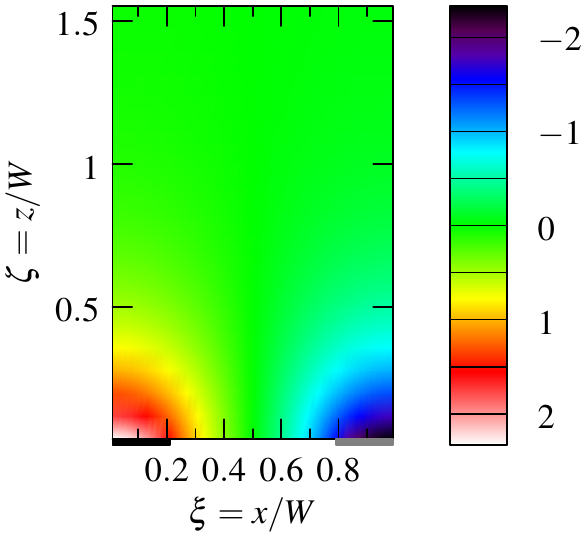}\hspace{2em}
    \label{ejemplo-concentracion:fig:distribucion-estacionaria-grande}
  }
  \caption{Relative concentration $\gamma_{\especieDet} = (c_{\especieDet}-\bar{c}_{\especieDet,0})/\bar{c}_{\especieDet,0}$ of the species $\especieDet$ for diferent aspect ratios when $\tau = \pi^{2}Dt/W^2 = 10$ (steady state), $w/W = \num{0,2}$ and $\abs{\widehat{\phi}_{\especieDet}} = \abs{W\phi_{\especieDet}}/(\pi^{2}D\bar{c}_{\especieDet,0}) = 1$ on the surface of the electrodes. Here $\gamma_{\max}$ stands for the maximum relative concentration in the whole cell and $-\gamma_{\max}$ for the minimum. In both pictures, the minimum relative concentration is below $-1$, which is a consequence of driving the cell at too high current.}
  \label{ejemplo-concentracion:fig:distribucion-estacionaria}
\end{figure*}

Fig. \ref{ejemplo-concentracion:fig:distribucion-estacionaria} shows the relative concentration in the whole unit cell for two different aspect ratios \marca{when steady state has been reached (approximated by $\tau = 10$)}. In the case of low aspect ratio $H/W=\num{0,2}/\pi$, the concentration never reaches the bulk value and seems not to depend on the vertical position, meaning that there is almost no vertical diffusion of the species. In contrast, there is a clear dependence on the horizontal position which resembles a $\cos(\pi x/W)$ as suggested previously, implying a high horizontal diffusion of species. 
In the case of high aspect ratio $H/W=5/\pi$, the concentration clearly reaches its bulk value far from the electrodes and also vertical and horizontal gradients are clearly shown. The presence of both gradients promotes radial diffusion of the species from/to the electrodes, thus allowing higher currents.

The maximum and minimum relative concentrations for the unit cells in Fig. \ref{ejemplo-concentracion:fig:distribucion-estacionaria} are located on each electrode, and have the same value but different sign due to symmetry. The minimum concentration must be non-negative $c_\especieDet(W,0,t) \geq 0$, therefore the relative concentration must be $\gamma_\especieDet(1,0,\tau) \geq -1$. This means, due to linearity, that the unit cells can handle a `maximum uniform generation rate' (current density) given by
\begin{equation}
  \abs{\widehat{\phi}_\especieDet^{\max}(\xi,\tau)} =
    \marca{\abs{\widehat{\phi}_\especieDet^{\max}(\xi)}} = \frac{1}{\gamma_{\max}}
  \label{ejemplo-concentracion:ec:phi-maximo}
\end{equation}
where $\gamma_{\max}$ corresponds to the maximum relative concentration obtained when $\abs{\widehat{\phi}_\especieDet(\xi,\tau)}=1$, and $-\gamma_{\max}$ corresponds to the minimum. Thus, the maximum uniform generation rates for the unit cells with aspect ratio $H/W=\num{0,2}/\pi$ and $H/W=5/\pi$ are $1/\num{12,6}$ and $1/\num{2,7}$ respectively, confirming once more that higher aspect ratios allows higher currents.

\begin{figure*}
  \centering
  \includegraphics{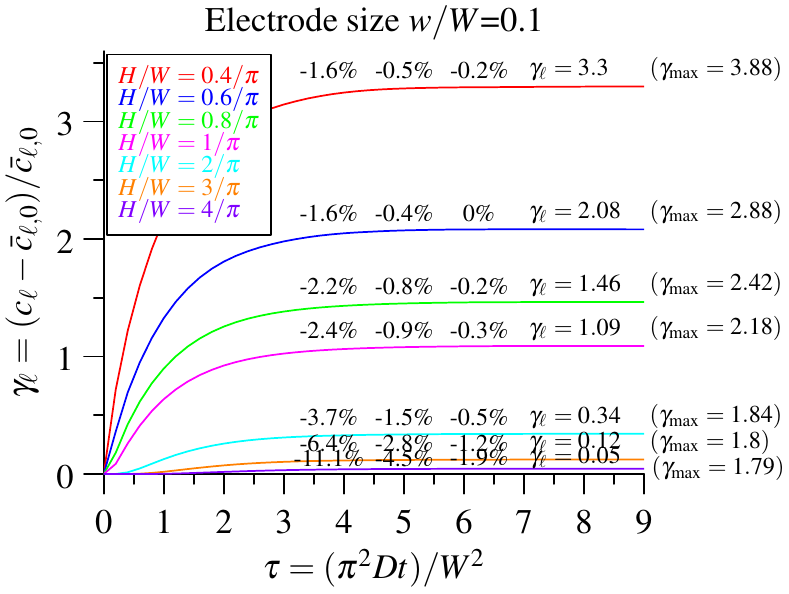}\hspace{1em}
  \includegraphics{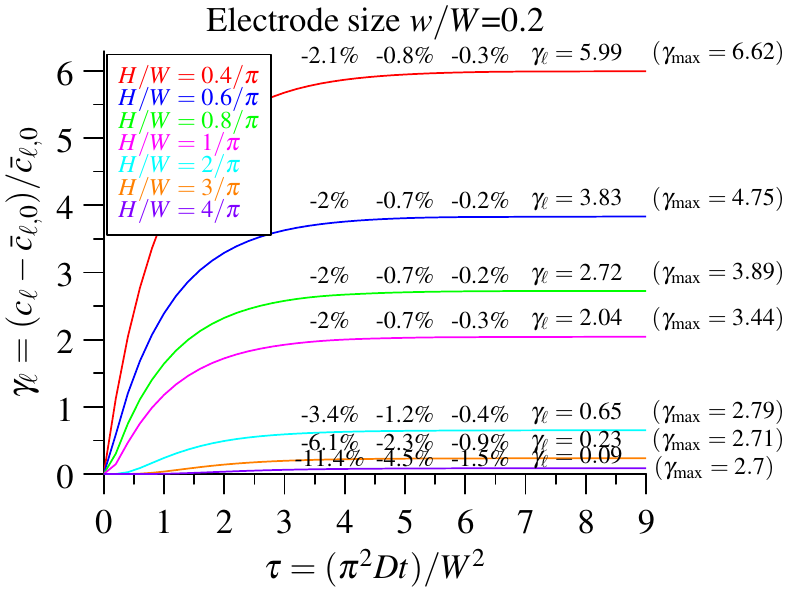}
  
  \includegraphics{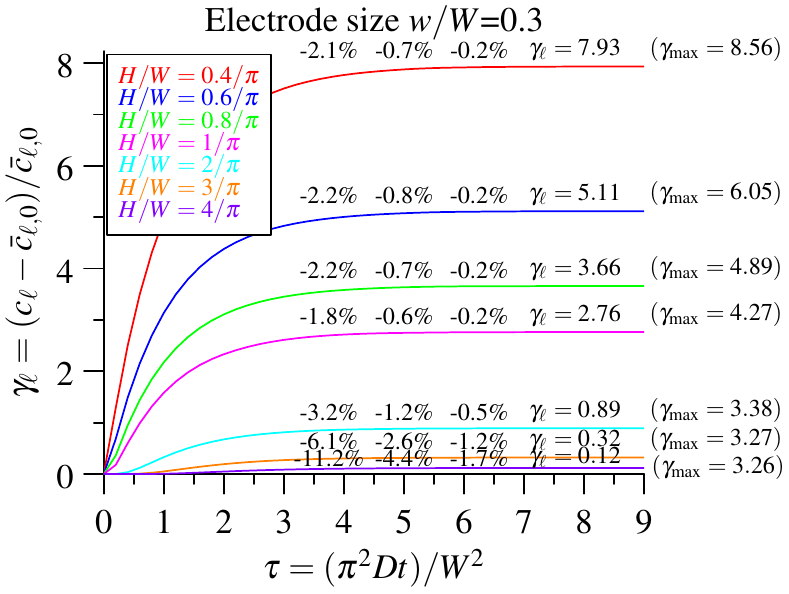}\hspace{1em}
  \includegraphics{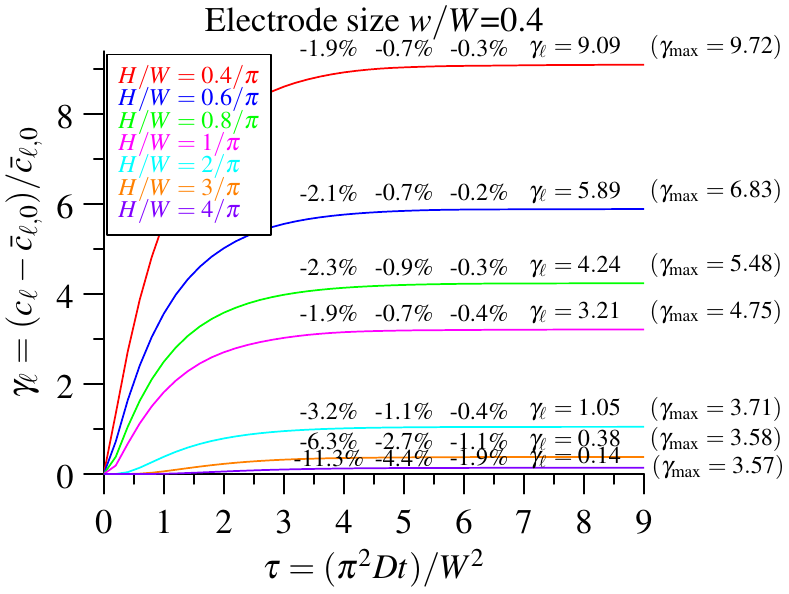}
  \caption{Relative concentration of the species $\especieDet$ at the furthest location from the electrodes $(x,z)=(0,H)$ for $\abs{\widehat{\phi}_{\especieDet}} = \abs{W\phi_{\especieDet}}/(\pi^{2}D\bar{c}_{\especieDet,0}) = 1$ on the surface of the electrodes, and considering different electrode sizes and cell aspect ratios. The values over each curve represent: the percentage of the concentration respect to the steady state at $(x,z)=(0,H)$ for $\tau = \pi^{2}Dt/W^2 = \set{4,\,5,\,6}$ and $\gamma_{\especieDet}$ stands for the relative concentration in steady state at $(x,z)=(0,H)$. The value of $\gamma_{\max}$ shown in brackets stands for the maximum relative concentration in the whole cell obtained at the surface of the electrodes $(x,z)=(0,0)$.}
  \label{ejemplo-concentracion:fig:concentracion-lejana}
\end{figure*}

Simulations in Fig. \ref{ejemplo-concentracion:fig:concentracion-lejana} show the evolution in time of the relative concentration at the furthest vertical position from the electrodes, which corresponds to $(x,z)=(0,H)$, for a variety of electrode sizes and aspect ratios. The furthest position was chosen because it can clearly reflect the change in the response time of the cell as the aspect ratio increases. For low aspect ratios a faster response is expected due to smaller diffusion distances, and conversely, for high aspect ratios a slower response is expected.

All graphs in Fig. \ref{ejemplo-concentracion:fig:concentracion-lejana} show that the time response of the unit cell effectively gets slower when the aspect ratio of the unit cell $H/W$ increases. Quantitatively, it can be observed that for low aspect ratios $H/W\leq 1/\pi$ the relative concentration is around $\num{-2}\%$, $\num{-0,7}\%$ and $\num{-0,2}\%$ lower than the steady state for $t=4\tau_\phi$, $t=5\tau_\phi$ and $t=6\tau_\phi$. This agrees with the \marca{theoretical values $\num{-1,8}\%$, $\num{-0,7}\%$ and $\num{-0,2}\%$ given at the end of} Section \ref{densidad-conocida}. For high aspect ratios $H/W\geq 3/\pi$ around $\num{-0,9}\%$ to $\num{-1,9}\%$ lower than the steady state is obtained for $t=6\tau_\phi$.

\begin{table}
  \centering
  \subtable[][Simulated values]{
    \begin{tabular}{c|cccc}
      & \multicolumn{4}{c}{$w/W$}\\
      $H/W$ & \num{0,1} & \num{0,2} & \num{0,3} & \num{0,4}\\ \hline
      $3/\pi$ & \num{6,7}\% & \num{8,5}\% & \num{9,8}\% & \num{11}\% \\
      $4/\pi$ & \num{2,8}\% & \num{3,3}\% & \num{3,7}\% & \num{3,9}\% 
    \end{tabular}
  }
  
  \subtable[][\marca{Theoretical bounds (Theorems \ref{semiinfinita:teo:error-cota} and \ref{semiinfinita:teo:condicion-semiinfinta})}]{
    \begin{tabular}{c|ccc}
      & \multicolumn{3}{c}{$w/W$}\\
      $H/W$ & \num{0,1} & \num{0,2} & \num{0,25} \\ \hline
      $3/\pi$ & \marca{$\leq \num{7,8}\%$} & \marca{$\leq \num{10,8}\%$} & \marca{$\leq \num{12,3}\%$} \\
      $4/\pi$ & \marca{$\leq \num{2,9}\%$} & \marca{$\leq \num{4}\%$} & \marca{$\leq \num{4,5}\%$}
    \end{tabular}
  }
  \caption{Steady state value of the relative concentration $\gamma_{\especieDet} = (c_{\especieDet}-\bar{c}_{\especieDet,0})/\bar{c}_{\especieDet,0}$ at the furthest location from the electrodes when applying the `maximum uniform generation rate' $\abs{\widehat{\phi}_\especieDet^{\max}} = \abs{W\phi_{\especieDet}^{\max}}/(\pi^{2}D \bar{c}_{\especieDet,0}) = 1/\gamma_{\max}$ in cells with high aspect ratio.}
  \label{ejemplo-concentracion:tab:semi-infinita}
\end{table}
The effect of semi-infinite geometries can also be seen in Fig. \ref{ejemplo-concentracion:fig:concentracion-lejana}, since for high aspect ratios the concentration far from the electrodes remains close to the bulk concentration. Quantitatively, when applying $\abs{\widehat{\phi}_{\especieDet}^{\max}(\xi,\tau)} = \abs{\widehat{\phi}_{\especieDet}(\xi,\tau)}/\gamma_{\max} = 1/\gamma_{\max}$, the steady state value of $\gamma_{\especieDet}$ in the plots must be rescaled to $\gamma_{\especieDet}/\gamma_{\max}$. Therefore, taking the case of $w/W=\num{0,4}$ and $H/W=3/\pi$ as an example, $\gamma_{\especieDet}$ and $\gamma_{\max}$ are given by $\num{0,38}$ and $\num{3,58}$ respectively, so concentration in steady state is just $\num{0,38}/\num{3,58}=\num{11}\%$ higher than the bulk concentration. More precision can be obtained when considering $H/W=4/\pi$, since the deviation from the bulk concentration is $\num{0,14}/\num{3,57}=\num{3,9}\%$ (see Table \ref{ejemplo-concentracion:tab:semi-infinita} for more values). These results agree with \marca{the bound presented in Theorem \ref{semiinfinita:teo:error-cota} and the} criterion established in Theorem \ref{semiinfinita:teo:condicion-semiinfinta}.

\subsection{Effect of the cell geometry in the limiting current}
\begin{figure*}
  \centering
  \subfigure[][]{
    \includegraphics{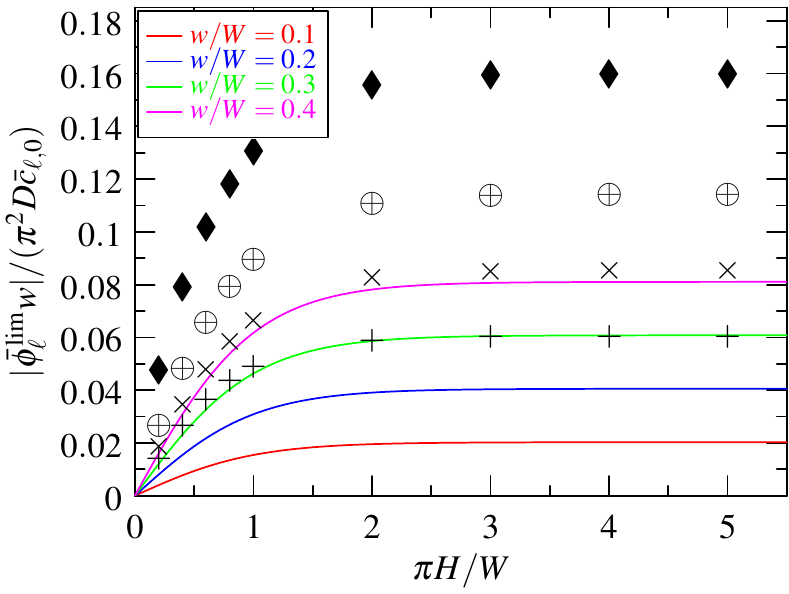}
    \label{ejemplo-corriente:fig:corriente-exhaustiva-simulada}
  }
  \subfigure[][]{
    \includegraphics{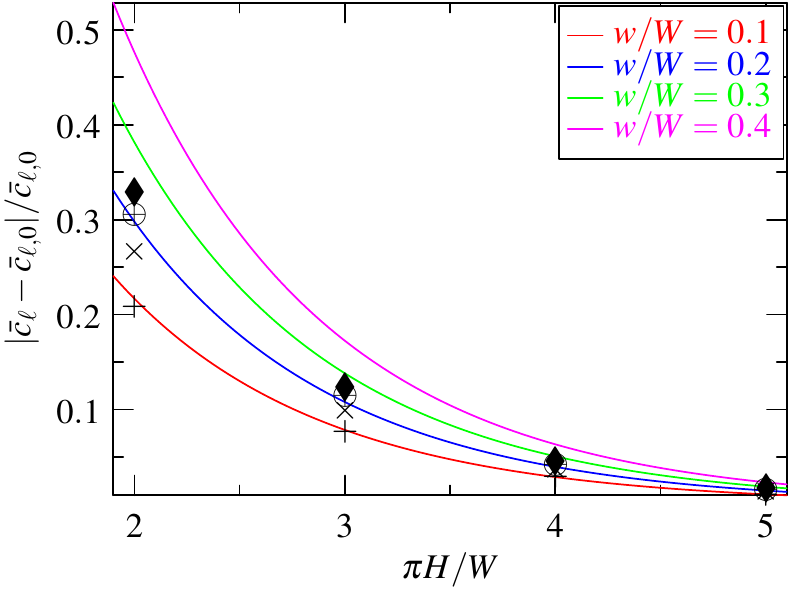}
    \label{ejemplo-corriente:fig:corriente-exhaustiva-lejos}
  }
  \caption{The symbols $+$, $\times$, $\oplus$ and $\blacklozenge$ are the simulated results obtained for $w/W=\set{\num{0,1},\ldots,\num{0,4}}$ respectively. The colored lines correspond to the theoretical bounds. 
  \subref{ejemplo-corriente:fig:corriente-exhaustiva-simulada} Simulation and theoretical lower bound in Eq. (\ref{corriente-limite:ec:phi-cota}) for $\abs{\bar{\phi}_{\especieDet}^{\lim}w}/(\pi^2 D \bar{c}_{\especieDet,0})$\marca{, which is proportional to the steady state limiting current}. 
  \subref{ejemplo-corriente:fig:corriente-exhaustiva-lejos} Relative concentration in steady state at $(x,z)=(0,H)$, comparison between simulation and the theoretical bound in Eq. (\ref{semiinfinita:ec:error-cota}).}
  \label{ejemplo-corriente:fig:corriente-exhaustiva}
\end{figure*}

In order to test the performance of Eq. (\ref{corriente-limite:ec:phi-cota}), several simulations were carried out using the scale transformations in Eq. (\ref{ejemplo-concentracion:ec:escala}). Here it is assumed that the determinant species of the cell $\especieDet\in\set{\mathcal{O},\mathcal{R}}$ has initial concentration $\bar{c}_{\especieDet,0}$ and also that the concentrations on the working and counter electrodes are the limiting concentrations $2\bar{c}_{\especieDet,0}$ and $0$ respectively. These limiting concentrations are due to extreme potentials at the electrodes, and they deviate equally from the initial concentration (but in opposite directions) since the currents on the electrodes are assumed of equal magnitude but opposite sign $\forall t$ (Kirchhoff's current law is satisfied inside the unit cell $\forall t$).

Like before, the simulations were carried out using an exponential mapped mesh, in order to provide higher resolution near the edges of the electrodes. The mesh was incrementaly refined until the first two decimal places of the limiting generation rate agreed with Eqs. (\ref{corriente-limite:ec:aoki}) and (\ref{corriente-limite:ec:morf}), see \ref{apendice:s33} for more details on the simulation setup. 

The results of the simulations were obtained for only one of the species, the determinant species $\especieDet\in\set{\mathcal{O},\mathcal{R}}$, while the results for the other species can be obtained by applying Eq. (\ref{problema:ec:reaccion}) for the generation rate and Remark \ref{problema:obs:ctotal-cte} for the concentration.

Fig. \ref{ejemplo-corriente:fig:corriente-exhaustiva-simulada} shows that the simulated limiting current \marca{in steady state} $\abs{i_{\lim}}\propto\abs{\bar{\phi}_\especieDet^{\lim}w}$ is around 2 to 3 times higher than the  lower bound in Eq. (\ref{corriente-limite:ec:phi-cota}) for $w/W \leq \num{0,4}$, which is a quite reasonable bounding. Also the simulation has a saturation effect with respect to $H/W$, accurately predicted by the $\tanh(\cdot)$ term in Eq. (\ref{corriente-limite:ec:phi-cota}). This shows the effect of semi-infinite geometry as the ratio $H/W$ increases. For small aspect ratios, only horizontal diffusion occurs and almost no vertical diffusion, which leads to lower limiting currents. When the aspect ratio is about $H/W=3/\pi$, bulk concentration is present only near the upper wall ($z=H$), providing the highest vertical concentration gradient and thus the highest limiting current. For $H/W > 3/\pi$ the region of bulk concentration is bigger, spanning $3/\pi \leq z/W \leq H/W$, but the diffusion layer in $0\leq z/W < 3/\pi$ remains the same, as well as the limiting current.

High aspect ratio unit cells provide the maximum limiting current available, since the region of bulk concentration helps to maintain a radial diffusion flow from/to the electrodes. In contrast, constrained diffusion (not radial) in low aspect ratio unit cells produces lower limiting currents \cite{Strutwolf2005}. This fact confirms that the limiting generation rate for semi-infinite geometries $\lim_{H/W\to\infty}\abs{\bar{\phi}_{\especieDet}^{\lim}}$ obtained by Aoki in \cite{Aoki1988}, and corrected by Morf \cite{Morf2006}, is actually an upper bound for lower aspect ratio unit cells, as stated in Remark \ref{corriente-limite:nota:cota-superior}. 

Once again, Fig. \ref{ejemplo-corriente:fig:corriente-exhaustiva-lejos} confirms that geometries satisfying $H/W>3/\pi$ can be considered as semi-infinite, since the concentration far from the electrodes remains similar to the bulk concentration. When limiting current is circulating through the cell, the steady state concentration at $(x,z)=(0,H)$ obtained for $H/W=3/\pi$ is only $\num{7,7}\%$ to $\num{12,4}\%$ higher than the bulk concentration. For $H/W=4/\pi$, the concentration is just $\num{2,8}\%$ to $\num{4,5}\%$ higher than the bulk value. In all cases the simulated results are bounded from above by the colored lines, and the bounds tend to be closer to the simulated results for electrodes satisfying $w/W\leq\num{0,2}$ as predicted in Eq. (\ref{semiinfinita:ec:error-cota}) and Theorem \ref{semiinfinita:teo:condicion-semiinfinta}.

\begin{table}
  \centering
  \begin{tabular}{llccl}
    \multicolumn{5}{c}{$H/W = \num{0,4}/\pi$}\\
    \hline
    $w/W$ & LB & \multicolumn{1}{c}{simulation} & \% of UB & UB\\
    \hline
    \num{0,1}  & \num{0,01}  & \num{0,03} & 50\% & \num{0,06}\\
    \num{0,2}  & \num{0,02}  & \num{0,03} & 33\% & \num{0,09}\\
    \num{0,25} & \num{0,02}  & \num{0,04} & 40\% & \num{0,10}\\
    \num{0,3}  & \num{0,02}  & \num{0,05} & 42\% & \num{0,12}$^*$\\
    \num{0,4}  & \num{0,03}  & \num{0,08} & 50\% & \num{0,16}$^*$\\
    \hline
  \end{tabular}
  
  \vspace{1em}
  \begin{tabular}{llccl}
    \multicolumn{5}{c}{$H/W = 1/\pi$}\\
    \hline
    $w/W$ & LB & \multicolumn{1}{c}{simulation} & \% of UB & UB\\
    \hline
    \num{0,1}  & \num{0,02} & \num{0,05} & 83\% & \num{0,06}\\
    \num{0,2}  & \num{0,03} & \num{0,07} & 78\% & \num{0,09}\\
    \num{0,25} & \num{0,04} & \num{0,08} & 80\% & \num{0,10}\\
    \num{0,3}  & \num{0,05} & \num{0,09} & 75\% & \num{0,12}$^*$\\
    \num{0,4}  & \num{0,06} & \num{0,13} & 81\% & \num{0,16}$^*$\\
    \hline
  \end{tabular}
  
  \vspace{1em}
  \begin{tabular}{llccl}
    \multicolumn{5}{c}{$H/W = 3/\pi$}\\
    \hline
    $w/W$ & LB & \multicolumn{1}{c}{simulation} & \% of UB & UB\\
    \hline
    \num{0,1}  & \num{0,02} & \num{0,06} & $\geq \marca{83}\%$ & \num{0,06}\\
    \num{0,2}  & \num{0,04} & \num{0,08} & $\phantom{\geq\:} \marca{89}\%$ & \num{0,09}\\
    \num{0,25} & \num{0,05} & \num{0,10} & $\geq 90\%$ & \num{0,10}\\
    \num{0,3}  & \num{0,06} & \num{0,11} & $\phantom{\geq\:} \marca{92}\%$ & \num{0,12}$^*$\\
    \num{0,4}  & \num{0,08} & \num{0,16} & $\geq \marca{94}\%$ & \num{0,16}$^*$\\
    \hline
  \end{tabular}
  
  \caption{Bounds and values of $\abs{\bar{\phi}_{\especieDet}^{\lim} w}/(\pi^2 D \bar{c}_{\especieDet,0})$ for different electrode widths $w/W$ and aspect ratios $H/W$. (LB) Lower bound in Eq. (\ref{corriente-limite:ec:phi-cota}), simulation value, (\% of UB) percentage of the simulation with respect to the Aoki-Morf upper bound and (UB) Aoki-Morf upper bound in Eqs. (\ref{corriente-limite:ec:aoki}) and (\ref{corriente-limite:ec:morf})$^*$. \marca{The asterisk indicates that Eq. (\ref{corriente-limite:ec:morf}) has been used instead of Eq. (\ref{corriente-limite:ec:aoki}).}}
  \label{ejemplo-corriente:tab:comparacion-cotas}
\end{table}

Table \ref{ejemplo-corriente:tab:comparacion-cotas} shows a comparison between the lower bound value in Eq. (\ref{corriente-limite:ec:phi-cota}), the simulation value and the upper bound obtained by Aoki-Morf in Eqs. (\ref{corriente-limite:ec:aoki}) and (\ref{corriente-limite:ec:morf}). The result obtained by Aoki-Morf is not longer precise for small aspect ratios such as $H/W = \num{0,4}/\pi$, but when used together with Eq. (\ref{corriente-limite:ec:phi-cota}), they can give a reasonable range for the actual value of the limiting generation rate and thus the limiting current.

\begin{figure*}
  \centering
  \includegraphics{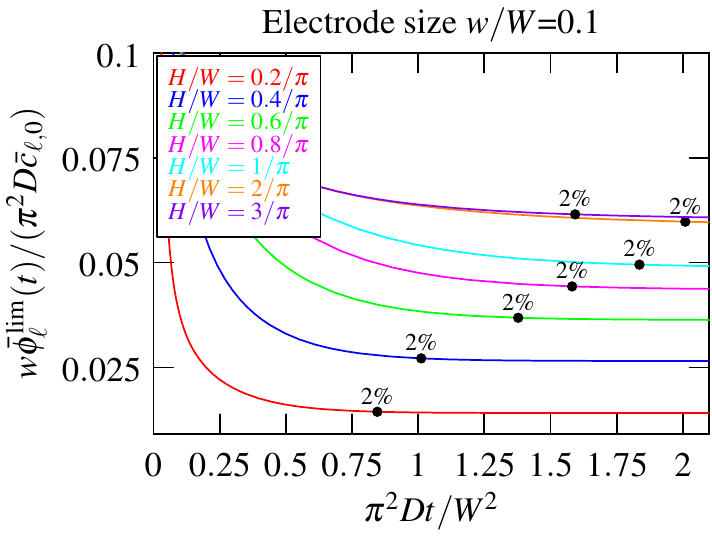}\hspace{1em}
  \includegraphics{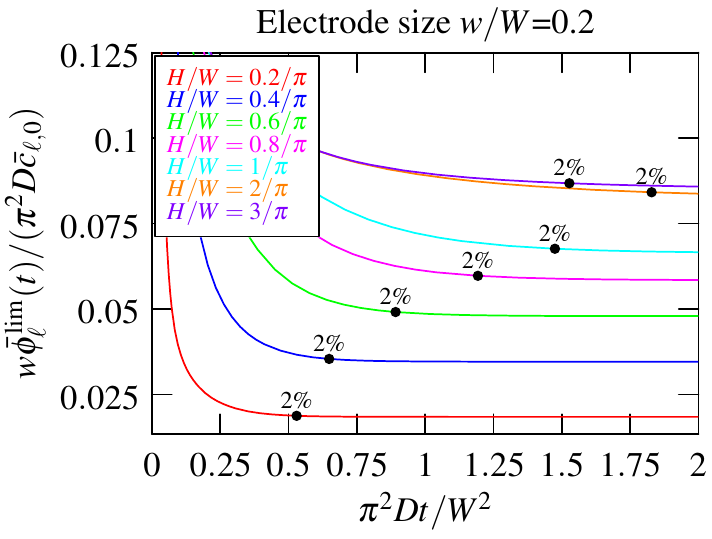}
  
  \includegraphics{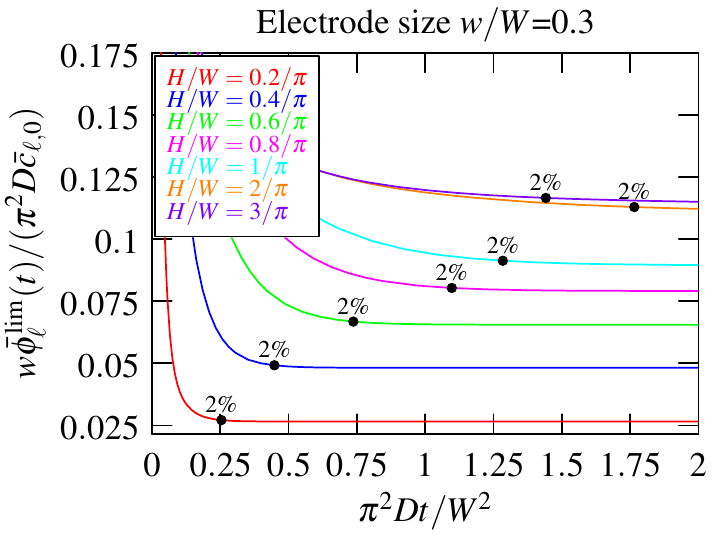}\hspace{1em}
  \includegraphics{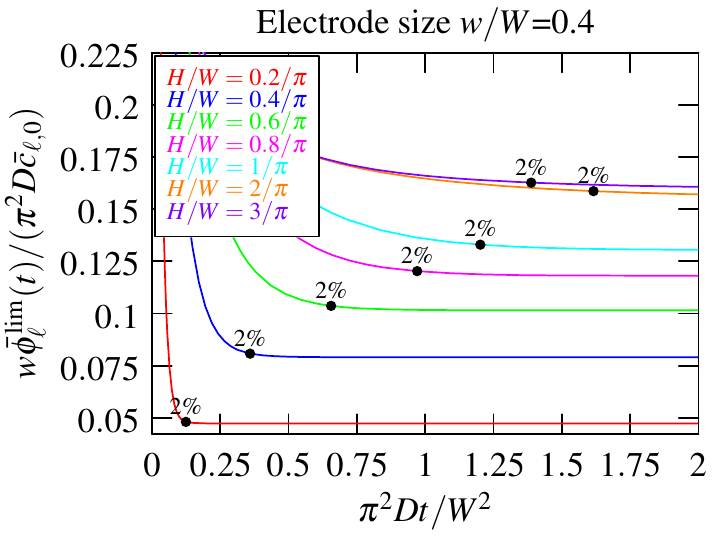}
  \caption{Time response of the average limiting generation rate $w\bar{\phi}_{\especieDet}^{\lim}(t)/(\pi^{2}D\bar{c}_{\especieDet,0})$ (limiting current) for a variety of electrode sizes and cell aspect ratios. \marca{On each curve it is indicated the time required to reach a 2\% difference with respect to} the steady state value.}
  \label{ejemplo-corriente:fig:phi_t}
\end{figure*}

Fig. \marca{\ref{ejemplo-corriente:fig:phi_t}} shows the time response of the average limiting generation rate (limiting current) for different electrode sizes and cell aspect ratios. \marca{On each curve it is shown the time required to reach a 2\% difference with respect to the steady state value.} It is interesting to notice that the time required for the current to reach steady state, when a step of concentrations has been applied to the electrodes ($2\bar{c}_{\especieDet}$ and $0$ to the working and counter respectively), is about \marca{2} to \marca{8} times lower than the time required by the concentration to reach steady state when a current step is applied, see Fig. (\ref{ejemplo-concentracion:fig:concentracion-lejana}) to compare. Therefore, the time to reach steady state $T_{ss}^{\phi}$ when a current step is applied (Eq. (\ref{densidad-conocida:ec:tau_phi})) could be used as an upper bound for the time required by the current to reach steady state when a concentration step is applied on the electrodes, which is likely to be the quantity recorded in an experiment.

\begin{figure}
  \centering
  \includegraphics{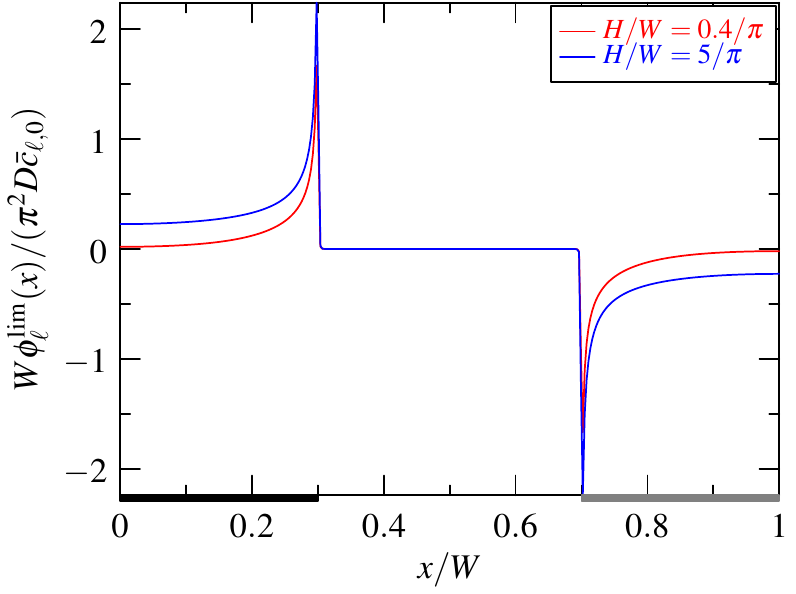}
  \caption{Shape of the limiting generation rate $W\phi_{\especieDet}^{\lim}(x)/(\pi^2 D \bar{c}_{\especieDet,0})$ (limiting current density) in steady state for $H/W=\num{0,4}/\pi$ and $H/W=5/\pi$ considering $w/W=\num{0,3}$.}
  \label{ejemplo-corriente:fig:corriente-forma}
\end{figure}

Finally, Fig. \marca{\ref{ejemplo-corriente:fig:corriente-forma}} shows the shape of the limiting generation rate (limiting current density) in steady state predicted by the simulation along the surfaces of the electrodes. As explained before, the edges of the electrodes are exposed to higher concentration gradients, allowing the species to escape/reach the edges easily. For this reason the current density needs to be very high at the edges of the electrodes, in order to maintain a uniform concentration along them. Also Fig. \ref{ejemplo-corriente:fig:corriente-forma} explicitly shows that the current density near the center of the electrodes increases as the aspect ratio $H/W$ increases, due to the presence of the region of bulk concentration far from the electrodes.

\section{Conclusions}
New time-dependent expressions were found for the concentration profile of an IDAE inside a finite geometry cell, when assuming a known current density and internal counter electrode. As immediate byproducts, a criterion \marca{defining the conditions for obtaining finite and semi-infinite cells with comparable behaviors}, as well as bounds for the limiting current in a finite cell, \marca{were obtained}. The results show that the exact expressions obtained by Aoki and Morf for the limiting current in semi-infinite geometries can be applied to finite geometries, if the new semi-infinite criterion is satisfied. In case the semi-infinite criterion is not satisfied, the new bounds for the limiting current can be applied and provide a reasonable estimation. The accuracy of the results was successfully validated through comparison of the theoretical expressions with finite-element numerical simulations.
These findings can be useful for designing finite geometry IDAE cells and help to understand the importance of the region of bulk concentration for obtaining higher limiting currents.

\section{Appendix}
\marca{Proofs and details of calculations for the results obtained here can be found in \ref{apendice:idae-resultados}. Details concerning the simulations can be found in \ref{apendice:simulaciones}. \ref{apendice:extension} extends the results obtained here and in \ref{apendice:idae-resultados} to a general cell with periodic (and non-periodic where possible) left/right boundary conditions. These Appendices are provided as supplementary information.}

\section{Aknowledgements}
The authors would like to thank Dr. Mithran Somasundrum for his help with the manuscript, also to the reviewers for their valuable comments and to acknowledge the \emph{National Research Council of Thailand} (NRCT). This project received financial support from the \emph{National Research University Project} (NRU) of Thailand's \emph{Office of Higher Education Commission}.

\biboptions{numbers, compress}
\bibliographystyle{model1a-num-names}
\bibliography{idae-corriente-en}{}

%% file: appendix.tex

\appendix

\onecolumn
\begin{center}
  \Large
  Supplementary information for:\\
  Mathematical Modeling of Interdigitated Electrode Arrays in Finite Electrochemical Cells\\ 
  \vspace{1em}\normalsize
  Cristian Guajardo, Sirimarn Ngamchana, Werasak Surareungchai\\
  \vspace{0.5em}\footnotesize
  \emph{King Mongkut's University of Technology Thonburi, 49 Soi Thianthale 25, Thanon Bangkhunthian Chaithale, Bangkok 10150, Thailand}
  \normalsize\vspace{3em}
\end{center}

\section{Results for an IDAE unit cell with finite height}
\label{apendice:idae-resultados}
\subsection{Results for any bottom boundary condition}
\label{apendice:idae-solucion_general}
Preliminary and very general results are found, which are independent of whether the potential or current density are known. This results have been also  extended for periodic and non-periodic left/right boundary conditions in \ref{apendice:c_total} and \ref{apendice:c-xzt}.

Consider a cell like the one described in section \ref{problema}. For sake of simplicity, Eqs. (\ref{problema:ec:c-PDE}) are subtracted with Eqs. (\ref{problema:ec:ci-PDE}). Later, by applying the Laplace transform in time $\mathcal{L}_{t}\set{\cdot}$ one obtains
\begin{eqnarray*}
  \frac{s}{D} \Delta C_{\especie}(x,z,s) &=& \parderiv{^2 \Delta C_{\especie}}{x^2}(x,z,s) + \parderiv{^2 \Delta C_{\especie}}{z^2}(x,z,s)\\
  \parderiv{\Delta C_{\especie}}{x}(0,z,s) &=& \parderiv{\Delta C_{\especie}}{x}(W,z,s) = 0\\
  \parderiv{\Delta C_{\especie}}{z}(x,H,s) &=& 0,\: \Delta F_{\especie}\bigpar{\Delta C_{\especie}, \parderiv{\Delta C_{\especie}}{z}, x, s}=0
\end{eqnarray*}
where $\Delta C_{\especie} := \mathcal{L}_{t}\set{\Delta c_{\especie}}$, $\Delta F_{\especie} := \mathcal{L}_{t}\set{\Delta f_{\especie}}$ and
\begin{eqnarray*}
  \Delta c_{\especie}(x,z,t) \hspace{-0.5em}&:=&\hspace{-0.5em}
    c_{\especie}(x,z,t) - c_{\especie,0}(x,z)\\
  \Delta f_{\especie}\bigpar{\Delta c_{\especie}, \parderiv{\Delta c_{\especie}}{z}, x, t} \hspace{-0.5em}&:=&\hspace{-0.5em}
    f_{\especie}\bigpar{c_{\especie}, \parderiv{c_{\especie}}{z}, x, t} - f_{\especie,0}\bigpar{c_{\especie,0}, \parderiv{c_{\especie,0}}{z}, x}
\end{eqnarray*}

This problem is solved by using the method of separation of variables, obtaining the `change in concentration' $\Delta c_{\especie}(x,z,t)$ in Laplace domain
\begin{subeqnarray}
  \label{apendice:ec:idae-Dc-general}
  \Delta C_{\especie}(x,z,s) \hspace{-0.5em}&=&\hspace{-0.5em}
    \sum_{n=0}^\infty \Delta{B}_{n}^{\especie}(z,s) \cos(n\pi x/W)\\
  \Delta{B}_{0}^{\especie}(z,s) \hspace{-0.5em}&=&\hspace{-0.5em}
    \Delta\bar{B}_{0}^{\especie}(s) \cosh\bigpar{\sqrt{\frac{s}{D}}(H-z)} \\
  \Delta{B}_{n}^{\especie}(z,s) \hspace{-0.5em}&=&\hspace{-0.5em}
    \Delta\bar{B}_{n}^{\especie}(s) \cosh\bigpar{\sqrt{\frac{s}{D} + \frac{n^{2}\pi^{2}}{W^{2}}}\,(H-z)}
\end{subeqnarray}
where $\Delta\bar{B}_{0}^{\especie}(z,s)$ and $\Delta\bar{B}_{n}^{\especie}(z,s)$ must be obtained from the bottom boundary condition.

Solving the problem in Eqs. (\ref{problema:ec:ci-PDE}) by using the method of separation of variables leads to analogous results for the initial concentration
\begin{subeqnarray}
  \label{apendice:ec:idae-ci-general}
  c_{\especie,0}(x,z) &=& \bar{c}_{\especie,0} + \sum_{n=1}^\infty b_{n}^{\especie,0}(z) \cos(n\pi x/W) \\
  \bar{c}_{\especie,0} &=& \frac{1}{W} \int_0^W c_{\especie,0}(x,z) \ud{x},\: \forall z\\
  b_{n}^{\especie,0}(z) &=& \bar{b}_{n}^{\especie,0} \cosh\bigpar{n\frac{\pi}{W}(H-z)}
\end{subeqnarray}
where $\bar{b}_{n}^{\especie,0}(z)$ must be obtained from the bottom boundary condition.

In case Kirchhoff's current law is satisfied in the unit cell $\forall t$ (for example when it includes a counter electrode), then the net current applied to the unit cell should be zero
\begin{displaymath}
  \int_0^W J(x,s)\, L\ud{x} \propto \int_0^W \parderiv{C_{\especie}}{z}(x,0,s)\, \ud{x} = 0
     \Leftrightarrow   \Delta{B}_{0}^{\especie}(z,s) = 0
\end{displaymath}
where $J=\mathcal{L}_{t}\set{j}$ and $j(x,t)$ is the current density on the bottom boundary.
In this case, the coefficient $\Delta{B}_{0}^{\especie}(z,s)$ must be zero
\begin{displaymath}
  \Delta{B}_{0}^{\especie}(z,s) = 0 \Leftrightarrow \frac{1}{W} \int_0^W \Delta C_{\especie}(x,z,s) \ud{x} = 0   
\end{displaymath}
therefore by adding
\begin{displaymath}
  \frac{1}{W} \int_0^W \Delta c_{\especie}(x,z,t) \ud{x} +
    \frac{1}{W} \int_0^W c_{\especie,0}(x,z) \ud{x} = 0 + \bar{c}_{\especie,0}
\end{displaymath}
the average concentration of the species $\especie$, along the $x$ axes, must remain uniform in $z$ and also constant
\begin{displaymath}
  \frac{1}{W} \int_0^W c_{\especie}(x,z,t) \ud{x} = \bar{c}_{\especie,0},\quad \forall z \mbox{ and } t\geq 0
\end{displaymath}

\begin{observacion}
  Note also that the total concentration satisfies the result in Eq. (\ref{apendice:ec:c_total-cte-uniforme})
  \begin{displaymath}
    c_{\mathcal{O}}(x,z,t) + c_{\mathcal{R}}(x,z,t) = c_0\quad \forall (x,z)\mbox{ and } t\geq 0
  \end{displaymath}
  This holds in the particular case of the unit cell described in Section \ref{problema}, since the unit cell can be extended periodically in $x$ with period $2W$ and therefore it allows Fourier transform in the $x$-coordinate. This periodic extension is possible due to the left/right symmetry/insulation boundary of the unit cell.
\end{observacion}

\subsection{Concentration for known current density}
\label{apendice:densidad-conocida}
In this section the initial concentration $c_{\especie,0}(x,z)$ and the change in concentration $\Delta c_{\especie}(x,z,t)$ are obtained as Fourier series, assuming that the current density inside the unit cell is known. An extension of the results to the cases of periodic and non-periodic left/right boundary conditions can be found in \ref{apendice:c-xzt}. These results will be useful to obtain the concentration profile in steady state and to calculate the time to reach steady state when applying a constant current.

By taking the bottom boundary for the initial concentration in Eqs. (\ref{densidad-conocida:ec:cb-abajo})
\begin{displaymath}
  f_{\especie,0}\bigpar{c_{\especie,0},\parderiv{c_{\especie,0}}{z},x} = D \parderiv{c_{\especie}}{z}(x,0) + \phi_{\especie,0}(x) = 0
\end{displaymath}
one obtains the Fourier coefficient of the initial concentration
\begin{equation}
  \label{apendice:ec:idae-ci-bn}
  b_{n}^{\especie,0}(z) =
    G_{\phi}\bigpar{H-z,n^2\frac{\pi^{2}}{W^{2}}} \cdot \mathcal{I}_n \set{\frac{\phi_{\especie,0}}{D}}
\end{equation}
where
\begin{subeqnarray}
  \label{apendice:ec:idae-G_phi-In}
  G_{\phi}(z,s) &=&
    \frac{\cosh(\sqrt{s}\,z)}{\sqrt{s} \sinh(\sqrt{s}\,H)}
    \slabel{apendice:ec:idae-G_phi}\\
  \mathcal{I}_n\set{\cdot} \hspace{-0.5em}&:=&\hspace{-0.5em}
    \frac{2}{W} \int_0^W \set{\cdot} \cos(n\pi x/W) \ud{x} 
    \slabel{apendice:ec:idae-In}
\end{subeqnarray}

Analogously by using Eqs. (\ref{densidad-conocida:ec:cb-abajo}), the Laplace equivalent for the bottom boundary condition of the change in concentration is obtained
\begin{displaymath}
  \Delta F_{\especie}\bigpar{\Delta C_{\especie},\parderiv{\Delta C_{\especie}}{z},x,s} = D \parderiv{\Delta C_{\especie}}{z}(x,0,s) + \Delta\Phi_{\especie}(x,s) = 0
\end{displaymath}
where $\Delta\Phi_{\especie} = \mathcal{L}\set{\Delta\phi_{\especie}}$ and $\Delta\phi_{\especie} = \phi_{\especie} - \phi_{\especie,0}$. Then the generation rate $\Delta\Phi_{\especie}(x,s)$ completely determines the coefficients of Eqs. (\ref{apendice:ec:idae-Dc-general}) as shown below
\begin{eqnarray*}
  \Delta{B}_{0}^{\especie}(z,s) \hspace{-0.5em}&=&\hspace{-0.5em}
    G_{\phi}\!\bigpar{H-z,\frac{s}{D}} \cdot \frac{1}{2}\, \mathcal{I}_0\!\set{\frac{\Delta\Phi_{\especie}}{D}}\!(s)\\
  \Delta{B}_{n}^{\especie}(z,s) \hspace{-0.5em}&=&\hspace{-0.5em}
    G_{\phi}\!\bigpar{H-z,\frac{s}{D}+\frac{n^{2}\pi^{2}}{W^{2}}} \cdot \mathcal{I}_n\!\set{\frac{\Delta\Phi_{\especie}}{D}}\!(s)
\end{eqnarray*}

If Kirchhoff's current law is satisfied in the unit cell $\forall t>0$, then the coefficient $\Delta B_{0}^{\especie}(z,s)$ must be zero, as already shown in \ref{apendice:idae-solucion_general}. Later, by applying the time-scaling and frecuency-shifting properties of the Laplace transform
\begin{displaymath}
  \Delta{B}_{n}^{\especie}(z,s) =
    G_{\phi}\!\bigpar{H-z,\bigcuad{s + \frac{n^{2}\pi^{2}D}{W^{2}}}\frac{1}{D}} \cdot \mathcal{I}_n\!\set{\frac{\Delta\Phi_{\especie}}{D}}\!(s)
\end{displaymath}
and by taking the Laplace inverse of $\Delta C_{\especie}(x,z,s)$ and $\Delta{B}_{n}^{\especie}(z,s)$, the change in concentration in time domain is obtained 
\begin{subeqnarray}
  \label{apendice:ec:idae-Dc-xzt}
  \Delta c_{\especie}(x,z,t) \hspace{-0.5em}&=&\hspace{-0.5em}
    \sum_{n=1}^{+\infty} \Delta b_n^{\especie}(z,t) \cos(n\pi x/W)\\
  \Delta b_n^{\especie}(z,t) \hspace{-0.5em}&=&\hspace{-0.5em}
    g_\phi(H-z,Dt)\, \mathrm{e}^{-n^{2}\frac{\pi^{2}}{W^{2}} Dt} D * \mathcal{I}_n\!\set{\!\frac{\Delta\phi_{\especie}}{D}\!}\!(t)
    \nonumber\\
  g_\phi(z,t) \hspace{-0.5em}&=&\hspace{-0.5em}
    \frac{1}{H} \bigcuad{1 + 2 \sum_{k=1}^\infty (-1)^k \mathrm{e}^{-k^2\frac{\pi^{2}}{H^{2}} t} \cos\bigpar{k\frac{\pi}{H}z}} 
\end{subeqnarray}
Here $\Delta b_n^{\especie} = \mathcal{L}^{-1} \set{\Delta{B}_n^{\especie}}$ and $g_\phi = \mathcal{L}^{-1} \set{G_\phi}$. The Laplace inverse $g_\phi$ can be obtained from tables, such as \cite[p.218]{Schiff1999} or \cite[Eq. (20.10.5)]{dlmf2010}, and it is given by the $4^{th}$ \emph{elliptic theta function}.

\subsection{Concentration for constant current density}
\label{solucion-densidad-constante}
The concentration profile in steady state and the time to reach this steady state are obtained, assuming that a constant\footnote{here \emph{constant} means that there is no time-dependence and \emph{uniform} means that there is no space-dependence, as it is usual when referring to fields and potentials with these characteristics.} current density is applied. These results have been also extended for periodic left/right boundary conditions in \ref{apendice:c-xzt-periodica}. The result for the steady state concentration will be useful later to obtain a lower bound for the limiting current.

In case the current density is constant in $t$, the generation rate of the species $\especie$ is also constant in $t$ $\Delta\phi_{\especie}(x,t)=\Delta\phi_{\especie}(x)$, and then the integral $\mathcal{I}_n\set{\Delta\phi_{\especie}/D}(t)=\mathcal{I}_n\set{\Delta\phi_{\especie}/D}$. By this mean the coefficient $\Delta b_n^{\especie}(z,t)$ can be obtained simply by integration
\begin{equation}
  \Delta b_n^{\especie}(z,t) = 
    \int_0^t g_\phi(H-z,D\tau)\, \mathrm{e}^{-n^{2}\pi^{2} D\tau/W^{2}} D \ud{\tau} \cdot \mathcal{I}_n\set{\frac{\Delta\phi_{\especie}}{D}}
  \label{apendice:ec:idae-Db_n-z8}
\end{equation}
and together with Eq. (\ref{apendice:ec:idae-Dc-xzt}), they determine the dynamics of the concentration profile for all $t\geq 0$.

After a `sufficiently long time' ($t\to +\infty$), the dynamics of the unit cell is complete and the concentration reaches the steady state 
\begin{eqnarray*}
  \Delta c_{\especie}(x,z,+\infty) &=& 
    \sum_{n=1}^{+\infty} \Delta b_n^{\especie}(z,+\infty) \cos(n\pi x/W)\\
  \Delta b_n^{\especie}(z,+\infty) &=&
     G_{\phi}\bigpar{H-z,n^2\frac{\pi^{2}}{W^{2}}} \cdot \mathcal{I}_{n}\set{\frac{\Delta\phi_{\especie}}{D}}
\end{eqnarray*}
where $G_{\phi}$ and $\mathcal{I}_{n}$ are defined in Eqs. (\ref{apendice:ec:idae-G_phi-In}). Then the total concentration of the species $\especie$ in steady state is obtained by adding $\Delta c_{\especie}(x,z,+\infty)$ and $c_{\especie,0}(x,z)$ (Eqs. (\ref{apendice:ec:idae-ci-general}) and (\ref{apendice:ec:idae-ci-bn}))
\begin{equation}
  \label{apendice:ec:idae-c-xz8_original}
  c_{\especie}(x,z,+\infty) =
    \sum_{n=1}^{+\infty} \mathcal{I}_n\!\set{\frac{\phi_{\especie}}{D}} G_{\phi}\bigpar{H-z,\frac{n^2\pi^{2}}{W^{2}}} \cos\bigpar{\frac{n\pi}{W}x} + \bar{c}_{\especie,0}
\end{equation}
This is valid when the Kirchhoff's current law is satisfied in the unit cell $\forall t$.

The time required to reach this steady state (after the current step has been applied) can be obtained by using Eqs. (\ref{apendice:ec:idae-Dc-xzt}) and (\ref{apendice:ec:idae-Db_n-z8}). Thus the concentration profile consists of a double summation (in the indexes $n$ and $k$) 
\begin{eqnarray*}
  && \Delta c_{\especie}(x,z,t) = \sum_{n=1}^\infty \Delta b_n^{\especie}(z,t) \cos(n\pi x/W)\\
  && \Delta b_n^{\especie}(z,t) = \mathcal{I}_n\set{\frac{\Delta\phi_{\especie}}{D}} \cdot \frac{D}{H} \left[ \frac{1-\mathrm{e}^{-n^2\pi^{2}Dt/W^{2}}}{n^2\pi^{2}D/W^{2}} + \right.\\
  && \left. 2 \sum_{k=1}^\infty (-1)^k \frac{1-\mathrm{e}^{-[n^2\pi^{2}/W^{2} + k^2\pi^{2}/H^{2}]Dt}}{[n^2\pi^{2}/W^{2} + k^2\pi^{2}/H^{2}]D} \cos(k\pi (H-z)/H) \right]
\end{eqnarray*}
which consists of exponential modes
\begin{displaymath}
  \exp\bigpar{-[n^2\pi^{2}/W^{2} + k^2\pi^{2}/H^{2}]Dt} = \exp\bigpar{-\bigcuad{n^2 + k^2\frac{W^{2}}{H^{2}}} \frac{\pi^{2}}{W^{2}} Dt}
\end{displaymath}
Note that the exponential modes with lower $n$ and $k$ indexes decay slowly with time, so it is enough to consider the slowest of these exponentials $\exp(-\pi^{2}Dt/W^{2})$ as an indicator for the time to reach the steady state $T_{ss}^\phi$
\begin{displaymath}
  T_{ss}^{\phi} \propto \tau_{\phi} := \frac{W^{2}}{\pi^{2}D}
\end{displaymath}
$T_{ss}^\phi$ can be chosen as $4\tau_\phi$, $5\tau_\phi$ or $6\tau_\phi$, since the dominating mode $\exp(-\pi^{2}Dt/W^{2})$ decays to approximately $\num{1,8}\%$, $\num{0,7}\%$ and $\num{0,2}\%$ respectively.

More precise results can be obtained for $T_{ss}^\phi$ when considering low aspect ratio configurations $H/W < 1/2$ and $t>\tau_\phi$. In this case the exponential modes with $n\geq 1$ and $k\geq 1$ may be considered extinct since they are bounded by
\begin{displaymath}
  \exp\bigpar{-\bigcuad{n^2 + k^2\frac{W^{2}}{H^{2}}} \frac{\pi^{2}}{W^{2}} Dt} < \exp(-[n^{2} + 4k^{2}]) \leq \mathrm{e}^{-5} \approx \num{0,7}\%
\end{displaymath}
Because of the fast convergence of the double summation (due to the squared indexes $n^2$ and $k^2$ in the exponentials), the terms with large $n$ and $k$ can be neglected so the error with respect to the steady state can be approximated by using only $n=1$ and neglecting all terms with $k$ index
\begin{displaymath}
  \Delta c_{\especie}(x,z,t) - \Delta c_{\especie}(x,z,+\infty) \approx -\mathcal{I}_1\set{\frac{\Delta\phi_{\especie}}{D}} \frac{e^{-\pi^{2}Dt/W^{2}}}{H\pi^{2}/W^{2}} \cos\bigpar{\frac{\pi x}{W}}
\end{displaymath}

In addition, when considering working and counter electrodes of identical size and located at the ends of the unit cell as in Fig. \ref{problema:fig:celdas}, the concentration profile in steady state can be roughly approximated by using $n=1$. This is because: (i) the concentration is a continuous function, this means that harmonics in the Fourier series with higher $n$ indexes have very low amplitude, therefore the concentration is mainly represented by lower harmonics. (ii) the location of both electrodes at the ends of the cell helps the concentration to have its maximum and minimum at the ends of the cell (like a cosine). (iii) electrodes of equal size help to have symmetry with respect to $W/2$, which is increased when both electrodes have widths $2w_{W} = 2w_{C} = W/2$ since they provide a concentration closer in shape to a cosine.
Also this profile can be further approximated for $H/W < 1/\pi$, since $\sinh(\pi H/W) \approx \pi H/W$, giving finally
\begin{displaymath}
  \Delta c_{\especie}(x,z,+\infty) \approx \mathcal{I}_1\set{\frac{\Delta\phi_{\especie}}{D}} \frac{\cosh(\pi(H-z)/W)}{H\pi^{2}/W^{2}} \cos\bigpar{\frac{\pi x}{W}}
\end{displaymath}
Due to these approximations, the relative error with respect to the steady state is given by
\begin{equation}
  \frac{\Delta c_{\especie}(x,z,t) - \Delta c_{\especie}(x,z,+\infty)}{\Delta c_{\especie}(x,z,+\infty)} \approx \frac{-\mathrm{e}^{-\pi^{2}Dt/W^{2}}}{\cosh(\pi(H-z)/W)}
\end{equation}
for $t>\tau_\phi$ and $H/W<1/\pi$ and follows exponential decay. Hence the relative error of the concentration (with respect to the steady state) is maximum at the furthest distance from the electrodes ($z=H$), and it is approximately $-\num{1,8}\%$, $-\num{0,7}\%$ and $-\num{0,2}\%$ for $t$ equal to $4\tau_\phi$, $5\tau_\phi$ and $6\tau_\phi$ respectively.

\subsection{Bounds for the limiting \marca{steady state} current}
\label{apendice:corrente-limite}
Consider the electrodes configuration of the unit cell in Fig. \ref{problema:fig:celda_unitaria}, where the working electrode (black) and the counter electrode (gray) have the same size and are located at the ends of the unit cell, and Kirchhoff's current law is satisfied in the unit cell $\forall t$ (meaning that there is no other external electrode). Consider also that the unit cell is working under steady state condition, therefore the steady state concentration obtained in Eq. (\ref{apendice:ec:idae-c-xz8_original}) holds
\begin{displaymath}
  c_{\especie}(x,z,+\infty) =  \bar{\phi}_{\especie} \sum_{n=1}^\infty \mathcal{I}_n \set{\frac{\varphi}{D}} G_{\phi}\bigpar{H-z,n^2\frac{\pi^{2}}{W^{2}}} \cos(n\pi x/W) + \bar{c}_{\especie,0}  
\end{displaymath}
where $\bar{\phi}_{\especie}$ is the \emph{average generation rate} of the species $\especie$ (on the surface of the working electrode) and $\varphi(x)$ is the \emph{normalized generation rate}, which are given by
\begin{displaymath}
  \bar{\phi}_{\especie} := \frac{1}{w} \int_0^w \phi_{\especie}(x) \ud{x}, \quad
  \varphi(x) := \frac{\phi_{\especie}(x)}{\bar{\phi}_{\especie}}
\end{displaymath}

Since the electrodes configuration is symmetric with respect to $x=W/2$, the concentration profile $c_{\especie}(x,z,+\infty)-\bar{c}_{\especie,0}$ and the current density (generation rate) are expected to be odd symmetric with respect to $x=W/2$. In this case the integral $\mathcal{I}_n\set{\varphi/D}$ can be reduced to
\begin{displaymath}
  \mathcal{I}_n\set{\frac{\varphi}{D}} =
    \frac{4}{W} \int_0^{W/2} \frac{\varphi(x)}{D} \cos(n\pi x/W) \ud{x}
\end{displaymath}
for odd $n$ and $\mathcal{I}_n\set{\varphi/D}=0$ for even $n$.

In the following subsection, a relation between the limiting generation rate and the concentration of the species with lowest initial average is obtained. With this result, a lower bound for the limiting current is obtained in the second subsection.

\subsubsection{Concentration of the determinant species}
Consider the species $\especieDet\in\set{\mathcal{O},\mathcal{R}}$ which has the \textbf{lowest} initial average concentration $\bar{c}_{\especieDet,0} = \min(\bar{c}_{\mathcal{O},0},\bar{c}_{\mathcal{R},0})$. The concentration of this species in steady state satisfies
\begin{equation}
  c_{\especieDet}(x,z,+\infty) - \bar{c}_{\especieDet,0} =  \bar{\phi}_{\especieDet} \sum_{n\:\mathrm{odd}} \mathcal{I}_n \set{\frac{\varphi}{D}} G_{\phi}\bigpar{H-z,n^2\frac{\pi^{2}}{W^{2}}} \cos(n\pi x/W)
  \label{apendice:ec:idae-c-zx8}
\end{equation}

Assume now that the limiting current is circulating in the cell\footnote{The limiting current can be achieved by applying extreme potentials at the electrodes}, in this case the species $\especieDet$ is generated/consumed at its limiting rate $\bar{\phi}_{\especieDet}^{lim}\varphi_{\lim}(x)$ on the surfaces of the electrodes. If $\bar{\phi}_{\especieDet}^{lim}>0$ on the surface of the working electrode, the species $\especieDet$ is being generated at the working and consumed at the counter. For this reason, the concentration of species $\especieDet$ must reach zero at the counter electrode. Due to the average property in Remark \ref{problema:obs:c-promedio} and the fact that $c_{\especieDet}(x,z,+\infty) - \bar{c}_{\especieDet,0}$ is odd symmetric with respect to $x=W/2$, the concentration of species $\especieDet$ at the working electrode must reach the saturation value $2\bar{c}_{\especieDet,0}$ (see Fig. \ref{corriente-limite:fig:concentracion-limite}). Then for all $x$ in the working electrode
\begin{displaymath}
  \bar{c}_{\especieDet,0} = \bar{\phi}_{\especieDet}^{\lim} \sum_{n\:\mathrm{odd}} \mathcal{I}_n \set{\frac{\varphi_{\lim}}{D}} G_\phi\bigpar{H,n^2\frac{\pi^{2}}{W^{2}}} \cos(n\pi x/W)
\end{displaymath}

Analogously, when $\bar{\phi}_{\especieDet}^{\lim} < 0$ on the working electrode, the species $\especieDet$ is consumed at the working electrode and generated at the counter electrode. For this reason, the concentration of species $\especieDet$ reaches $0$ at the working electrode (see Fig. \ref{corriente-limite:fig:concentracion-limite}). Then for all $x$ on the working electrode
\begin{displaymath}
  -\bar{c}_{\especieDet,0} = \bar{\phi}_{\especieDet}^{\lim} \sum_{n\:\mathrm{odd}} \mathcal{I}_n \set{\frac{\varphi_{\lim}}{D}} G_\phi\bigpar{H,n^2\frac{\pi^{2}}{W^{2}}} \cos(n\pi x/W)
\end{displaymath}

Note that in the limiting current case and $\bar{c}_{\especieDet,0} \leq c_{0}/2 = (\bar{c}_{\mathcal{O},0} + \bar{c}_{\mathcal{R},0})/2$, the concentration of species $\especieDet$ must reach zero on one of the electrodes, whereas the other species may not. The only way that both species can reach zero concentration at the electrodes is when $\bar{c}_{\especieDet,0} = \bar{c}_{\mathcal{O},0} = \bar{c}_{\mathcal{R},0}$, in this case $\bar{c}_{\especieDet,0} = c_{0}/2 = (\bar{c}_{\mathcal{O},0} + \bar{c}_{\mathcal{R},0})/2$.

\subsubsection{Calculation of the limiting current bounds}
In summary, when the limiting current circulates in the cell, the following relation holds
\begin{displaymath}
  \frac{\bar{c}_{\especieDet,0}}{\abs{\bar{\phi}_{\especieDet}^{\lim}}} = \sum_{n\:\mathrm{odd}} \mathcal{I}_n \set{\frac{\varphi_{\lim}}{D}} G_\phi\bigpar{H,n^2\frac{\pi^{2}}{W^{2}}} \cos(n\pi x/W)
\end{displaymath}
and after integrating along the working electrode, one obtains
\begin{equation}
  \frac{\bar{c}_{\especieDet,0}w}{\abs{\bar{\phi}_{\especieDet}^{\lim}}} = \sum_{n\:\mathrm{odd}} \mathcal{I}_n \set{\frac{\varphi_{\lim}}{D}} G_{\phi}\bigpar{H,n^2\frac{\pi^{2}}{W^{2}}} \frac{\sin(n\pi w/W)}{(n\pi/W)}
  \label{calculo-corriente-limite:ec:phi_lim-exacta}
\end{equation}
Note that the summation at the right hand side of the equation must be positive, since $\bar{c}_{\especieDet,0}\geq 0$ and $\abs{\bar{\phi}_{\especieDet}^{\lim}}\geq 0$.

By finding an upper bound for the previous summation, it is possible to obtain a lower bound for the generation rate of the species $\especieDet$, and thus, the limiting current. 

Considering that the integral $\mathcal{I}_n \set{\varphi_{\lim}/{D}}$ can be bounded by
\begin{equation}
  \abs{\mathcal{I}_n \set{\frac{\varphi_{\lim}}{D}}} < \frac{4}{W} \int_0^w \frac{\varphi_{\lim}(x)}{D} \ud{x} = \frac{4w}{DW}
  \label{calculo-corriente-limite:ec:integral-cota}
\end{equation}
$\forall n$ odd since $\varphi_{\lim}(x)\geq 0$ in the working electrode, and by taking into account the series
\begin{displaymath}
  \sum_{n\:\mathrm{odd}} \frac{1}{n^2} = \frac{3}{4} \zeta(2)
\end{displaymath}
where $\zeta(s)$ is the Riemann's zeta function and $\zeta(2) = \pi^2/6$, then the summation can be bounded by
\begin{eqnarray*}
  && \abs{ \sum_{n\:\mathrm{odd}} \mathcal{I}_n \set{\frac{\varphi_{\lim}}{D}} G_{\phi}\bigpar{H,n^2\frac{\pi^{2}}{W^{2}}} \frac{\sin(n\pi w/W)}{(n\pi/W)} }\\
  && < \frac{4w}{DW} \sum_{n\:\mathrm{odd}} \frac{\abs{\sin(n\pi w/W)}}{(n\pi/W)^2 \tanh(n\pi H/W)}\\
  && < \frac{4w}{DW} \frac{3/4\:\: \zeta(2)}{(\pi/W)^{2} \tanh(\pi H/W)} =
  \frac{wW}{2D\tanh(\pi H/W)}
\end{eqnarray*}

With the previous result, the limiting generation rate of the determinant species is bounded from below by 
\begin{displaymath}
  \abs{\bar{\phi}_{\especieDet}^{\lim}} > \frac{2D}{W}\tanh\bigpar{\pi\frac{H}{W}} \bar{c}_{\especieDet,0}
\end{displaymath}

\subsection{Relative error respect to bulk concentration}
\label{apendice:concentracion-relativa}
Here the relative error of the steady state concentration with respect to the bulk concentration is obtained, by using the steady state result in Eq. (\ref{apendice:ec:idae-c-zx8}). This relative error can provide a quantitative criterion to determine when a semi-infinite geometry cell can be approximated by a finite geometry cell.

Consider Eq. (\ref{apendice:ec:idae-c-zx8}) at the furthest location from the electrodes $z=H$, then the error of the concentration of the determinant species $\especieDet$ in steady state with respect to its bulk concentration is given by
\begin{displaymath}
  \abs{c_{\especieDet}(x,H,+\infty) - \bar{c}_{\especieDet,0}}
    = \abs{\bar{\phi}_{\especieDet} \sum_{n\:\mathrm{odd}} \mathcal{I}_n\set{\frac{\varphi}{D}} G_{\phi}\bigpar{0,n^2\frac{\pi^{2}}{W^{2}}} \cos(n\pi x/W)}
\end{displaymath}
By taking the upper bound for the limiting current $\abs{\bar{\phi}_{\especieDet}} \leq \lim_{H/W \to \infty} \abs{\bar{\phi}_{\especieDet}^{\lim}}$ in Remark \ref{corriente-limite:nota:cota-superior} and the bound for the integral $\abs{\mathcal{I}_{n}\set{\varphi/D}}<4w/(DW)$ in Eq. (\ref{calculo-corriente-limite:ec:integral-cota}), the deviation from the bulk concentration must be bounded by
\begin{eqnarray*}
  \abs{c_{\especieDet}(x,0,+\infty) - \bar{c}_{\especieDet,0}}
  &\leq& \lim_{H/W \to \infty} \abs{\bar{\phi}_{\especieDet}^{\lim}} \cdot
    \frac{4w}{DW} \sum_{n\:\mathrm{odd}} G_{\phi}\bigpar{0,n^2\frac{\pi^{2}}{W^{2}}}\\
  &\leq& \lim_{H/W \to \infty} \abs{\bar{\phi}_{\especieDet}^{\lim}} \cdot
    \frac{4w}{\pi D} \sum_{n\:\mathrm{odd}} \frac{\pi}{W} G_{\phi}\bigpar{0,n^2\frac{\pi^{2}}{W^{2}}}
\end{eqnarray*}

\begin{figure}
  \centering
  \subfigure[][]{
    \includegraphics{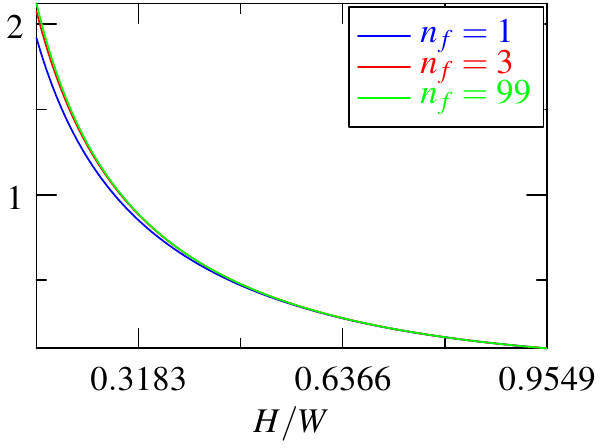}
    \label{apendice:fig:sum_Gphi_1}
  }
  \subfigure[][]{
    \includegraphics{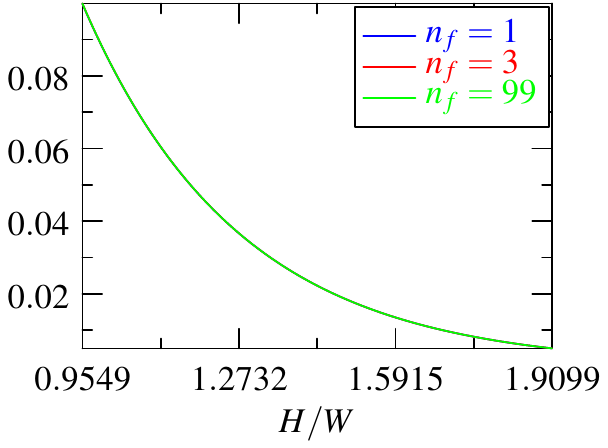}
    \label{apendice:fig:sum_Gphi_2}
  }
  \subfigure[][]{
    \includegraphics{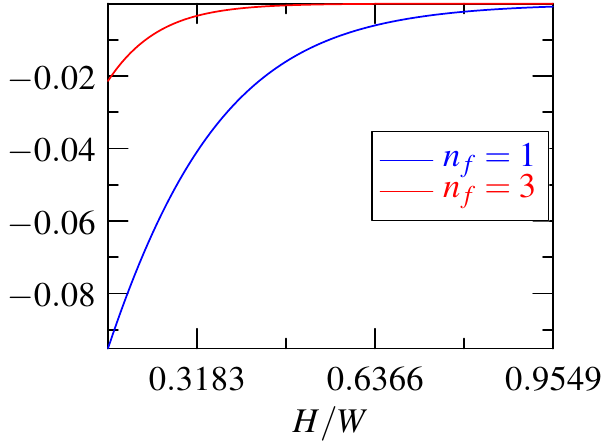}
    \label{apendice:fig:err_rel_sum_Gphi_1}
  }
  \subfigure[][]{
    \includegraphics{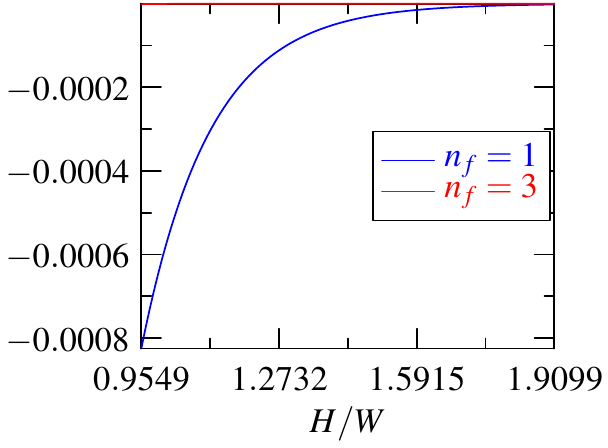}
    \label{apendice:fig:err_rel_sum_Gphi_2}
  }
  \caption{Top pictures: Plot of $\sum_{n=1}^{n_f} (n\sinh(n\pi H/W))^{-1}$ for odd $n$, \subref{apendice:fig:sum_Gphi_1} $H/W\in[\num{0,5}/\pi,\, 3/\pi]$ and \subref{apendice:fig:sum_Gphi_2} $H/W\in[3/\pi,\, 6/\pi]$. Bottom pictures: Plot of the relative error for $n_{f}=\set{1,3}$ with respect to $n_{f}=99$ considering \subref{apendice:fig:err_rel_sum_Gphi_1} $H/W\in[\num{0,5}/\pi,\, 3/\pi]$ and \subref{apendice:fig:err_rel_sum_Gphi_2} $H/W\in[3/\pi,\, 6/\pi]$.}
  \label{apendice:fig:sumGphi}
\end{figure}
Since the approximation in Eq. (\ref{corriente-limite:ec:morf}) has an accuracy within $1\%$
\begin{displaymath}
  \lim_{H/W\to +\infty} \abs{\bar{\phi}_{\especieDet}^{\lim}} \approx 
    \frac{\pi D\bar{c}_{\especieDet,0}}{2w\ln(\frac{4W}{\pi w})},\quad \mathrm{for}\: w/W\leq 1/4
\end{displaymath}
and since the following approximation has an accuracy within $1\%$ due to the Fig. \ref{apendice:fig:sumGphi}
\begin{displaymath}
  \frac{\pi}{W} G_{\phi}\bigpar{0,n^2\frac{\pi^{2}}{W^{2}}} = \sum_{n\:\mathrm{odd}} \frac{1}{n\sinh(n\pi H/W)} \approx \frac{1}{\sinh(\pi H/W)},\quad \mathrm{for}\: H/W\geq 1/\pi
\end{displaymath}
then, at the furthest location from the electrodes ($z=H$), the relative error of the concentration with respect to the bulk concentration is given by
\begin{displaymath}
  \abs{\frac{c_{\especieDet}(x,H,+\infty) - \bar{c}_{\especieDet,0}}{\bar{c}_{\especieDet,0}}}
    \lesssim 2\bigcuad{\ln\bigpar{\frac{4W}{\pi w}} \sinh\bigpar{\pi \frac{H}{W}}}^{-1},\quad \mathrm{for}\: w/W\leq 1/4\: \mathrm{and}\: H/W\geq 1/\pi
\end{displaymath}

\section{Details on the simulations}
\label{apendice:simulaciones}
Numerical solutions to the time dependent PDE in Eq. (\ref{problema:ec:c-PDE}) where found by using the software package \textsf{Comsol 3.5a}. The relative and the absolute tolerances of the solver were set to $\num{e-3}$ and $\num{e-4}$ respectively. The linear system solver was left as \texttt{Direct (UMFPACK)}. The time stepping method was configured to use the option \texttt{BDF} (backward differentiation formula). The option \texttt{Steps taken by the solver} was left as \texttt{Free} and the solver output times were stored.

\begin{figure}
  \centering
  \includegraphics{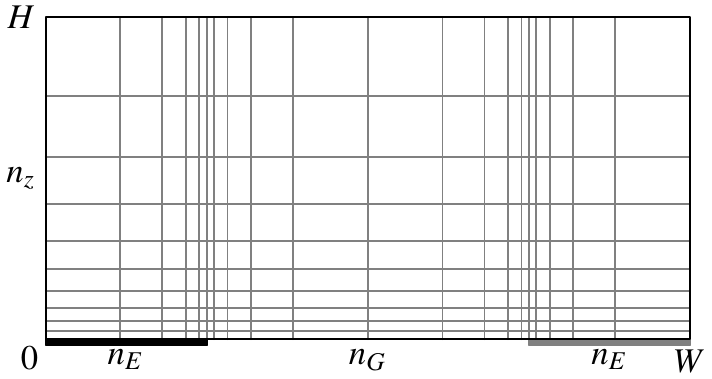}
  \caption{Sketch of the exponential mapped mesh applied to the unit cell for perfoming the simulations.}
  \label{apendice:fig:exp_malla}
\end{figure}
Due to the discontinuities present at the edges of the electrodes \cite{Strutwolf2005}, an exponential mapped mesh was selected to provide higher resolution near the electrodes and their edges \marca{(see \cite[Chapter 7]{Britz2005} for more details on this kind of mesh)}. \marca{The sketch in Fig. \ref{apendice:fig:exp_malla} shows a exponentially expanding mesh in the $x$ and $z$ dimensions with a total of $n_{x}\times n_{z}$ elements, such that $n_{x} = 2n_{E} + n_{G}$, where $n_{G}$ is the number of elements along the gap and $n_{E}$ is the number of elements along each electrode. For visualization purposes, the mesh shows $n_{x}\times n_{z} = 20\times 10$ elements ($2n_{E} = n_{G} = 10$).}

\subsection{Example of a current controlled electrochemical cell}
\label{apendice:s31}
The mesh used in this simulation was obtained by incrementaly refining its resolution until the first three decimal places of the maximum concentration in steady state did not change. The number of \marca{elements} used in the $x$-axis and the $z$-axis was $n_{x}=144$ and $n_{z}=71$ respectively, \marca{with} $2n_{E} = n_{G} = 72$. The smallest division used on both axes was $\approx\SI{0,5}{\micro\metre}$, and the ratio between the \marca{largest} and smallest \marca{elements} on both axes was $2$.

The errors in Figs. \ref{ejemplo-celda:fig:conc-electrodos-error} and \ref{ejemplo-celda:fig:conc-error} were obtained by taking the difference between the simulation results and their theoretical counterparts. For calculating the theoretical values used in Fig. \ref{ejemplo-celda:fig:conc-electrodos-error} partial sums approximations were used for the concentration in Eqs. (\ref{apendice:ec:idae-Dc-xzt}) and (\ref{apendice:ec:idae-Db_n-z8}) with $z=0$
\begin{subeqnarray}
  && c_{\especie}(x,z,t) \approx \\
  && \sum_{n=1}^{n_f} \mathcal{I}_{n}\set{\frac{\phi_{\especie}}{D}} 
    \bigcuad{\int_{0^{-}}^{Dt} g_{\phi}(H-z,u) \mathrm{e}^{-n^{2}\pi^{2}u/W^{2}} \ud{u}}
    \cos(n\pi x/W) + \bar{c}_{\especie,0} \nonumber\\
  && \mathcal{I}_{n}\set{\phi_{\especie}} =
    \frac{2j}{n\pi F n_{e}} [\sin(n\pi w/W) - \sin(n\pi (1-w/W))]\\
  && \int_{0^{-}}^{Dt} g_{\phi}(H-z,u) \mathrm{e}^{-n^{2}\pi^{2}u/W^{2}} \ud{u} \approx\\
  && \frac{1}{H}\bigcuad{\frac{1-\mathrm{e}^{-n^{2}\pi^{2}Dt/W^{2}}}{n^{2}\pi^{2}/W^{2}} + 2\sum_{k=1}^{kf} (-1)^{k} \frac{1-\mathrm{e}^{-(n^{2}/W^{2}+k^{2}/H^{2})\pi^{2}Dt}}{(n^{2}/W^{2}+k^{2}/H^{2})\pi^{2}} \cos(k\pi(1-z/H))} \nonumber
\end{subeqnarray}
where the initial concentration of species $\especie$ correspond to $c_{\especie,0}(x,z) = \bar{c}_{\especie,0}$, the lower boundary condition is $\abs{\Delta\phi_{\especie}(x)} = \abs{\phi_{\especie}(x)} = \marca{j/(F n_e)}$ on the electrodes and $j$ is the current density (constant in $t$ and uniform on the surface of the electrodes but with opposite sign). For calculating the steady state values used in Figs. \ref{ejemplo-celda:fig:conc-electrodos-error} and \ref{ejemplo-celda:fig:conc-error} partial sums approximations were used for the steady state concentration in Eq. (\ref{apendice:ec:idae-c-xz8_original}) and (\ref{apendice:ec:idae-G_phi})
\begin{subeqnarray}
  c_{\especie}(x,z,+\infty) &\approx&
    \sum_{n=1}^{n_f} \mathcal{I}_{n}\set{\frac{\phi_{\especie}}{D}} G_{\phi}(H-z,n^{2}\pi^{2}/W^{2}) \cos(n\pi x/W) + \bar{c}_{\especie,0}\\
  G_{\phi}(z,s) &=& \frac{\cosh(\sqrt{s}\,z)}{\sqrt{s}\sinh(\sqrt{s}\,H)}
\end{subeqnarray}

\begin{figure*}
  \centering
  \subfigure[][]{
    \includegraphics{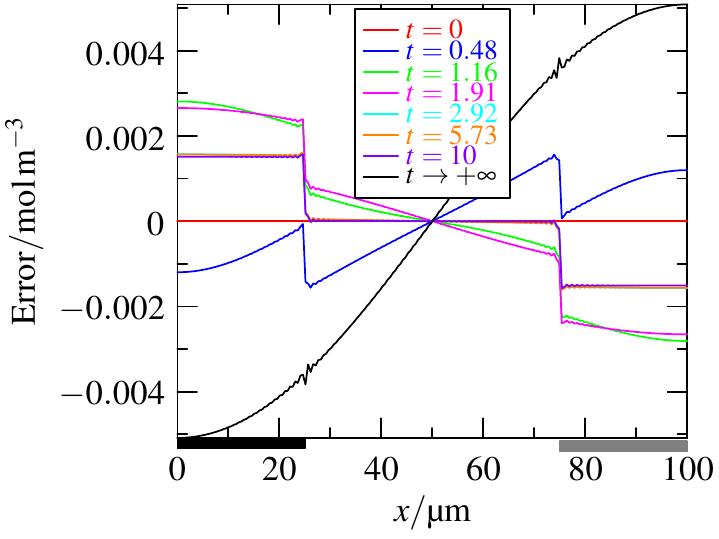}
    \label{apendice:fig:conc-electrodos-error_nf201_kf100}
  }
  \subfigure[][]{
    \includegraphics{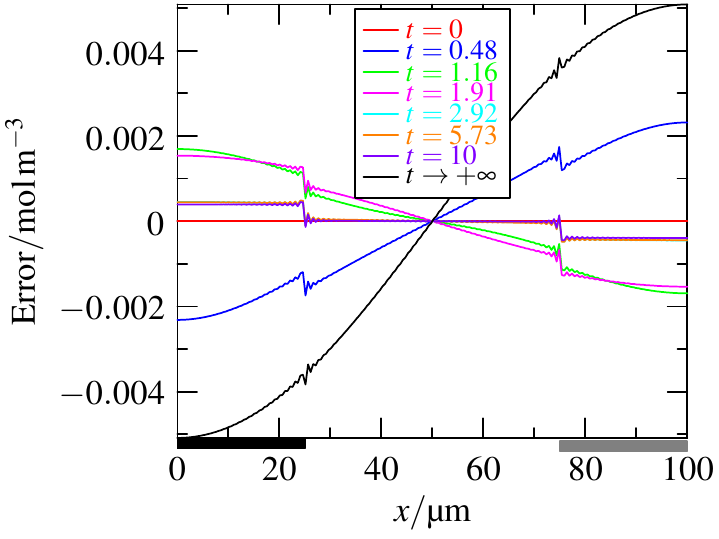}
    \label{apendice:fig:conc-electrodos-error_nf201_kf400}
  }
  \caption{Error of the simulated concentrations at $z=0$ with respect to their theoretical counterparts. \subref{apendice:fig:conc-electrodos-error_nf201_kf100} Colored lines: Differences for times between $t=0$ and $t=\SI{10}{\second}$ using partial sums up to 201 and 100 for $n$ and $k$ respectively. Black line: Difference between the simulation for $t=\SI{5,73}{\second}$ with respect to the theoretical steady state using partial sums up to $n=201$. \subref{apendice:fig:conc-electrodos-error_nf201_kf400} Colored lines: Differences for times between $t=0$ and $t=\SI{10}{\second}$ using partial sums up to 201 and 400 for $n$ and $k$ respectively. Black line: Difference between the simulation for $t=\SI{5,73}{\second}$ with respect to the theoretical steady state using partial sums up to $n=201$. \marca{Note that the line for the parameter value $t=\SI{2,92}{s}$ is overlapped by the parameter value $t=\SI{5,73}{s}$ in figures (a) and (b).}}
  \label{apendice:fig:conc-electrodos}
\end{figure*}

Fig. \ref{apendice:fig:conc-electrodos} shows the error of the simulated concentrations with respect to their theoretical counterparts for different values of $t$. When comparing Figs. \ref{apendice:fig:conc-electrodos}\subref{apendice:fig:conc-electrodos-error_nf201_kf100} and \ref{apendice:fig:conc-electrodos}\subref{apendice:fig:conc-electrodos-error_nf201_kf400}, it can be noticed that the magnitude of differences and the discontinuities depend strongly on the partial sums in the index $k$. The higher the upper value of the index $k$ in the partial sums, the smaller and more continuous are the differences.

\begin{figure*}
  \centering
  \subfigure[][]{
    \includegraphics{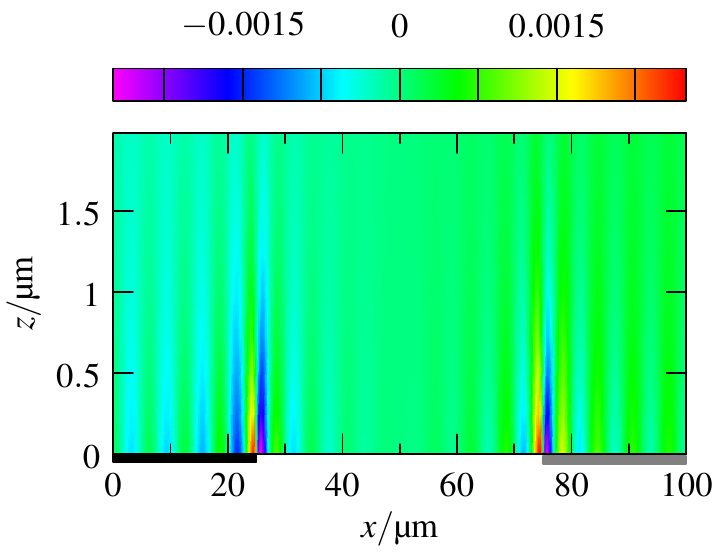}
    \label{apendice:fig:conc-error_nf_bajo}
  }
  \subfigure[][]{
    \includegraphics{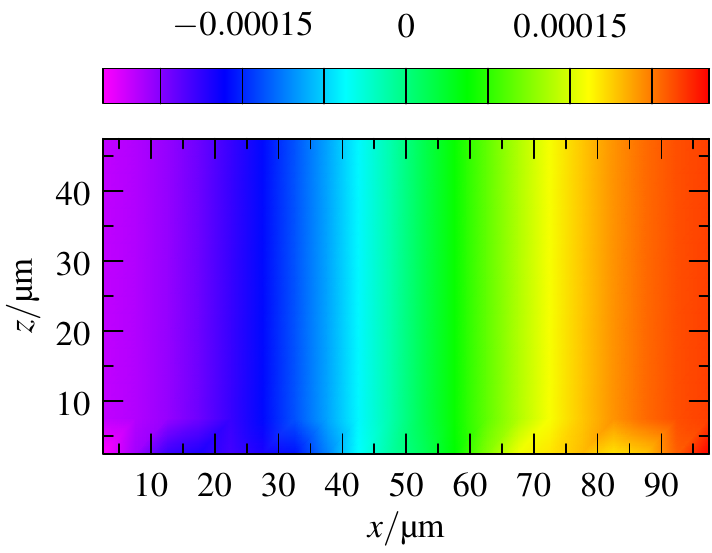}
    \label{apendice:fig:conc-error_nf_bajo_completo}
  }
  
  \subfigure[][]{
    \includegraphics{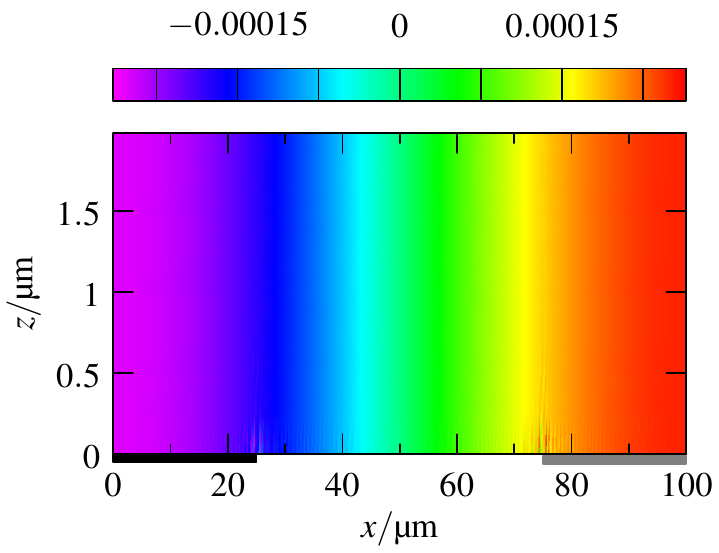}
    \label{apendice:fig:conc-error_nf_alto}
  }
  \subfigure[][]{
    \includegraphics{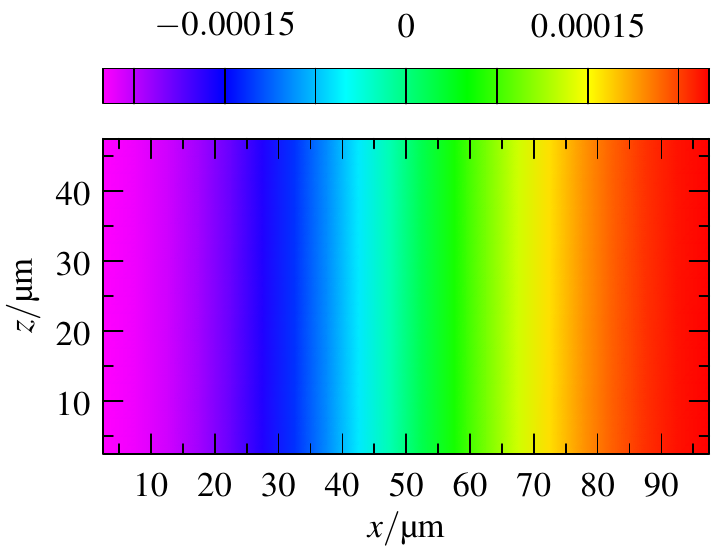}
    \label{apendice:fig:conc-error_nf_alto_completo}
  }
  \caption{Error of the simulated concentration at $t=\SI{10}{\second}$ with respect to the theoretical concentration in steady state. \subref{apendice:fig:conc-error_nf_bajo} and \subref{apendice:fig:conc-error_nf_bajo_completo} Error when using partial sums up to $n=31$ \marca{for the regions \subref{apendice:fig:conc-error_nf_bajo} $z\leq\SI{2}{\micro\metre}$ and \subref{apendice:fig:conc-error_nf_bajo_completo} $z>\SI{2}{\micro\metre}$}. \subref{apendice:fig:conc-error_nf_alto} and \subref{apendice:fig:conc-error_nf_alto_completo} Error when using partial sums up to \marca{$n=301$ for the regions \subref{apendice:fig:conc-error_nf_alto} $z\leq\SI{2}{\micro\metre}$ and \subref{apendice:fig:conc-error_nf_alto_completo} $z>\SI{2}{\micro\metre}$}.}
  \label{apendice:fig:conc-error}
\end{figure*}

Fig. \ref{apendice:fig:conc-error} shows the error of the simulated concentration at $t=\SI{10}{\second}$ with respect to the theoretical concentration in steady state. \marca{Figs. \ref{apendice:fig:conc-error}\subref{apendice:fig:conc-error_nf_bajo} and \ref{apendice:fig:conc-error}\subref{apendice:fig:conc-error_nf_bajo_completo} clearly show that the error near the surface of the electrodes ($z\leq\SI{2}{\micro\metre}$) is higher than far from the electrodes ($z>\SI{2}{\micro\metre}$) when using partial sums up to $31$ for $n$. Figs. \ref{apendice:fig:conc-error}\subref{apendice:fig:conc-error_nf_alto} and \ref{apendice:fig:conc-error}\subref{apendice:fig:conc-error_nf_alto_completo} show that the error looks similar in both regions of the unit cell when using partial sums up to $301$ for $n$, however it is still possible to find small perturbations very near the edges of the electrodes.} When comparing \ref{apendice:fig:conc-error}\subref{apendice:fig:conc-error_nf_bajo} and \ref{apendice:fig:conc-error}\subref{apendice:fig:conc-error_nf_alto}\marca{, that is when $z\leq\SI{2}{\micro\metre}$}, it is noticed that \marca{there are} oscillations located mainly near the surface of the electrodes and they depend on the partial sums in the index $n$. The higher the upper value of the index $n$, the smaller the amplitude of the oscillations and the smaller its wavelength.

\subsection{Effect of the cell geometry in the concentration profile}
\label{apendice:s32}

As shown in Fig. \ref{ejemplo-concentracion:fig:concentracion-lejana}, this example required a combination of values for $H/W\in\{\num{0,4}/\pi,\, \num{0,6}/\pi,\, \num{0,8}/\pi,\, 1/\pi,\,$ $ 2/\pi,\, 3/\pi,\, 4/\pi\}$ and $w/W\in\set{\num{0,1},\, \num{0,2},\, \num{0,3},\, \num{0,4}}$ requiring a total of $7\times 4 = 28$ simulations. 

\marca{The exponential meshes used in all the simulations are detailed in Table \ref{apendice:tab:exp_malla_info_s32}. The smallest element near the electrode edge (in the $x$ and $z$ dimensions) was selected as $\num{0,005}$, the ratio between the largest and smallest elements in the $x$-axis was chosen as $4$ and the growth factor along the $z$-axis was fixed to $\approx\num{1,015}$.}

Finally, accuracy was checked by incrementaly refining around the meshes obtained for $H/W=3/\pi$, leading to 3 decimal places when comparing the maximum concentration in steady state.
\begin{table}
  \centering
  \subtable[][Partition of the $x$-axis]{
    \begin{tabular}{l|rlcccc}
      $w/W$ & $n_{x}$ & $\delta_{x}$ & $n_{E}$ & $R_{E}$ & $n_{G}$ & $R_{G}$\\ \hline
      \num{0,1}  & \num{104} & \num{0,005} & \num{15} & \num{1,71} & $37\times 2$ & \num{4}\\
      \num{0,2}  & \num{100} & \num{0,005} & \num{22} & \num{3}    & $28\times 2$ & \num{4}\\
      \num{0,25} & \num{92}  & \num{0,005} & \num{23} & \num{4}    & $23\times 2$ & \num{4}\\
      \num{0,3}  & \num{100} & \num{0,005} & \num{28} & \num{4}    & $22\times 2$ & \num{3}\\
      \num{0,4}  & \num{104} & \num{0,005} & \num{37} & \num{4}    & $15\times 2$ & \num{1,71}
    \end{tabular}
  }
  \subtable[][Partition of the $z$-axis]{
    \begin{tabular}{c|ccl}
      $H/W$ & $\delta_{z}$ & $n_{z}$ & $R_{z}$\\ \hline
      $\num{0,2}/\pi$ & \num{0,005} & 12  & \num{1,18}\\
      $\num{0,4}/\pi$ & \num{0,005} & 22  & \num{1,38}\\
      $\num{0,6}/\pi$ & \num{0,005} & 30  & \num{1,55}\\
      $\num{0,8}/\pi$ & \num{0,005} & 38  & \num{1,75}\\
      $\num{1}/\pi$   & \num{0,005} & 45  & \num{1,95}\\
      $\num{2}/\pi$   & \num{0,005} & 71  & \num{2,9}\\
      $\num{3}/\pi$   & \num{0,005} & 92  & \num{4}\\
      $\num{4}/\pi$   & \num{0,005} & 104 & \num{4,8}\\
      $\num{5}/\pi$   & \num{0,005} & 116 & \num{5,77}
    \end{tabular}
  }
  \caption{Information for partitioning the $x$ and $z$ axes. $\delta_{i}$, $n_{i}$ and $R_{i}$ stands for the size of the smallest element, number of elements and ratio between the largest and smallest element respectively. The index $i\in\set{x,z,E,G}$ stands for the $x$ and $z$ axes, and electrode and gap respectively.}
  \label{apendice:tab:exp_malla_info_s32}
\end{table}

\subsection{Effect of the cell geometry in the limiting current}
\label{apendice:s33}
\marca{Meshes} with $H/W=3/\pi$ and different values of $w/W$ where first constructed and refined incrementally until obtaining two decimal places of accuracy for the generation rate on the electrodes. The simulated generation rate was contrasted with its theoretical value given in Eqs. (\ref{corriente-limite:ec:aoki}) and (\ref{corriente-limite:ec:morf}) to ensure the accuracy of two decimal places. \marca{Under the same conditions}, the concentration presented three decimal places of accuracy, checked by looking the maximum concentration on the electrodes when doing the incremental refinement of the mesh.

\marca{The smallest element in the $x$ and $z$ axes thus obtained was $\num{0,0025}$, the ratio between the largest and smallest elements in the $x$-axis was fixed to $4$ and the growth factor along the $z$-axis was fixed to $\approx\num{1,008}$.} The information in Table \ref{apendice:tab:exp_malla_info_s33} sumarizes the meshes used for all the simulations in this example.

\begin{table}
  \centering
  \subtable[][Partition of the $x$-axis]{
    \begin{tabular}{l|rlcccc}
      $w/W$ & $n_{x}$ & $\delta_{x}$ & $n_{E}$ & $R_{E}$ & $n_{G}$ & $R_{G}$\\ \hline
      \num{0,1}  & \num{208} & \num{0,0025} & \num{30} & \num{1,73} & $74\times 2$ & \num{4}\\
      \num{0,2}  & \num{198} & \num{0,0025} & \num{44} & \num{3}    & $55\times 2$ & \num{4}\\
      \num{0,25} & \num{184} & \num{0,0025} & \num{46} & \num{4}    & $46\times 2$ & \num{4}\\
      \num{0,3}  & \num{198} & \num{0,0025} & \num{55} & \num{4}    & $44\times 2$ & \num{3}\\
      \num{0,4}  & \num{208} & \num{0,0025} & \num{74} & \num{4}    & $30\times 2$ & \num{1,73}
    \end{tabular}
  }
  \subtable[][Partition of the $z$-axis]{
    \begin{tabular}{c|ccl}
      $H/W$ & $\delta_{z}$ & $n_{z}$ & $R_{z}$\\ \hline
      $\num{0,2}/\pi$ & \num{0,0025} & 23  & \num{1,2}\\
      $\num{0,4}/\pi$ & \num{0,0025} & 43  & \num{1,4}\\
      $\num{0,6}/\pi$ & \num{0,0025} & 60  & \num{1,6}\\
      $\num{0,8}/\pi$ & \num{0,0025} & 75  & \num{1,8}\\
      $\num{1}/\pi$   & \num{0,0025} & 88  & \num{2}\\
      $\num{2}/\pi$   & \num{0,0025} & 140 & \num{3}\\
      $\num{3}/\pi$   & \num{0,0025} & 176 & \num{4}\\
      $\num{4}/\pi$   & \num{0,0025} & 204 & \num{5}\\
      $\num{5}/\pi$   & \num{0,0025} & 228 & \num{6}
    \end{tabular}
  }
  \caption{Information for partitioning the $x$ and $z$ axes. $\delta_{i}$, $n_{i}$ and $R_{i}$ stands for the size of the smallest element, number of elements and ratio between the largest and smallest element respectively. The index $i\in\set{x,z,E,G}$ stands for the $x$ and $z$ axes, and electrode and gap respectively.}
  \label{apendice:tab:exp_malla_info_s33}
\end{table}

\section{General results for a finite height cell}
\label{apendice:extension}
\marca{The results shown in this appendix intend to extend the results of \ref{apendice:idae-resultados} to a cell of which its concentration profile has periodic (and non-periodic where possible) left/right boundary conditions. The main tool used to extend these results is the Fourier transform which has been applied in the $x$-coordinate. The extended results include uniformity properties for the total and average concentrations and expressions for the concentration profile in terms of the generation of the species.}
\subsection{Model of a finite height cell}
\label{apendice:modelo}
Consider an electrochemical cell that contains the electrochemical species $\especie\in\set{\mathcal{O},\mathcal{R}}$, which react at the surface of the electrodes according to
\begin{equation}
  \mathcal{O} + n_{e}\ce{e- <=>} \mathcal{R}
  \label{apendice:ec:reaccion}
\end{equation} 
This cell extends horizontally between $-\infty<x<\infty$, vertically between $0 \leq z \leq H$ and in depth between $0\leq y \leq L$. The concentration of the species is assumed not to depend on the $y$-coordinate, therefore the diffusive transport of the species can be modeled in 2D by
\begin{subeqnarray}
  \label{apendice:ec:c-PDE}
  \frac{1}{D} \parderiv{c_{\especie}}{t}(x,z,t) &=& \parderiv{^2c_{\especie}}{x^2}(x,z,t) + \parderiv{^2c_{\especie}}{z^2}(x,z,t)
    \slabel{apendice:ec:c-difusion}\\
  c_{\especie}(x,z,0^{-}) &=& c_{\especie,0}(x,z)
    \slabel{apendice:ec:c-ci}\\
  \parderiv{c_{\especie}}{z}(x,H,t) &=& 0
    \slabel{apendice:ec:c-arriba}
\end{subeqnarray}
Where $D$ is the diffusion coeficient of both species.

The left and right boundaries of the cell have been left unspecified, but it is assumed that the concentrations of both species have Fourier transform in the variable $x$.

The electrodes in this cell are located at the bottom boundary and they can have any configuration or arrangement provided that the 2D symmetry is maintained. The behaviour of this bottom boundary is written here in terms of the generation rate (flux) of the species $\phi_{\especie}(x,t)$
\begin{equation}
  -D\parderiv{c_{\especie}}{z}(x,0,t) = \phi_{\especie}(x,t)
    \label{apendice:ec:c-abajo}
\end{equation}

The generation rate of the species is such that $\phi_{\mathcal{R}}(x,t) = -\phi_{\mathcal{O}}(x,t)$ on the surface of the electrodes and $\phi_{\mathcal{R}}(x,t) = \phi_{\mathcal{O}}(x,t) = 0$ out of the electrodes.

In this model it is assumed that the initial concentration $c_{\especie,0}(x,z)$ of species $\especie$ comes from a previous steady state, and thus it must satisfy
\begin{subeqnarray}
  \label{apendice:ec:ci-PDE}
  0 &=& \parderiv{^2c_{\especie,0}}{x^2}(x,z) + \parderiv{^2c_{\especie,0}}{z^2}(x,z)\\
  \parderiv{c_{\especie,0}}{z}(x,H) &=& 0\\
  -D\parderiv{c_{\especie,0}}{z}(x,0) &=& \phi_{\especie,0}(x)
\end{subeqnarray}
Where $\phi_{\mathcal{R},0}(x) = -\phi_{\mathcal{O},0}(x)$ on the surface of the electrodes and $\phi_{\mathcal{R},0}(x) = \phi_{\mathcal{O},0}(x) = 0$ out of the electrodes.

The left and right boundaries of the initial concentration have been left unspecified, but it is assumed that the initial concentrations of both species have Fourier transform in the variable $x$.

\subsection{Total concentration in a finite height cell}
\label{apendice:c_total}
Here it is shown that the total concentration in a finite height cell with 2D symmentry is constant in $t$ and uniform in $x$, provided that the diffusion coefficient of both species is the same and that the sum of the generation rates of both species is zero on the electrodes.

Consider the cell described in \ref{apendice:modelo}. Due Eq. (\ref{apendice:ec:ci-PDE}) and the fact that $\phi_{\mathcal{R},0}(x) = -\phi_{\mathcal{O},0}(x)$ on the surface of the electrodes, the total initial concentration in the cell $c(x,z,0^{-}) = c_{\mathcal{O},0}(x,z) + c_{\mathcal{R},0}(x,z)$ must satisfy
\begin{eqnarray*}
  0 &=& \parderiv{^2c}{x^2}(x,z,0^{-}) + \parderiv{^2c}{z^2}(x,z,0^{-})\\
  \parderiv{c}{z}(x,0,0^{-}) &=& \parderiv{c}{z}(x,H,0^{-}) = 0
\end{eqnarray*}
\marca{By applying Fourier transform $\mathcal{F}_{x}\set{\cdot}$ to the $x$-coordinate one obtains}
\begin{eqnarray*}
  0 &=& -\omega_{x}^{2}\hat{c}(\omega_{x},z,0^{-}) + \parderiv{^2\hat{c}}{z^2}(\omega_{x},z,0^{-})\\
  \parderiv{\hat{c}}{z}(\omega_{x},0,0^{-}) &=& \parderiv{\hat{c}}{z}(\omega_{x},H,0^{-}) = 0
\end{eqnarray*}
\marca{where $\omega_{x}$ is the Fourier variable and $\hat{c} = \mathcal{F}_{x}\set{c}$. The solution of the equation in $\omega_{x}$-domain is $\hat{c}(\omega_{x},z,0^{-}) = 0$ for $\omega_{x}\neq 0$, therefore $c(x,z,0^{-}) = c(z,0^{-})$ must not depend on $x$. Now using $c(z,0^{-})$ to solve the original equation in $x$-domain, one obtains that} the total initial concentration must be a real constant $c(x,z,0^{-}) = c(z,0^{-}) = c_0$ \marca{which does not depend on $(x,z)$}.

Due to the previous result, Eqs. (\ref{apendice:ec:c-PDE}), (\ref{apendice:ec:c-abajo}) and the fact that $\phi_{\mathcal{R}}(x,t) = -\phi_{\mathcal{O}}(x,t)$ on the electrodes; the total concentration $c(x,z,t) = c_{\mathcal{O}}(x,z,t) + c_{\mathcal{R}}(x,z,t)$ must satisfy
\begin{eqnarray*}
  \frac{1}{D} \parderiv{c}{t}(x,z,t) &=& \parderiv{^2c}{x^2}(x,z,t) + \parderiv{^2c}{z^2}(x,z,t)\\
  c(x,z,0^{-}) &=& c_0\\
  \parderiv{c}{z}(x,0,t) &=& \parderiv{c}{z}(x,H,t) = 0
\end{eqnarray*}
\marca{Taking the difference $\Delta c(x,z,t) = c(x,z,t) - c_0$ and applying Fourier transform\footnote{A circumflex is used to denote a Fourier transform: $\hat{f} = \mathcal{F}_{x}\set{f}$} $\mathcal{F}_{x}\set{\cdot}$ in $x$ and Laplace transform\footnote{A capital letter is used to denote a Laplace transform: $F = \mathcal{L}_{t}\set{f}$} $\mathcal{L}_{t}\set{\cdot}$ in $t$ one obtains}
\begin{eqnarray*}
  \bigpar{\frac{s}{D} + \omega_{x}^{2}} \Delta\hat{C}(\omega_{x},z,s) &=& \parderiv{^2 \Delta\hat{C}}{z^2}(\omega_{x},z,s)\\
  \parderiv{\Delta\hat{C}}{z}(\omega_{x},0,s) &=& \parderiv{\Delta\hat{C}}{z}(\omega_{x},H,s) = 0
\end{eqnarray*}
\marca{where $\omega_{x}$ is the Fourier variable, $s$ is the Laplace variable and $\Delta\hat{C} = \mathcal{L}_{t}\mathcal{F}_{x}\set{\Delta c}$. The solution of this equation is given by $\Delta\hat{C}(\omega_{x},z,s) = 0$ $\forall \omega_{x}$, therefore } the total concentration in the cell must be a real constant
\begin{equation}
  c(x,z,t)=c_{0}, \mbox{ for all } x,z \mbox{ and } t\geq 0
  \label{apendice:ec:c_total-cte-uniforme}
\end{equation}

\begin{teorema}
  The result in Eq. (\ref{apendice:ec:c_total-cte-uniforme}) holds for any bottom boundary \marca{condition} (concentration, generation rate or a combination of both), provided that an electrochemical cell like the one described in \ref{apendice:modelo} is considered. This is because both electrochemical species have the same diffusion coefficient $D$ and because the cell must satisfy $\phi_{\mathcal{R},0}(x) = -\phi_{\mathcal{O},0}(x)$ due to Eq. (\ref{apendice:ec:reaccion}), independently on how the bottom boundary \marca{condition} is chosen.
\end{teorema}

\subsection{Concentration in a finite height cell in terms of the generation rate of species}
\label{apendice:c-xzt}
The concentration of species $\especie$ in a finite height cell with 2D symmetry is obtained in terms of the generation rate of species. As byproduct, properties for the `horizontal average concentration' of species $\especie$ are obtained.

\subsubsection{Initial concentration}
Consider the cell described in \ref{apendice:modelo}. An expression for the initial concentration $c_{\especie,0}(x,z) = c_{\especie,0}^{h}(x,z) + c_{\especie,0}^{p}(x,z)$ can be found when obtaining separately the homogeneous and particular solutions $c_{\especie,0}^{h}(x,z)$ and $c_{\especie,0}^{p}(x,z)$.

The homogeneous solution $c_{\especie,0}^{h}(x,z)$ must satisfy Eq. (\ref{apendice:ec:ci-PDE}) with $\phi_{\especie,0}(x) = 0$. \marca{By applying Fourier transform $\mathcal{F}_{x}\set{\cdot}$ to the $x$-coordinate one obtains}
\begin{eqnarray*}
  0 &=& -\omega_{x}^{2}\hat{c}_{\especie,0}^{h}(\omega_{x},z) + \parderiv{^2\hat{c}_{\especie,0}^{h}}{z^2}(\omega_{x},z)\\
  \parderiv{\hat{c}_{\especie,0}^{h}}{z}(\omega_{x},0) &=& \parderiv{\hat{c}_{\especie,0}^{h}}{z}(\omega_{x},H) = 0
\end{eqnarray*}
\marca{where $\hat{c}_{\especie,0}^{h} = \mathcal{F}_{x}\set{c_{\especie,0}^{h}}$. The solution of the equation in $\omega_{x}$-domain is $\hat{c}(\omega_{x},z) = 0$ for $\omega_{x}\neq 0$, therefore $c_{\especie,0}^{h}(x,z) = c_{\especie,0}^{h}(z)$ must not depend on $x$. Using $c_{\especie,0}^{h}(z)$ to solve the equation in $x$-domain, one obtains that} $c_{\especie,0}^{h}(x,z)$ must be  a real constant $\bar{c}_{\especie,0}$
\begin{displaymath}
  c_{\especie,0}^{h}(x,z) = \bar{c}_{\especie,0}
\end{displaymath}

The particular solution $c_{\especie,0}^{p}(x,z)$ must satisfy Eq. (\ref{apendice:ec:ci-PDE}) with $\phi_{\especie,0}(x) \neq 0$. This solution can be found by applying the Fourier transform $\mathcal{F}_{x}\set{\cdot}$ in the $x$-coordinate
\begin{eqnarray*}
  0 &=& -\omega_{x}^{2} \hat{c}_{\especie,0}^{p}(\omega_{x},z) + \parderiv{^{2}\hat{c}_{\especie,0}^{p}}{^{2}z}(\omega_{x},z)\\
  \parderiv{\hat{c}_{\especie,0}^{p}}{z}(\omega_{x},H) = 0, &&
    -D\parderiv{\hat{c}_{\especie,0}^{p}}{z}(\omega_{x},0) = \hat{\phi}_{\especie,0}(\omega_{x})
\end{eqnarray*}
Then the solution in Fourier domain is obtained by analyzing separately the cases for $\omega_{x}=0$ and $\omega_{x}\neq 0$. In the case $\omega_{x}\neq 0$ the result is
\begin{displaymath}
  \hat{c}_{\especie,0}^{p}(\omega_{x},z) = G_{\phi}(H-z,\omega_{x}^{2})\, \frac{\hat{\phi}_{\especie,0}}{D}(\omega_{x})
\end{displaymath}
Where $\omega_{x}$ is the Fourier variable, $\hat{c}_{\especie,0}^{p} = \mathcal{F}_{x}\set{c_{\especie,0}^{p}}$, $\hat{\phi}_{\especie,0} = \mathcal{F}_{x}\set{\phi_{\especie,0}}$ and
\begin{equation}
  G_{\phi}(z,s) = \frac{\cosh(\sqrt{s}\, z)}{\sqrt{s}\, \sinh(\sqrt{s}\, H)}
  \label{apendice:ec:G_phi}
\end{equation}
In the case $\omega_{x} = 0$, the particular solution must be a real constant and the generation rate must be zero
\begin{displaymath}
  \hat{c}_{\especie,0}^{p}(0,z) = k_{\especie,0},\quad \hat{\phi}_{\especie,0}(0) = 0
\end{displaymath}
If $\lim_{\omega_{x}\to 0} \hat{c}_{\especie,0}^{p}(\omega_{x},z)$ exists, then $k_{\especie,0}$ can be chosen to allow $\hat{c}_{\especie,0}^{p}(\omega_{x},z)$ to be continuous in $\omega_{x} = 0$
\begin{displaymath}
   \hat{c}_{\especie,0}^{p}(0,z) = 
   \lim_{\omega_{x}\to 0} G_{\phi}(H-z,\omega_{x}^2)\frac{\hat{\phi}_{\especie,0}}{D}(\omega_{x}) = 
   \lim_{\omega_{x}\to 0} \frac{\hat{\phi}_{\especie,0}(\omega_{x})/D}{\omega_{x}\sinh(\omega_{x} H)}
\end{displaymath}

Since the Fourier transform of the homogeneous solution is $\hat{c}_{\especie,0}^{h}(\omega_{x},z) = 2\pi\delta(\omega_{x})\, \bar{c}_{\especie,0}$, then the results for the Fourier transform of the initial concentration $\hat{c}_{\especie,0}(\omega_{x},z) = \hat{c}_{\especie,0}^{h}(\omega­_{x},z) + \hat{c}_{\especie,0}^{p}(\omega_{x},z)$ can be sumarized in the following theorems

\begin{teorema}
  Consider a finite height cell like the one described in \ref{apendice:modelo}. If the following limit exists
  \begin{displaymath}
    \lim_{\omega_{x}\to 0} \frac{\hat{\phi}_{\especie,0}(\omega_{x})/D}{\omega_{x}\sinh(\omega_{x} H)}
  \end{displaymath}
  then the initial concentration of species $\especie$ can be expressed by its Fourier transform 
  \begin{equation}
    \hat{c}_{\especie,0}(\omega_{x},z) = 2\pi\delta(\omega_{x})\,\bar{c}_{\especie,0} + G_{\phi}(H-z,\omega_{x}^2)\, \frac{\hat{\phi}_{\especie,0}}{D}(\omega_{x})
    \label{apendice:ec:ci-wz}
  \end{equation}
\end{teorema}

\begin{teorema}
  If a cell like the one described in \ref{apendice:modelo} is considered then the horizontal integral of the initial generation rate is zero
  \begin{equation}
    \hat{\phi}_{\especie,0}(0) = \int_{-\infty}^{\infty} \phi_{\especie,0}(x) \ud{x} = 0
    \label{apendice:ec:phi-w0}
  \end{equation}
  and if the following limit exists
  \begin{displaymath}
    \lim_{\omega_{x}\to 0} \frac{\hat{\phi}_{\especie,0}(\omega_{x})/D}{\omega_{x}\sinh(\omega_{x} H)}
  \end{displaymath}
  then the horizontal integral of the initial concentration does not depend on $z$  and corresponds to
  \begin{equation}
    \int_{-\infty}^{\infty} c_{\especie,0}(x,z) - \bar{c}_{\especie,0} \ud{x} = \lim_{\omega_{x}\to 0} \frac{\hat{\phi}_{\especie,0}(\omega_{x})/D}{\omega_{x}\sinh(\omega_{x} H)}
    \label{apendice:ec:ci-w0z}
  \end{equation}
  This result is obtained by evaluating the following Fourier transform in $\omega_{x}=0$
  \begin{displaymath}
    \hat{c}_{\especie,0}(\omega_{x},z) - 2\pi\delta(\omega_{x})\,\bar{c}_{\especie,0} = \int_{-\infty}^{\infty} [c_{\especie,0}(x,z) - \bar{c}_{\especie,0}]\, \mathrm{e}^{-i \omega_{x} x} \ud{x}
  \end{displaymath}
\end{teorema}

\begin{observacion}
  Note that if $\phi_{\especie,0}(x)$ is periodic with period $P$, then the following limit is immediately satisfied
  \begin{displaymath}
    \lim_{\omega_{x}\to 0} \frac{\hat{\phi}_{\especie,0}(\omega_{x})/D}{\omega_{x}\sinh(\omega_{x} H)} = 0
  \end{displaymath}
  This is because $\hat{\phi}_{\especie,0}(\omega_{x}) = 0$ for all $\omega_{x}\neq n\, 2\pi/P$ (with integer $n$) due to the fact that $\phi_{\especie,0}(x)$ is periodic. Therefore, the product 
  \begin{displaymath}
    \frac{\hat{\phi}_{\especie,0}(\omega_{x})/D}{\omega_{x}\sinh(\omega_{x} H)} = 0
  \end{displaymath}
  for all $\omega_{x}\neq n\, 2\pi/P$ (with integer $n$).
\end{observacion}

\subsubsection{Change in concentration}
Similar results can be found for the change in concentration $\Delta c_{\especie}(x,z,t)$ of species $\especie$ when considering the cell defined in \ref{apendice:modelo}.

By substracting Eqs. (\ref{apendice:ec:c-PDE}), (\ref{apendice:ec:c-abajo}) with (\ref{apendice:ec:ci-PDE}), and later by applying the Fourier transform $\mathcal{F}_{x}\set{\cdot}$ in $x$ and Laplace transform $\mathcal{L}_{t}\set{\cdot}$ in $t$, one obtains
\begin{subeqnarray}
  \label{apendice:ec:Dc-PDE}
  \bigpar{\frac{s}{D} + \omega_{x}^2} \Delta\hat{C}_{\especie}(\omega_{x},z,s) &=& 
    \parderiv{^2 \Delta\hat{C}_{\especie}}{z^2}(\omega_{x},z,s)\\
  \parderiv{\Delta\hat{C}_{\especie}}{z}(\omega_{x},H,s) &=& 0\\
  -D\parderiv{\Delta\hat{C}_{\especie}}{z}(\omega_{x},0,s) &=& \Delta\hat{\Phi}(\omega_{x},s)
\end{subeqnarray}
where $\omega_{x}$ is the Fourier variable, $s$ is the Laplace variable and
\begin{eqnarray*}
  \Delta\hat{C}_{\especie} = \mathcal{L}_{t}\mathcal{F}_{x}\set{\Delta c_{\especie}} &&
    \Delta c_{\especie}(x,z,t) = c_{\especie}(x,z,t) - c_{\especie,0}(x,z)\\
  \Delta\hat{\Phi}_{\especie} = \mathcal{L}_{t}\mathcal{F}_{x}\set{\Delta \phi_{\especie}} &&
    \Delta \phi_{\especie}(x,t) = \phi_{\especie}(x,t) - \phi_{\especie,0}(x)
\end{eqnarray*}

This system of equations has no singularity in $\omega_{x}=0$ and therefore requires no special analysis. Then the change in concentration of species $\especie$ is given by
\begin{equation}
  \Delta\hat{C}_{\especie}(\omega_{x},z,s) = G_{\phi}\bigpar{H-z,\frac{s}{D} + \omega_{x}^2}\, \frac{\Delta\hat{\Phi}_{\especie}}{D}(\omega_{x},s)
  \label{apendice:ec:Dc-wzs}
\end{equation}
By using the time-scaling and frecuency-shifting properties of the Laplace transform and by taking the inverse Laplace transform, the solution is obtained only in the Fourier domain
\begin{equation}
  \Delta\hat{c}_{\especie}(\omega_{x},z,t) = g_{\phi}(H-z,Dt)\mathrm{e}^{-\omega_{x}^{2}Dt}D * \frac{\Delta\hat{\phi}_{\especie}}{D}(\omega_{x},t)
  \label{apendice:ec:Dc-wzt}
\end{equation}
where $*$ represents the time convolution, $\Delta\hat{c}_{\especie} = \mathcal{F}_{x}\set{\Delta c_{\especie}}$ and $g_{\phi} = \mathcal{L}_{t}^{-1}\set{G_{\phi}}$ equals to
\begin{equation}
  g_{\phi}(z,t) = \frac{1}{H} \bigcuad{1 + 2\sum_{k=1}^{\infty} (-1)^{k} \mathrm{e}^{-k^2\pi^2 t/H^2} \cos(k\pi z/H)}
  \label{apendice:ec:g_phi}
\end{equation}
and it is given by the Laplace inverse of $G_{\phi}$ in Eq. (\ref{apendice:ec:G_phi}). This Laplace inverse corresponds to the $4^{th}$ \emph{elliptic theta function} and it can be obtained from tables such as \cite[p.218]{Schiff1999} or \cite[Eq. (20.10.5)]{dlmf2010}.

\begin{teorema}
  The concentration of the species $\especie\in\set{\mathcal{O},\mathcal{R}}$ in a finite height cell, as described in \ref{apendice:modelo}, is given by
  \begin{displaymath}
    c_{\especie}(x,z,t) = \Delta c_{\especie}(x,z,t) + c_{\especie,0}(x,z)
  \end{displaymath}
  where Fourier transform (in the $x$-coordinate) of the change in concentration $\Delta c_{\especie}$ is determined by Eqs. (\ref{apendice:ec:Dc-wzt}) and (\ref{apendice:ec:g_phi}), and the Fourier transform (in the $x$-coordinate) of the initial concentration $c_{\especie,0}$ is given by Eq. (\ref{apendice:ec:ci-wz}), provided that the following limit exists
  \begin{displaymath}
    \lim_{\omega_{x}\to 0} \frac{\hat{\phi}_{\especie,0}(\omega_{x})/D}{\omega_{x}\sinh(\omega_{x} H)}
  \end{displaymath}
\end{teorema}

\begin{teorema}
  If a cell like the one described in \ref{apendice:modelo} is considered and the following limit exists
  \begin{displaymath}
    \lim_{\omega_{x}\to 0} \frac{\hat{\phi}_{\especie,0}(\omega_{x})/D}{\omega_{x}\sinh(\omega_{x} H)}
  \end{displaymath}
  then the horizontal integral of the concentration $c_{\especie}(x,z,t) = \Delta c_{\especie}(x,z,t) + c_{\especie,0}(x,z)$ is given by
  \begin{eqnarray}
    && \int_{-\infty}^{\infty} c_{\especie}(x,z,t) - \bar{c}_{\especie,0} \ud{x} =
    \label{apendice:ec:c-w0zt}\\
    && \lim_{\omega_{x}\to 0} \frac{\hat{\phi}_{\especie,0}(\omega_{x})/D}{\omega_{x}\sinh(\omega_{x} H)} + g_{\phi}(H-z,Dt) * \int_{-\infty}^{\infty} \phi_{\especie}(x,t) \ud{x}
    \nonumber
  \end{eqnarray}
  This result is due to evaluation of Eq. (\ref{apendice:ec:Dc-wzt}) in $\omega_{x}=0$ and the incorporation of Eqs. (\ref{apendice:ec:phi-w0}) and (\ref{apendice:ec:ci-w0z}).
\end{teorema}

\subsection{Concentration in a periodic finite height cell in terms of the generation rate of species}
\label{apendice:c-xzt-periodica}
The results in the previous sections hold for any finite height cell that allows Fourier transform in the $x$-coordinate of the concentrations. In this section the results are restricted only to periodic finite height cells (periodic in the $x$-coordinate), but could be extended to the non-periodic case provided a careful analysis is done for $\omega_{x}=0$ and its neighbourhood.

If a finite height cell as described in \ref{apendice:modelo} is considered, where the concentrations of the species are periodic in the $x$-coordinate with period $P$, then the generation rate at the bottom boundary $\phi_{\especie}(x,t) = \Delta\phi_{\especie}(x,t) + \phi_{\especie,0}(x)$ must be periodic as well and can be written in terms of its Fourier series (note that the property in Eq. (\ref{apendice:ec:phi-w0}) has been considered)
\begin{subeqnarray}
  \label{apendice:ec:phi-xzt}
  \Delta\phi_{\especie}(x,t) &=& \sum_{n=-\infty}^{\infty} \mathcal{E}_{n}\set{\Delta\phi_{\especie}}(t)\, \mathrm{e}^{\mathrm{i}n2\pi x/P}\\
  \phi_{\especie,0}(x) &=& \sum_{n\neq 0} \mathcal{E}_{n}\set{\phi_{\especie,0}}\, \mathrm{e}^{\mathrm{i}n2\pi x/P}\\
  \mathcal{E}_{n}\set{\cdot} &=& \frac{1}{P} \int_{-P/2}^{P/2} \set{\cdot}\, \mathrm{e}^{-\mathrm{i}n2\pi x/P} \ud{x}
\end{subeqnarray}
where the Fourier transforms are given by
\begin{subeqnarray}
  \label{apendice:ec:phi-wzt}
  \Delta\hat{\phi}_{\especie}(\omega_{x},t) &=& \sum_{n=-\infty}^{\infty} \mathcal{E}_{n}\set{\Delta\phi_{\especie}}(t)\, 2\pi\delta\bigpar{\omega_{x} - n\frac{2\pi}{P}}\\
  \hat{\phi}_{\especie,0}(\omega_{x}) &=& \sum_{n\neq 0} \mathcal{E}_{n}\set{\phi_{\especie,0}}\, 2\pi\delta\bigpar{\omega_{x} - n\frac{2\pi}{P}}
\end{subeqnarray}
and the Laplace-Fourier transform is given by
\begin{equation}
  \Delta\hat{\Phi}_{\especie}(\omega_{x},s) = \sum_{n=-\infty}^{\infty} \mathcal{E}_{n}\set{\Delta\Phi_{\especie}}(s)\, 2\pi\delta\bigpar{\omega_{x} - n\frac{2\pi}{P}}
  \label{apendice:ec:Dphi-wzs}
\end{equation}

\subsubsection{Concentration in steady state}
For the case of a periodic finite height cell, the values in steady state ($t\to +\infty$) can be computed from the Eqs. (\ref{apendice:ec:ci-wz}) and (\ref{apendice:ec:Dc-wzs}).

In the case of the initial concentration, Eq. (\ref{apendice:ec:ci-wz}) is considered and the generation rate in Eq. (\ref{apendice:ec:phi-wzt}) is applied. Later, by using the Fourier inverse the following result is obtained
\begin{eqnarray*}
  c_{\especie,0}(x,z) &=& \bar{c}_{\especie,0} + \sum_{n\neq 0} a_{n}^{\especie,0}(z)\, \mathrm{e}^{\mathrm{i}n2\pi x/P}\\
  a_{n}^{\especie,0}(z) &=& G_{\phi}\bigpar{H-z,n^{2}\frac{4\pi^{2}}{P^{2}}} \mathcal{E}_{n}\set{\frac{\phi_{\especie,0}}{D}}
\end{eqnarray*}

In the case of the change in concentration, take first Eq. (\ref{apendice:ec:Dc-wzs}) and apply the periodic generation rate in Eq. (\ref{apendice:ec:Dphi-wzs})
\begin{eqnarray*}
  \Delta\hat{C}_{\especie}(\omega_{x},z,s) &=& \sum_{n=-\infty}^{\infty} \Delta A_{n}^{\especie}(z,s)\, 2\pi\delta\bigpar{\omega_{x} - n\frac{2\pi}{P}}\\
  \Delta A_{0}^{\especie}(z,s) &=& G_{\phi}\bigpar{H-z,\frac{s}{D}} \mathcal{E}_{0}\set{\frac{\Delta\Phi_{\especie}}{D}}(s)\\
  \Delta A_{n}^{\especie}(z,s) &=& G_{\phi}\bigpar{H-z,\frac{s}{D} + n^{2}\frac{4\pi^{2}}{P^{2}}} \mathcal{E}_{n}\set{\frac{\Delta\Phi_{\especie}}{D}}(s)
\end{eqnarray*}
later, apply the final value theorem of the Laplace transform and take the Fourier inverse to obtain
\begin{eqnarray*}
  \Delta c_{\especie}(x,z,+\infty) &=& \sum_{n=-\infty}^{\infty} \Delta a_{n}^{\especie}(z,+\infty)\, \mathrm{e}^{\mathrm{i}n2\pi x/P}\\
  \Delta a_{0}^{\especie}(z,+\infty) &=& \lim_{s\to 0} s\, G_{\phi}\bigpar{H-z,\frac{s}{D}} \lim_{s\to 0} s\, \frac{1}{s} \mathcal{E}_{0}\set{\frac{\Delta\Phi_{\especie}}{D}}(s)\\
  \Delta a_{n}^{\especie}(z,+\infty) &=& G_{\phi}\bigpar{H-z,n^{2}\frac{4\pi^{2}}{P^{2}}} \mathcal{E}_{n}\set{\frac{\Delta\phi_{\especie}}{D}}(+\infty)
\end{eqnarray*}
Notice that
\begin{eqnarray*}
  \lim_{s\to 0} s\, G_{\phi}\bigpar{H-z,\frac{s}{D}} &=& \frac{D}{H}\\
  \lim_{s\to 0} s\, \frac{1}{s} \mathcal{E}_{0}\set{\frac{\Delta\Phi_{\especie}}{D}}(s) &=&
  \int_{0-}^{+\infty} \frac{1}{P} \int_{-P/2}^{P/2} \frac{\Delta\phi_{\especie}}{D}(x,t) \ud{x} \ud{t}
\end{eqnarray*}
and due to Eq. (\ref{apendice:ec:phi-w0})
\begin{displaymath}
  \int_{-P/2}^{P/2} \phi_{\especie,0}(x) \ud{x} = \int_{-\infty}^{+\infty} \phi_{\especie,0}(x) \ud{x} = 0
\end{displaymath}

\begin{teorema}
  If a cell like the one described in \ref{apendice:modelo} is considered, the concentration is periodic with period $P$ and the following integral converges
  \begin{displaymath}
    \frac{1}{H} \int_{0-}^{+\infty} \frac{1}{P} \int_{-P/2}^{P/2} \phi_{\especie}(x,t) \ud{x} \ud{t}
  \end{displaymath}
  then the steady state concentration of species $\especie$ $c_{\especie}(x,z,+\infty) = \Delta c_{\especie}(x,z,+\infty) + c_{\especie,0}(x,z)$ is given by
  \begin{eqnarray*}
    c_{\especie}(x,z,+\infty) &=& \bar{c}_{\especie,0} + \sum_{n=-\infty}^{\infty} a_{n}^{\especie}(z,+\infty)\, \mathrm{e}^{\mathrm{i}n2\pi x/P}\\
    a_{0}^{\especie}(z,+\infty) &=& \frac{1}{H} \int_{0-}^{+\infty} \frac{1}{P} \int_{-P/2}^{P/2} \phi_{\especie}(x,t) \ud{x} \ud{t}\\
    a_{n}^{\especie}(z,+\infty) &=& G_{\phi}\bigpar{H-z,n^{2}\frac{4\pi^{2}}{P^{2}}} \mathcal{E}_{n}\set{\frac{\phi_{\especie}}{D}}(+\infty)
  \end{eqnarray*}
  where the following limit is a necessary condition
  \begin{displaymath}
    \lim_{t\to +\infty} \frac{1}{P} \int_{-P/2}^{P/2} \phi_{\especie}(x,t) \ud{x} = 0
  \end{displaymath}
\end{teorema}

\subsubsection{Concentration for constant generation rate}
For the case of a periodic finite height cell, the response in time domain can be obtained from the Eq. (\ref{apendice:ec:Dc-wzt}). In order to avoid problems of convergence, it is assumed that the following integral is zero
\begin{displaymath}
  \frac{1}{P} \int_{-P/2}^{P/2} \phi_{\especie}(x) \ud{x} = \frac{1}{P} \int_{-P/2}^{P/2} \Delta \phi_{\especie}(x) \ud{x} = \mathcal{E}_{0}\set{\Delta \phi_{\especie}} = 0
\end{displaymath}

After evaluating Eq. (\ref{apendice:ec:Dc-wzt}) by using Eq. (\ref{apendice:ec:phi-wzt}) and taking the Fourier inverse, one obtains the following result in time domain
\begin{eqnarray*}
  \Delta c_{\especie}(x,z,t) &=& \sum_{n\neq 0} \Delta a_{n}^{\especie}(z,t) \mathrm{e}^{\mathrm{i}n2\pi x/P}\\
  \Delta a_{n}^{\especie}(z,t) &=& g_{\phi}(H-z,Dt) \mathrm{e}^{-n^{2}4\pi^{2}Dt/P^{2}} D * \mathcal{E}_{n}\set{\frac{\Delta \phi_{\especie}}{D}}(t)
\end{eqnarray*}
where $*$ is the time convolution and $g_{\phi}$ is defined in Eq. (\ref{apendice:ec:g_phi}).

If the generation rate of the species $\especie$ is constant in $t$ $\Delta\phi_{\especie}(x,t)=\Delta\phi_{\especie}(x)$, then the following integral is also constant in $t$ $\mathcal{E}_n\set{\Delta\phi_{\especie}/D}(t)=\mathcal{E}_n\set{\Delta\phi_{\especie}/D}$. By this mean the coefficient $\Delta a_n^{\especie}(z,t)$ is obtained simply by integration
\begin{displaymath}
  \Delta a_{n}^{\especie}(z,t) = \int_{0^{-}}^{Dt} g_{\phi}(H-z,u) \mathrm{e}^{-n^{2}4\pi^{2}u/P^{2}} \ud{u} \cdot \mathcal{E}_{n}\set{\frac{\Delta \phi_{\especie}}{D}}
\end{displaymath}

After a `sufficiently long time' ($t\to +\infty$), the dynamics of the periodic cell is complete and the concentration reaches the steady state described by the coefficient
\begin{displaymath}
  \Delta a_{n}^{\especie}(z,+\infty) = G_{\phi}\bigpar{H-z,n^{2}\frac{4\pi^{2}}{P^{2}}} \cdot \mathcal{E}_{n}\set{\frac{\Delta \phi_{\especie}}{D}}
\end{displaymath}

\begin{teorema}
  Consider a cell like the one described in \ref{apendice:modelo}, where the concentration is periodic with period $P$, the generation rate $\phi_{\especie}(x)$ is constant in $t$ and the following integral holds
  \begin{displaymath}
    \frac{1}{P} \int_{-P/2}^{P/2} \phi_{\especie}(x) \ud{x} = 0
  \end{displaymath}
  then the time response of the change in concentracion is given by
  \begin{eqnarray*}
    \Delta c_{\especie}(x,z,t) &=& \sum_{n\neq 0} \Delta a_{n}^{\especie}(z,t) \mathrm{e}^{\mathrm{i}n2\pi x/P}\\
    \Delta a_{n}^{\especie}(z,t) &=& \int_{0^{-}}^{Dt} g_{\phi}(H-z,u) \mathrm{e}^{-n^{2}4\pi^{2}u/P^{2}} \ud{u} \cdot \mathcal{E}_{n}\set{\frac{\Delta \phi_{\especie}}{D}}
  \end{eqnarray*}
  And after a sufficiently long time ($t\to +\infty$), it converges to steady state.
\end{teorema}

The time required to reach the steady state (after the step in generation rate has been applied) can be calculated from $\Delta c_{\especie}(x,z,t)$, which consists of a double summation (in the indexes $n$ and $k$) 
\begin{eqnarray*}
  && \Delta c_{\especie}(x,z,t) = \sum_{n\neq 0} \Delta a_n^{\especie}(z,t) \mathrm{e}^{\mathrm{i}n2\pi x/P}\\
  && \Delta a_n^{\especie}(z,t) = \mathcal{E}_n\set{\frac{\Delta\phi_{\especie}}{D}} \cdot \frac{D}{H} \left[ \frac{1-\mathrm{e}^{-n^{2}4\pi^{2}Dt/P^{2}}}{n^{2}4\pi^{2}D/P^{2}} + \right.\\
  && \left. 2 \sum_{k=1}^\infty (-1)^k \frac{1-\mathrm{e}^{-[n^{2}4\pi^{2}/P^{2}+k^{2}\pi^{2}/H^{2}]Dt}}{[n^{2}4\pi^{2}/P^{2}+k^{2}\pi^{2}/H^{2}]D} \cos(k\pi (H-z)/H) \right]
\end{eqnarray*}
Note that this double summation consists of exponential modes
\begin{displaymath}
  \exp\bigpar{-\bigcuad{n^{2}\frac{4\pi^{2}}{P^{2}} + k^{2}\frac{\pi^{2}}{H^{2}}}Dt} = \exp\bigpar{-\bigcuad{n^2+k^2\frac{P^{2}}{4H^{2}}} \frac{4\pi^{2}}{P^{2}} Dt}
\end{displaymath}
and their evolution in time is determined by the time parameter
\begin{displaymath}
  \tau_{\phi} = \frac{P^{2}}{4\pi^{2}D}
\end{displaymath}
The exponential modes with lower $n$ and $k$ indexes decay slowly with time, and together with $\tau_{\phi}$ they are determinant in the time required to reach steady state.

Consider the case when $2H/P<1/2$ and $t>\tau_\phi$. \marca{Here the exponential modes with $\abs{n}\geq 1$ and $\abs{k}\geq 1$ may be considered extinct since they are bounded by}
\begin{displaymath}
  \exp\bigpar{-\bigcuad{n^{2} + k^{2}\frac{P^{2}}{4H^{2}}} \frac{4\pi^{2}}{P^{2}}Dt} < \exp(-[n^{2} + 4k^{2}]) \geq \mathrm{e}^{-5} \approx \num{0,7}\%
\end{displaymath}
Because of the fast convergence of the previous double summation (due to the squared indexes $n^2$ and $k^2$), the terms with large $n$ and $k$ can be neglected so the error with respect to the steady state can be approximated by using only $\abs{n}=1$
\begin{displaymath}
  \Delta c_{\especie}(x,z,t) - \Delta c_{\especie}(x,z,\infty) \approx -\frac{1}{H} \frac{\mathrm{e}^{-4\pi^{2}Dt/P^{2}}}{4\pi^{2}/P^{2}} \sum_{\abs{n}=1} \mathcal{E}_{n}\set{\frac{\Delta\phi_{\especie}}{D}} \mathrm{e}^{\mathrm{i}n2\pi x/P}
\end{displaymath}
This means that the error with respect to the steady state decays following the exponential mode $\exp(-4\pi^{2}Dt/P^{2})$, therefore this can be used as an indicator for the time to reach the steady state $T_{ss}^\phi$
\begin{displaymath}
  T_{ss}^{\phi} \propto \tau_{\phi}
\end{displaymath}
$T_{ss}^\phi$ can be chosen as $4\tau_\phi$, $5\tau_\phi$ or $6\tau_\phi$, since the exponential mode $\exp(-4\pi^{2}Dt/P^{2})$ decays to approximately $\num{1,8}\%$, $\num{0,7}\%$ and $\num{0,2}\%$ respectively.

\begin{teorema}
  Consider a cell like the one described in \ref{apendice:modelo}, where the concentration is periodic with period $P$, the generation rate $\phi_{\especie}(x)$ is constant in $t$ and the following integral holds
  \begin{displaymath}
    \frac{1}{P} \int_{-P/2}^{P/2} \phi_{\especie}(x) \ud{x} = 0
  \end{displaymath}
  If the ratio $2H/P<1/2$ is satisfied, then for $t>\tau_\phi$ the error with respect to the steady state decays following the exponential mode $\exp(-4\pi^{2}Dt/P^{2})$ and it is given by
  \begin{displaymath}
    \Delta c_{\especie}(x,z,t) - \Delta c_{\especie}(x,z,\infty) \approx -\frac{1}{H} \frac{\mathrm{e}^{-4\pi^{2}Dt/P^{2}}}{4\pi^{2}/P^{2}} \sum_{\abs{n}=1} \mathcal{E}_{n}\set{\frac{\Delta\phi_{\especie}}{D}} \mathrm{e}^{\mathrm{i}n2\pi x/P}
  \end{displaymath}
  and thus the time to reach steady state $T_{ss}^{\phi}$ is proportional to
  \begin{displaymath}
    T_{ss}^{\phi} \propto \tau_{\phi} = \frac{P^{2}}{4\pi^{2}D}
  \end{displaymath}
\end{teorema}